\documentclass[prd,preprint,superscriptaddress,showpacs,amsmath,amssymb]{revtex4}
\ifx\pdfoutput\undefined
\usepackage[dvips]{graphicx}
\else
\usepackage[pdftex]{graphicx}
\usepackage{epstopdf}
\fi

\numberwithin{equation}{section}

\setkeys{Gin}{width=.9\columnwidth, height=.37\textheight, keepaspectratio}
\usepackage{dcolumn}% Align table columns on decimal point
\usepackage{bm}% bold math
\usepackage{url}
\usepackage{xspace}

\newenvironment{cfigure}[1][tbp]{\begin{figure}[#1]\centering}{\end{figure}}

\newcommand{\fig}[1]{Fig.~\ref{#1}}
\newcommand{\tab}[1]{Table~\ref{#1}}
\newcommand{\Fig}[1]{Figure~\ref{#1}}
\newcommand{\Tab}[1]{Table~\ref{#1}}

\newcommand{\twotag}  {2-tag\xspace}
\newcommand{\onetagt} {1-tag(T)\xspace}
\newcommand{\onetagl} {1-tag(L)\xspace}
\newcommand{\zerotag} {0-tag\xspace}

\newcommand{\ppbar}  {\ensuremath{p\bar{p}}\xspace}
\newcommand{\ttbar}  {\ensuremath{t\bar{t}}\xspace}
\newcommand{\bbbar}  {\ensuremath{b\bar{b}}\xspace}
\newcommand{\ccbar}  {\ensuremath{c\bar{c}}\xspace}
\newcommand{\qqbar}  {\ensuremath{q\bar{q}}\xspace}

\newcommand{\zee}    {\ensuremath{Z\rightarrow e^{+}e^{-}}}

\newcommand{\pte}    {\ensuremath{p_{T}^{e}}}
\newcommand{\ptmu}   {\ensuremath{p_{T}^{\mu}}}
\newcommand{\ptnu}   {\ensuremath{p_{T}^{\nu}}}

\newcommand{\ptjet}  {\ensuremath{p_{T}^{\text{jet}}}}

\newcommand{\ptquark}{\ensuremath{p_{T}^{\text{quark}}}}

\newcommand{\etajet} {\ensuremath{\eta^{\text{jet}}}\xspace}
\newcommand{\etjet}  {\ensuremath{E_{T}^{\text{jet}}}\xspace}
\newcommand{\etaevt} {\ensuremath{\eta_{\text{evt}}}\xspace}

\newcommand{\et}     {\ensuremath{E_{T}}\xspace}

\newcommand{\pt}     {\ensuremath{p_{T}}\xspace}

\newcommand{\pz}     {\ensuremath{p_{z}}\xspace}
\newcommand{\met}    {\mbox{$\protect \raisebox{.3ex}{$\not$}\et$}\xspace}

\newcommand{\wjets}  {\ensuremath{W+\text{jets}}\xspace}
\newcommand{\chisq}  {\ensuremath{\chi^{2}}\xspace}
\newcommand{\chisqmin}{\ensuremath{\chi^{2}_{\text{min}}}\xspace}

\newcommand{\mtop}   {\ensuremath{\mathrm{M}_{\text{top}}}\xspace}
\newcommand{\mreco}  {\ensuremath{m_{t}^{\text{reco}}}\xspace}
\newcommand{\mrecok} {\ensuremath{m_{t,k}^{\text{reco}}}\xspace}

\newcommand{\mterrp} {\ensuremath{\delta_{\mtop}^{+}}\xspace}
\newcommand{\mterrm} {\ensuremath{\delta_{\mtop}^{-}}\xspace}
\newcommand{\errp}   {\ensuremath{\delta^{+}}\xspace}
\newcommand{\errm}   {\ensuremath{\delta^{-}}\xspace}
\newcommand{\runi}   {Run~I\xspace}
\newcommand{\runii}  {Run~II\xspace}
\newcommand{\cdfii}  {CDF~II\xspace}
\newcommand{\DZero}  {D\O\xspace}

\newcommand{\mjj}    {\ensuremath{m_{\mathrm{jj}}}\xspace}
\newcommand{\mjjk}   {\ensuremath{m_{\mathrm{jj},k}}\xspace}
\newcommand{\jes}    {\ensuremath{\mathrm{JES}}\xspace}

\newcommand{\genunit}[2]{\ensuremath{#1~\mathrm{#2}}\xspace}
\newcommand{\degrees}[1]{\ensuremath{#1^{\mathrm{o}}}\xspace}

\newcommand{\mev}[1] {\ensuremath{#1~\mathrm{MeV}}}

\newcommand{\tev}[1] {\ensuremath{#1~\mathrm{TeV}}}
\newcommand{\gev}[1] {\ensuremath{#1~\mathrm{GeV}}}
\newcommand{\gevnoarg}{\ensuremath{\mathrm{GeV}}}
\newcommand{\gevc}[1] {\ensuremath{#1~\mathrm{GeV}/c}}
\newcommand{\gevcnoarg}{\ensuremath{\mathrm{GeV}/c}}
\newcommand{\gevcc}[1]{\ensuremath{#1~\mathrm{GeV}/c^{2}}}
\newcommand{\gevccnoarg}{\ensuremath{\mathrm{GeV}/c^{2}}}
\newcommand{\invpb}[1]{\ensuremath{#1~\mathrm{pb}^{-1}}}
\newcommand{\invpbnoarg}{\ensuremath{\mathrm{pb}^{-1}}}
\newcommand{\sigunit}[1]{\ensuremath{#1~\sigma}\xspace}
\newcommand{\sigcunit}[1]{\ensuremath{#1~\sigma_c}\xspace}

\newcommand{\measErr}[2]{\ensuremath{#1 \pm #2}\xspace}
\newcommand{\measAErr}[3]{\ensuremath{#1~^{+#2}_{-#3}}\xspace}
\newcommand{\measStat}[2]{\ensuremath{#1 \pm #2~(\text{stat.})}\xspace}
\newcommand{\measStatJES}[2]{\ensuremath{#1 \pm #2~(\text{stat.}+\jes)}\xspace}
\newcommand{\measAStat}[3]{\ensuremath{#1~^{+#2}_{-#3}~(\text{stat.})}\xspace}
\newcommand{\measAStatJES}[3]{\ensuremath{#1~^{+#2}_{-#3}~(\text{stat.}+\jes)}\xspace}
\newcommand{\measAStatJESBold}[3]{\ensuremath{#1~^{+#2}_{-#3}~(\textbf{stat.}+\jes)}\xspace}
\newcommand{\measAStatSepJES}[4]{\ensuremath{#1~^{+#2}_{-#3}~(\text{stat.}) \pm #4~(\jes)}\xspace}

\newcommand{\measAStatSyst}[4]{\ensuremath{#1~^{+#2}_{-#3}~(\text{stat.})\pm #4~(\text{syst.})}\xspace}
\newcommand{\measAStatJESSyst}[4]{\ensuremath{#1~^{+#2}_{-#3}~(\text{stat.}+\jes)\pm #4~(\text{other syst.})}\xspace}

\newcommand{\lshapemreco}{\ensuremath{{\mathcal L}^{\mreco}_{\text{shape}}}\xspace}
\newcommand{\lshapemjj}{\ensuremath{{\mathcal L}^{\mjj}_{\text{shape}}}\xspace}

\newcommand{\lnev}{\ensuremath{{\mathcal L}_{\text{nev}}}\xspace}
\newcommand{\lbg}{\ensuremath{{\mathcal L}_{\text{bg}}}\xspace}
\newcommand{\ljes}{\ensuremath{{\mathcal L}_{\jes}}\xspace}

\newcommand{\kinblah}{for 73 signal candidate events
(\twotag and \onetagt subsamples),
compared to the prediction from {\sc herwig} \ttbar signal events
(with generated top quark mass of \gevcc{172.5})
and simulated background events.}

\begin{document}

\title{Top Quark Mass Measurement Using the Template Method in the
Lepton + Jets Channel at CDF II}
\affiliation{Institute of Physics, Academia Sinica, Taipei, Taiwan 11529, Republic of China} 
\affiliation{Argonne National Laboratory, Argonne, Illinois 60439} 
\affiliation{Institut de Fisica d'Altes Energies, Universitat Autonoma de Barcelona, E-08193, Bellaterra (Barcelona), Spain} 
\affiliation{Baylor University, Waco, Texas  76798} 
\affiliation{Istituto Nazionale di Fisica Nucleare, University of Bologna, I-40127 Bologna, Italy} 
\affiliation{Brandeis University, Waltham, Massachusetts 02254} 
\affiliation{University of California, Davis, Davis, California  95616} 
\affiliation{University of California, Los Angeles, Los Angeles, California  90024} 
\affiliation{University of California, San Diego, La Jolla, California  92093} 
\affiliation{University of California, Santa Barbara, Santa Barbara, California 93106} 
\affiliation{Instituto de Fisica de Cantabria, CSIC-University of Cantabria, 39005 Santander, Spain} 
\affiliation{Carnegie Mellon University, Pittsburgh, PA  15213} 
\affiliation{Enrico Fermi Institute, University of Chicago, Chicago, Illinois 60637} 
\affiliation{Joint Institute for Nuclear Research, RU-141980 Dubna, Russia} 
\affiliation{Duke University, Durham, North Carolina  27708} 
\affiliation{Fermi National Accelerator Laboratory, Batavia, Illinois 60510} 
\affiliation{University of Florida, Gainesville, Florida  32611} 
\affiliation{Laboratori Nazionali di Frascati, Istituto Nazionale di Fisica Nucleare, I-00044 Frascati, Italy} 
\affiliation{University of Geneva, CH-1211 Geneva 4, Switzerland} 
\affiliation{Glasgow University, Glasgow G12 8QQ, United Kingdom} 
\affiliation{Harvard University, Cambridge, Massachusetts 02138} 
\affiliation{Division of High Energy Physics, Department of Physics, University of Helsinki and Helsinki Institute of Physics, FIN-00014, Helsinki, Finland} 
\affiliation{University of Illinois, Urbana, Illinois 61801} 
\affiliation{The Johns Hopkins University, Baltimore, Maryland 21218} 
\affiliation{Institut f\"{u}r Experimentelle Kernphysik, Universit\"{a}t Karlsruhe, 76128 Karlsruhe, Germany} 
\affiliation{High Energy Accelerator Research Organization (KEK), Tsukuba, Ibaraki 305, Japan} 
\affiliation{Center for High Energy Physics: Kyungpook National University, Taegu 702-701; Seoul National University, Seoul 151-742; and SungKyunKwan University, Suwon 440-746; Korea} 
\affiliation{Ernest Orlando Lawrence Berkeley National Laboratory, Berkeley, California 94720} 
\affiliation{University of Liverpool, Liverpool L69 7ZE, United Kingdom} 
\affiliation{University College London, London WC1E 6BT, United Kingdom} 
\affiliation{Massachusetts Institute of Technology, Cambridge, Massachusetts  02139} 
\affiliation{Institute of Particle Physics: McGill University, Montr\'{e}al, Canada H3A~2T8; and University of Toronto, Toronto, Canada M5S~1A7} 
\affiliation{University of Michigan, Ann Arbor, Michigan 48109} 
\affiliation{Michigan State University, East Lansing, Michigan  48824} 
\affiliation{Institution for Theoretical and Experimental Physics, ITEP, Moscow 117259, Russia} 
\affiliation{University of New Mexico, Albuquerque, New Mexico 87131} 
\affiliation{Northwestern University, Evanston, Illinois  60208} 
\affiliation{The Ohio State University, Columbus, Ohio  43210} 
\affiliation{Okayama University, Okayama 700-8530, Japan} 
\affiliation{Osaka City University, Osaka 588, Japan} 
\affiliation{University of Oxford, Oxford OX1 3RH, United Kingdom} 
\affiliation{University of Padova, Istituto Nazionale di Fisica Nucleare, Sezione di Padova-Trento, I-35131 Padova, Italy} 
\affiliation{University of Pennsylvania, Philadelphia, Pennsylvania 19104} 
\affiliation{Istituto Nazionale di Fisica Nucleare Pisa, Universities of Pisa, Siena and Scuola Normale Superiore, I-56127 Pisa, Italy} 
\affiliation{University of Pittsburgh, Pittsburgh, Pennsylvania 15260} 
\affiliation{Purdue University, West Lafayette, Indiana 47907} 
\affiliation{University of Rochester, Rochester, New York 14627} 
\affiliation{The Rockefeller University, New York, New York 10021} 
\affiliation{Istituto Nazionale di Fisica Nucleare, Sezione di Roma 1, University of Rome ``La Sapienza," I-00185 Roma, Italy} 
\affiliation{Rutgers University, Piscataway, New Jersey 08855} 
\affiliation{Texas A\&M University, College Station, Texas 77843} 
\affiliation{Istituto Nazionale di Fisica Nucleare, University of Trieste/\ Udine, Italy} 
\affiliation{University of Tsukuba, Tsukuba, Ibaraki 305, Japan} 
\affiliation{Tufts University, Medford, Massachusetts 02155} 
\affiliation{Waseda University, Tokyo 169, Japan} 
\affiliation{Wayne State University, Detroit, Michigan  48201} 
\affiliation{University of Wisconsin, Madison, Wisconsin 53706} 
\affiliation{Yale University, New Haven, Connecticut 06520} 
\author{A.~Abulencia}
\affiliation{University of Illinois, Urbana, Illinois 61801}
\author{D.~Acosta}
\affiliation{University of Florida, Gainesville, Florida  32611}
\author{J.~Adelman}
\affiliation{Enrico Fermi Institute, University of Chicago, Chicago, Illinois 60637}
\author{T.~Affolder}
\affiliation{University of California, Santa Barbara, Santa Barbara, California 93106}
\author{T.~Akimoto}
\affiliation{University of Tsukuba, Tsukuba, Ibaraki 305, Japan}
\author{M.G.~Albrow}
\affiliation{Fermi National Accelerator Laboratory, Batavia, Illinois 60510}
\author{D.~Ambrose}
\affiliation{Fermi National Accelerator Laboratory, Batavia, Illinois 60510}
\author{S.~Amerio}
\affiliation{University of Padova, Istituto Nazionale di Fisica Nucleare, Sezione di Padova-Trento, I-35131 Padova, Italy}
\author{D.~Amidei}
\affiliation{University of Michigan, Ann Arbor, Michigan 48109}
\author{A.~Anastassov}
\affiliation{Rutgers University, Piscataway, New Jersey 08855}
% ...
\collaboration{CDF Collaboration}
\noaffiliation

\date{\today}

\begin{abstract}
This article presents a measurement of the top quark mass using the
\cdfii detector at Fermilab.  Colliding beams of protons and
anti-protons at Fermilab's Tevatron ($\sqrt{s}=\tev{1.96}$) produce
top/anti-top pairs, which decay to $W^+W^-\bbbar$; events are selected
where one $W$ decays to hadrons, and the other $W$ decays to either $e$ or
$\mu$ plus a neutrino.
The data sample corresponds to an integrated luminosity of
approximately \invpb{318}.  A total of 165 \ttbar events are separated
into four subsamples based on jet transverse energy thresholds and the
number of $b$ jets identified by reconstructing a displaced vertex.
In each event, the reconstructed top quark invariant mass is
determined by minimizing a \chisq for the overconstrained kinematic
system.  At the same time, the mass of the hadronically decaying $W$
boson is measured in the same event sample.  The observed $W$ boson
mass provides an \emph{in situ} improvement in the determination of
the hadronic jet energy scale, \jes.
A simultaneous likelihood fit of the reconstructed top quark masses
and the $W$ boson invariant masses in the data sample to distributions
from simulated signal and background events gives a top quark mass of
\gevcc{\measAStatJESSyst{173.5}{3.7}{3.6}{1.3}}, or
\gevcc{\measAErr{173.5}{3.9}{3.8}}.
\end{abstract}

% activate the following line for publication

\maketitle

\newpage
\section{Introduction}
\label{sec:intro}

The top quark is the heaviest observed elementary particle, with a
mass roughly 40 times larger than the mass of the $b$ quark. This
property of the top quark produces large contributions to electroweak
radiative corrections, making more accurate measurements of the top
quark mass important for precision tests of the standard model and
providing tighter constraints on the mass of the putative Higgs
particle. The near-unity Yukawa coupling of the top quark also hints
at a role for the particle in electroweak symmetry breaking. Improved
measurements of the top quark mass are key not only for completing our
current description of particle physics, but also for understanding
possible physics beyond the standard model.

The top quark was first observed in 1995 during the first run of the
Fermilab Tevatron, by CDF~\cite{r_observation_cdf} and
\DZero~\cite{r_observation_d0}. By the end of \runi, the combined
measurement of the top quark mass was
\gevcc{\measErr{178.0}{4.3}}~\cite{r_run1_comb} using 100--\invpb{125}
of data per experiment.  This article reports a
measurement of the top quark mass in the lepton + jets decay channel
using the upgraded CDF II detector at Fermilab, with $\invpb{318}$ of
\ppbar data collected between February 2002 and August 2004.  A brief
overview of the analysis is as follows.

We scrutinize the data for events where a \ttbar pair has been
produced and has decayed to two $W$ bosons and two $b$ quarks, where
subsequently one $W$ boson decayed to two quarks, and the other $W$
boson decayed to an electron or muon and a neutrino. Thus we look for
a high-energy electron or muon, momentum imbalance in the detector
representing the neutrino, two jets of particles corresponding to the
$b$ quarks, and two additional jets corresponding to the hadronic $W$
decay.

Our measurement uses an observable that is strongly correlated with
the top quark pole mass, namely the reconstructed top quark mass.
This quantity is determined for each event by minimizing a \chisq
function in a kinematic fit to a \ttbar final
state~\cite{r_evidence_prd}.  In this fit, we apply energy and
momentum conservation, constrain both sets of $W$ decay daughters to
have the invariant mass of the $W$ boson, and constrain both $Wb$
states to have the same mass.  The mass reconstruction is complicated
by an ambiguity as to which jet represents each quark in the final state.
However, since the above procedure yields an overconstrained system,
we can choose which jet to assign to each quark based on the fit
quality.  In addition, some jets are experimentally identified as
arising from $b$ quarks by utilizing the relatively long lifetime of
the $b$ quark, reducing the number of allowed jet-quark assignments.

The method we use to measure the top quark mass is similar in concept
to an analysis performed at CDF using data from
\runi~\cite{Affolder:2000vy}.  We compare the distribution of the
reconstructed mass from events in the data with the distributions
derived from events simulated at various values of the top quark
mass. We also simulate events from the expected background
processes. Our measured value is the top quark mass for which the
simulated events, when combined with the background, best describe the
distribution in the data. We improve the power of the method by
separating the events into four subsamples that have different
background contamination and different sensitivity to the top quark
mass.

An important uncertainty in top mass measurements arises from the
uncertainty in the jet energy scale, particularly for the two jets
from $b$ quarks that are direct decay products of the top quarks.  To
reduce this uncertainty, we have developed a technique exploiting the
fact that the daughters of the hadronically decaying $W$ boson should
form an invariant mass consistent with the precisely known $W$ boson
mass.  We constrain the jet energy scale by comparing the distribution
of observed dijet invariant mass for candidate $W$ boson daughter jets
with simulated distributions assuming various shifts in the jet energy
scale with respect to our nominal scale.  We show that this improves
the jet energy scale information and is largely independent of the top
quark mass.  Furthermore, since this information applies in large part
to $b$ jets as well, it can be used to significantly reduce the
uncertainties in the overall top quark mass measurement.  A
measurement of the top quark mass without this additional information
gives consistent results, albeit with larger overall uncertainties.

A brief outline of this article is as follows: In
Section~\ref{sec:detBkgdSel}, we describe the CDF II detector used for
the analysis and our event selection for \ttbar candidates in the
lepton + jets channel, and give background
estimates. Section~\ref{sec:jetCorrSyst} explains the corrections we
make to the jets measured in our detector, as well as the systematics
associated with these corrections that dominate top quark mass
measurements. Also described in this section is how we reduce these
systematics using the $W$ dijet mass. The machinery for reconstructing
distributions of top quark masses and dijet masses is explained in
Section~\ref{sec:massRecon}, and our method for fitting these
distributions is described in
Section~\ref{sec:fitting}. Section~\ref{sec:results} gives the results
of fits to the data, as well as cross-checks for our measurement. The
remaining systematics are detailed in Section~\ref{sec:sys}, and we
conclude in Section~\ref{sec:conc}.

\section{Detector, Backgrounds, and Event Selection}
\label{sec:detBkgdSel}

This section begins with an explanation of the \ttbar event signature
along with a summary of the background processes that can mimic it.
The relevant parts of the CDF II detector are briefly described, as
well as the Monte Carlo generation and simulation procedures. The
event selection and the separation into disjoint subsamples are
defined next. Finally, the expected number of background events is
discussed.

\subsection{Event Signature}
\label{ssec:signature}

In the standard model, the top quark decays with a very short lifetime
($\tau\approx4\times10^{-25}~\mathrm{s}$) and with $\sim\!\!100\%$
branching ratio into a $W$ boson and a $b$ quark.  The \ttbar event
signature is therefore determined by the decay products of the two $W$
bosons, each of which can produce two quarks or a charged lepton and a
neutrino.  This analysis considers events in the lepton + jets
channel, where one $W$ decays to quarks and the other $W$ decays to
$e\nu_{e}$ or $\mu\nu_{\mu}$. In the following, ``lepton'' will refer
exclusively to a candidate electron or muon.  Thus, events of interest
to this measurement have an energetic $e$ or $\mu$, a neutrino, and
four jets, two of which are $b$ jets.  More jets may be present due to
hard gluon radiation from an incoming parton (initial state radiation,
ISR) or from a final-state quark (final state radiation, FSR).  Events
where a $W$ boson decays to $\tau\nu_{\tau}$ can also enter the event
sample when a secondary electron or muon from the tau decay passes the
lepton cuts---about 6\% of identified \ttbar events have this decay
chain.

There are several non-\ttbar processes that have similar signatures
and enter into the event sample for this analysis. Events where a
leptonically-decaying $W$ boson is found in association with QCD
production of at least four additional jets, sometimes including a
\bbbar pair, have the same signature and are an irreducible
background.  Singly-produced top quarks, e.g. $q\bar{q}\rightarrow
t\bar{b}$, with a leptonic $W$ decay and additional jets produced via
QCD radiation, also have the same signature.  Additional background
events enter the sample when the \ttbar signature is faked. For
example, a jet can fake an isolated lepton, albeit with small
probability, a neutrino can be mistakenly inferred when the missing
energy in the event is mismeasured, and a leptonically decaying $Z$
boson can look like a $W$ if one lepton goes undetected.

\subsection{Detector}
\label{ssec:detector}

The Collider Detector at Fermilab is a general-purpose detector
observing \ppbar collisions at Fermilab's Tevatron.  The detector
geometry is cylindrical, with the $z$ axis pointing along a tangent to
the Tevatron ring, in the direction of proton flight in the
accelerator.  Transverse quantities such as $\et$ and $\pt$ are
magnitudes of projections into the plane perpendicular to the $z$
axis. The coordinates $x$, $y$, $r$, and $\phi$ are defined in this
transverse plane, with the $x$ axis pointing outward from the
accelerator ring, and the $y$ axis pointing straight up.  The angle
$\theta$ is the polar angle measured from the proton direction, and
$\eta=-\ln(\tan\frac{\theta}{2})$ is the pseudorapidity.  When $\eta$
is calculated using the reconstructed interaction point, it is
referred to as $\etaevt$. \Fig{f_CDFpic} shows an elevation view of
the CDF detector.  The relevant subdetectors are described briefly
below.  A more complete description of the CDF \runii detector is
provided elsewhere~\cite{r_cdfDetector}.

\begin{cfigure}
\includegraphics{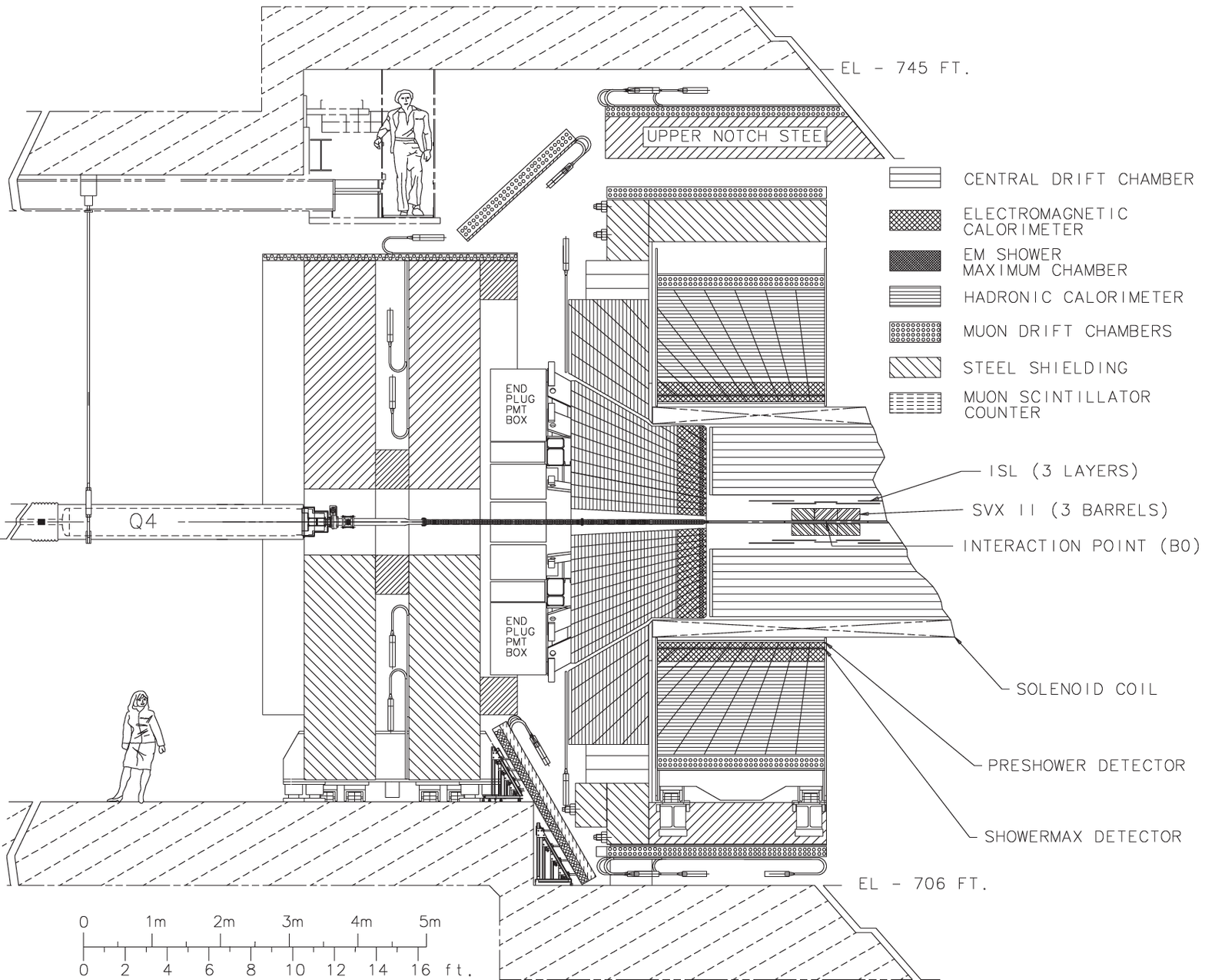}
\caption[Elevation view of the CDF detector.]
{An elevation view of the CDF \runii detector. From the collision region
outwards,
CDF consists of a silicon strip detector, a tracking drift chamber, an
electromagnetic calorimeter, a hadronic calorimeter, and muon
chambers.}
\label{f_CDFpic}
\end{cfigure}

The CDF tracking system is the first detector element crossed by a
particle leaving the interaction point in the central region.  The
silicon detectors~\cite{r_NIMSVX} provide three-dimensional position measurements with
very good resolution for charged particles close to the interaction
region, allowing extrapolation of tracks back to the collision point
and reconstruction of secondary, displaced vertices. There are a total of 722,432 channels, with a typical strip
pitch of 55--$65~\mu\mathrm{m}$ for axial strips,
60--$75~\mu\mathrm{m}$ for \degrees{1.2} small-angle stereo strips,
and 125--$145~\mu\mathrm{m}$ for \degrees{90} stereo strips. The
silicon detector is divided into three separate subdetectors. The
layer 00 (L00) is a single-sided layer of silicon mounted directly on
the beampipe (made of beryllium), at a radius of 1.4--\genunit{1.6}{cm},
providing an axial measurement close to the collision point.  The
SVXII detector is \genunit{90}{cm} long and contains 12 wedges in
$\phi$, each with 5 layers of silicon at radii from \genunit{2.5}{cm}
to \genunit{10.6}{cm}. One side of each layer contains strips oriented
in the axial direction, and the other side contains \degrees{90}
stereo strips in three cases, and \degrees{1.2} small-angle stereo
strips in two cases.  The Intermediate Silicon Layers (ISL) comprise
three additional layers of double-sided silicon at larger radii:
at \genunit{22}{cm} for $|\eta|<1$, and at \genunit{20}{cm} and
\genunit{28}{cm} for $1<|\eta|<2$.  Each layer of the ISL provides
axial and small-angle stereo measurements.

The Central Outer Tracker (COT)~\cite{r_NIMCOT} measures particle
locations over a large radial distance, providing precise measurements
of track curvature up to about $|\eta|=1$.  It is a large open-cell
drift chamber with 8 ``superlayers'' (4 axial and 4 with a \degrees{2}
stereo angle), each of which contains 12 wire layers, for a total of
96 layers.  There are 30,240 wires in total. The COT active volume is
\genunit{310}{cm} in length and covers \genunit{43}{cm} to
\genunit{132}{cm} in radius.  An axial magnetic field of
$1.4~\mathrm{T}$ is provided by a superconducting solenoid surrounding
the silicon detectors and central drift chamber.

Particle energies are measured using sampling calorimeters. The
calorimeters are segmented into towers with projective geometry.  The
segmentation of the CDF calorimeters is rather coarse, so that often
several particles contribute to the energy measured in one tower.

In the central region, i.e.\ $|\eta|<1.1$, the calorimeter is divided
into wedges subtending \degrees{15} in $\phi$. Each wedge has ten
towers, of roughly equal size in $\eta$, on each side of $\eta=0$.
The central electromagnetic calorimeter (CEM)~\cite{r_NIMCEM} contains
alternating layers of lead and scintillator, making 18 radiation
lengths of material. The transverse energy resolution for high-energy
electrons and photons is
$\frac{\sigma(\et)}{\et}=\frac{13.5\%}{\sqrt{\et[\gevnoarg]}}\oplus2\%$.
Embedded in the CEM is a shower maximum detector, the CES, which
provides good position measurements of electromagnetic showers at a
depth of six radiation lengths and is used in electron identification.
The CES consists of wire proportional chambers with wires and cathode
strips providing stereo position information.  The central hadronic
calorimeter (CHA) and the end wall hadronic calorimeter
(WHA)~\cite{r_NIMCHAWHA} are of similar construction, with alternating
layers of steel and scintillator (4.7 interaction lengths). The WHA
fills a gap in the projective geometry between the CHA and the plug
calorimeter.

The calorimetry~\cite{r_NIMPCal} in the end plugs ($1<|\eta|<3.6$) has
a very complicated tower geometry, but the \degrees{15} wedge pattern
is respected.  The plug electromagnetic calorimeter (PEM) has lead
absorber and scintillating tile read out with wavelength shifting
fibers. An electron traversing the PEM passes through 23.2 radiation
lengths of material.  The energy resolution for high-energy electrons
and photons is
$\frac{\sigma(E)}{E}=\frac{14.4\%}{\sqrt{E[\gevnoarg]}}\oplus0.7\%$.
There is a shower maximum detector (PES), whose scintillating strips
measure the position of electron and photon showers. The plug hadronic
calorimeter (PHA) has alternating layers of iron and scintillating
tile, for a total of 6.8 interaction lengths.

Muon identification is performed by banks of single-wire drift cells
four layers deep. The central muon detector (CMU)~\cite{r_NIMCMU} is
located directly behind the hadronic calorimeter in a limited portion
of the central region ($|\eta|<0.6$).  The central muon upgrade (CMP)
adds additional coverage in the central region and reduces background
with an additional \genunit{60}{cm} of steel shielding, corresponding
to 2.4 interaction lengths at \degrees{90}.  The central muon
extension (CMX) covers the region $0.6<|\eta|<1.0$, and contains eight
layers of drift tubes, with the average muon passing through six.

A three-level trigger system is used to select interesting events to
be recorded to tape at $\sim\genunit{75}{Hz}$ from the bunch crossing
rate of $\genunit{1.7}{MHz}$.  This analysis uses data from triggers
based on high-$\pt$ leptons, which come from the leptonically decaying
$W$ in the event.  The first two trigger levels perform limited
reconstruction using dedicated hardware, including the eXtremely Fast
Tracker (XFT), which reconstructs tracks from the COT in the
$r$-$\phi$ plane with a momentum resolution of better than
$2\%\pt[\gevcnoarg]$~\cite{Thomson:2002xp}.  The electron trigger
requires a coincidence of an XFT track with an electromagnetic cluster
in the central calorimeter, while the muon trigger requires that an
XFT track point toward a set of hits in the muon chambers.  The third
level is a software trigger that performs full event
reconstruction. Electron and muon triggers at the third level require
fully reconstructed objects as in the event selection described below,
but with looser criteria.

\subsection{Monte Carlo Simulation}
\label{ssec:mc}

This analysis relies on the use of Monte Carlo (MC) event generation
and detector simulation.  Event generation is performed by {\sc
herwig} v6.505~\cite{r_herwig} for \ttbar signal samples, and {\sc
herwig}, {\sc pythia} v6.216~\cite{r_pythia}, and {\sc alpgen}
v1.3~\cite{r_alpgen} for background and control samples.

A detailed description of the CDF detector is used in a simulation
that tracks the interactions of particles in each subdetector and
fills data banks whose format is the same as the raw
data~\cite{Gerchtein:2003ba}. The {\sc geant} package~\cite{r_geant3}
provides a good description of most interactions, and detailed models
are developed and tuned to describe other aspects (for example, the
COT ionization and drift properties) so that high-level quantities
like tracking efficiency and momentum resolution from the data can be
reproduced. The calorimeter simulation is performed using a
parameterized shower simulation ({\sc
gflash}~\cite{Grindhammer:1989zg}) tuned to single particle energy
response and shower shapes from the data.

\subsection{Event Selection}
\label{ssec:selection}

A data sample enriched in \ttbar events in the lepton + jets channel
is selected by looking for events with an electron (muon) with
$\et>\gev{20}$ ($\pt>\gevc{20}$), missing transverse energy
$\met>\gev{20}$, at least three jets with $\et>\gev{15}$, and a
fourth jet with $\et>\gev{8}$. This section describes the
event selection in detail.

Selected events must contain exactly one well identified lepton
candidate in events recorded by the high-$\pt$ lepton triggers. The
lepton candidate can be a central electron (CEM), or a muon observed
in the CMU and CMP detectors (CMUP) or a muon observed in the CMX
detector (CMX).  The trigger efficiencies for leptons in the final
sample are high, $\sim96\%$ for electrons and $\sim90\%$ for muons,
and show negligible $\pt$ dependence.

Electrons are identified by a high-momentum track in the tracking
detectors matched with an energy cluster in the electromagnetic
calorimeter with $\et>\gev{20}$. The rate of photons and hadronic
matter faking electrons is reduced by requiring the ratio of
calorimeter energy to track momentum to be no greater than 2 (unless
$\pt>\gevc{50}$, in which case this requirement is not imposed), and
by requiring the ratio of hadronic to electromagnetic energy in the
calorimeter towers to be less than $0.055+0.00045\cdot E_{EM}$.
Isolated electrons from $W$ decays are preferentially selected over
electrons from $b$ or $c$ quark semi-leptonic decays by requiring the
additional calorimeter energy in a cone of $\Delta
R=\sqrt{\Delta\phi^2+\Delta\etaevt^2}=0.4$ around the cluster to be
less than 10\% of the cluster energy.  Electrons are rejected if they
come from photon conversions to $e^+e^-$ pairs that have been
explicitly reconstructed.

Muons are identified by a high-momentum track in the tracking
detectors ($\pt>\gevc{20}$) matched with a set of hits in the muon
chambers. The calorimeter towers to which the track points must
contain energy consistent with a minimum ionizing particle.  An
isolation cut is imposed, requiring the total calorimeter energy in a
cone of $\Delta R=0.4$ around the muon track (excluding the towers
through which the muon passed) to be less than 10\% of the track
momentum. Cosmic ray muons explicitly identified are rejected. A
complete description of electron and muon selection, including all
additional cuts used, can be found elsewhere~\cite{r_lepID}.

A neutrino from the leptonic $W$ boson decay is inferred when the
observed momentum in the transverse plane does not balance.  The
missing transverse energy, $\met$, is formed by projecting each tower
energy in the central, wall, and plug calorimeters into the plane
transverse to the beams and summing: $\met =
-\left\|\sum_{i}\et^{i}\mathbf{n_{i}}\right\|$, where $\mathbf{n_{i}}$
is the unit vector in the transverse plane that points to the $i$th
calorimeter tower.  The $\met$ is corrected using the muon track
momentum when a muon is identified in the event. For clusters of
towers that have been identified as jets, we apply an additional
correction to the $\met$ due to different detector response relative to
the fiducial central region and due the effects of multiple \ppbar
interactions.  We require the $\met$ to be at least $\gev{20}$.

Jets are identified by looking for clusters of energy in the
calorimeter using a cone algorithm, {\tt {\sc jetclu}}, where the cone
radius is $\Delta R=0.4$.  Towers with $\et>\gev{1}$ are used as a
seed for the jet search, then nearby towers are added to the clusters,
out to the maximum radius of 0.4.  A final step of splitting and
merging is performed such that a tower does not contribute to more
than one jet.  More details about the jet clustering are available
elsewhere~\cite{r_jetclu}.  Jet energies are corrected for relative
detector response and for multiple interactions, as described in
Section~\ref{ssec:jetCorr}.

Jets can be identified as $b$ jets using a displaced vertex tagging
algorithm, which proceeds as follows.  The primary event vertex is
identified using a fit to all prompt tracks in the event and a
beamline constraint.  The beamline is defined as a linear fit to the
collection of primary vertices for particular running periods.
The luminous region described by the beamline has a width of approximately
\genunit{30}{\mu m} in the transverse view and \genunit{29}{cm} in the $z$ direction.  Jets
with $\et>\gev{15}$ are checked for good-quality tracks with both COT
and silicon information. When a secondary vertex can be reconstructed
from at least two of those tracks, the signed distance between the primary
and secondary vertices along the jet direction
in the plane transverse to the beams ($L_{2D}$)
is calculated, along with its uncertainty ($\sigma(L_{2D})$).  If
$L_{2D}/\sigma(L_{2D})>7.5$, the jet is considered tagged.  The
per-jet efficiency for $b$ jets in the central region is shown as a
function of jet $\et$ in \fig{f:btageff}; the algorithm has an
efficiency of about 60\% for tagging at least one $b$ jet in a \ttbar
event.  More information concerning $b$ tagging is available
elsewhere~\cite{r_run2topxsec}.

\begin{cfigure}
\includegraphics{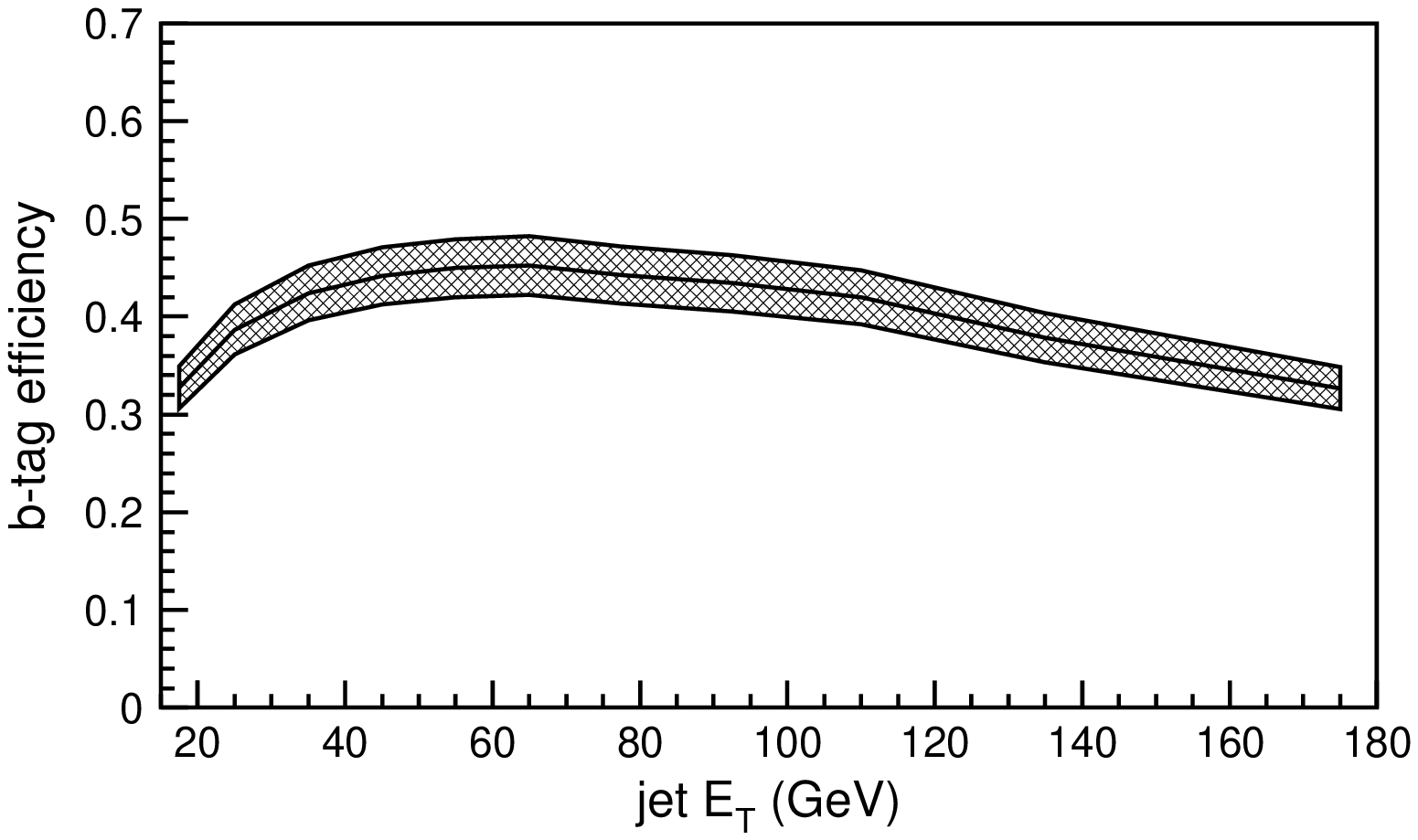}
\caption[Efficiency of $b$-tagging algorithm for central jets.]
{The efficiency of the secondary vertex $b$-tagging algorithm is shown
as a function of jet $\et$ for $b$ jets in the central region of the
detector ($|\eta|<1$), where the tracking efficiency is high.  The
shaded band gives the $\pm\sigunit{1}$ range for $b$-tagging
efficiency.  The curve is measured using a combination of data and
Monte Carlo simulated samples.}
\label{f:btageff}
\end{cfigure}

An additional $b$ tagging algorithm is used only in a cross-check of
this analysis, described in Section~\ref{ssec:otherfits}.  The Jet
Probability (JPB) tagger~\cite{Abe:1994st,Buskulic:1993ka} calculates
the probability of observing the $r$-$\phi$ impact parameters of the tracks in
the jet with respect to the primary interaction vertex, under the
hypothesis that the jet does \emph{not} arise from a heavy-flavor
quark. In the check described later, a jet is identified as a $b$ jet
if it has a JPB value less than 5\%.  Since it uses much of the same
information, the JPB tag efficiency is correlated with the displaced
vertex tag efficiency.

We require at least four jets in the event with $|\eta|<2.0$ in order
to reconstruct the \ttbar system. In events with more than 4 jets,
only the 4 jets with highest $\et$ (the leading 4 jets) are used in
jet-quark assignments. The events are separated into four subsamples
based on the jet activity. These four categories of events are found
to have different background content and different shapes in the
reconstruction of the top quark mass for signal events.  By treating
the subsamples separately, the statistical power of the method is
improved. Double-tagged (\twotag) events have two $b$-tagged jets in
the event.  These events have low background contamination, as well as
excellent mass resolution, since the number of allowed jet-quark
assignments is small. In this category, we require three jets with
$\et>\gev{15}$ and the fourth jet with $\et>\gev{8}$. Tight
single-tagged (\onetagt) events have exactly one $b$-tagged jet in the
event, and all four jets with $\et>\gev{15}$. Loose single-tagged
(\onetagl) events also have exactly one $b$ tag, but the fourth jet
has $\gev{8}<\et<\gev{15}$. These two categories have good mass
resolution, but \onetagl events have a higher background content than
\onetagt events. Finally, \zerotag events have no $b$ tags, and thus a
high background contamination. To increase the signal to background ratio
(S:B), a tighter $\et$ cut is required: all four jets must have $\et>\gev{21}$.

We find 165 \ttbar candidates in \invpb{318} of data selected for good
quality in all relevant subdetectors.  The jet selection requirements
for each of the four event types are summarized in \tab{t:eventTypes},
which also lists the expected signal to background ratio and the
number of each event type found in the data.  The expected S:B assumes
a standard model top quark with a mass of \gevcc{178} (the \runi world
average) and a corresponding \ttbar theoretical cross section of
\genunit{6.1}{pb}.  Since in the \zerotag category we do not have an
independent background estimate, no estimate of S:B is given; about 22
\ttbar events are expected.

\begin{table}
\caption[Selection requirements for the four event types.]
{The selection requirements for the four types of events are
given. The subsamples have different background content and
reconstructed mass shapes. The jet $\et$ requirements apply to the
leading four jets in the event, but additional jets are
permitted. Also shown are the number of events observed in
\invpb{318} of data, and, for purposes of illustration, the expected signal to
background ratio (S:B) assuming a \ttbar cross section of
\genunit{6.1}{pb}. The \zerotag sample category has no independent
background estimate.}
\label{t:eventTypes}
\begin{ruledtabular}
\begin{tabular}{lrcccc}
\multicolumn{2}{c}{Category} & \twotag & \onetagt & \onetagl & \zerotag \\
\hline
Jet $\et$ & j1--j3 & $\et>15$ & $\et>15$ &
	$\et>15$ & $\et>21$ \\
cuts (\gevnoarg) & j4 & $\et>8$ & $\et>15$ &
 	$15>\et>8$ & $\et>21$ \\
\hline
\multicolumn{2}{c}{$b$-tagged jets} & 2 & 1 & 1 & 0 \\
\hline
\multicolumn{2}{c}{Expected S:B} & 10.6:1 & 3.7:1 & 1.1:1 & N/A \\
\hline
\multicolumn{2}{c}{Number of events} & 25 & 63 & 33 & 44 \\
\end{tabular}
\end{ruledtabular}
\end{table}

\subsection{Background Estimation}
\label{ssec:backgrounds}

Wherever possible, we obtain an estimate of the background
contamination in each subsample that is nearly independent of the observed
number of events in that subsample; adding this information as a
constraint in the likelihood fit \emph{a priori} improves the result.

The amount and composition of the background contamination depends
strongly on the number of jets with $b$ tags. In the double $b$-tagged
sample, the background contribution is very small.  In the single
$b$-tagged sample, the dominant backgrounds are $W$ + multijet events
and non-$W$ QCD events where the primary lepton is not from a $W$
decay. The $W$ + multijet events contain either a heavy
flavor jet or a light flavor jet mistagged as a heavy flavor jet. In
the events with no $b$ tag, $W$ + multijet production dominates, and
the jets are primarily light flavor since there are no $b$ tags.

\Tab{t:backgroundnumber} gives estimates for the background composition
in each tagged subsample. Note that some of the estimates in
\tab{t:backgroundnumber} for the various background processes are
correlated, so the uncertainty on the total background is not simply
the sum in quadrature of the component uncertainties.  The procedures
for estimating each background type are described in the following
sections, and are detailed elsewhere~\cite{r_run2topxsec}.

\begin{table}
\caption[Background composition by process.]
{The sources and expected numbers of background events in the three
subsamples with $b$ tags.}
\label{t:backgroundnumber}
\begin{ruledtabular}
\begin{tabular}{lccc}
Source & \multicolumn{3}{c}{Expected Background} \\
\cline{2-4}
 & \twotag & \onetagt & \onetagl \\
\hline
Non-$W$ (QCD)       & \measErr{0.31}{0.08}  & \measErr{2.32}{0.50} & \measErr{2.04}{0.54}\\
$W\bbbar$+$W\ccbar$+$Wc$
                    & \measErr{1.12}{0.43}  & \measErr{3.91}{1.23} & \measErr{6.81}{1.85}\\
$W$ + light jets    & \measErr{0.40}{0.08}  & \measErr{3.22}{0.41} & \measErr{4.14}{0.53}\\
$WW/WZ$             & \measErr{0.05}{0.01}  & \measErr{0.45}{0.10} & \measErr{0.71}{0.13}\\
Single top          & \measErr{0.008}{0.002}& \measErr{0.49}{0.09} & \measErr{0.60}{0.11}\\
\hline
Total               & \measErr{1.89}{0.52}  & \measErr{10.4}{1.72} & \measErr{14.3}{2.45}\\
\end{tabular}
\end{ruledtabular}
\end{table}

\subsubsection{Non-$W$ (QCD) background}

For the non-$W$ background (QCD multijet events), a data-driven
technique estimates the contribution to the signal sample.  The
sideband regions of the lepton isolation ($>0.2$) vs $\met$
($<\gev{15}$) plane (after subtracting the expected \ttbar and $W$ +
multijet contributions) are used to predict the number of QCD multijet
events in the signal region, assuming no correlation between the
isolation and $\met$.

\subsubsection{$W$ + multijet backgrounds}

Simulated samples of $W$ + multijet backgrounds are obtained using the
{\sc alpgen} generator, which produces multiple partons associated
with a $W$ boson using an exact leading order matrix element
calculation.  The generator is interfaced with {\sc herwig} to
simulate parton showering and hadronization.  {\sc alpgen} describes
the kinematics of events with high jet multiplicity very well, but
suffers from a large theoretical uncertainty in the normalization due
to the choice of $Q^2$ scale and next-to-leading order (NLO) effects.
Thus, the normalization for these backgrounds is taken from the data.
The normalization for the $W$ + multijet background in the subsamples
requiring $b$ tags comes from the $W$ + multijet events before
tagging, after subtracting the expected contributions for \ttbar and
non-$W$ processes.  Due to this procedure, the tagged background
predictions are weakly coupled to the observed numbers of events in
the tagged subsamples.  Using the same procedure, the \zerotag
background estimate would be strongly coupled to the number of
observed \zerotag events.  In order to avoid this correlation in the
likelihood fit, no background constraint is used for the \zerotag
sample.

The major contributions for the $W$ + heavy flavor backgrounds, i.e.\
events with a $b$ tag on a real $b$ or $c$ jet, come from the
$W\bbbar$, $W\ccbar$, and $Wc$ processes.  The fractions of inclusive
$W$ + multijet events that contain $\bbbar$ pairs, $\ccbar$ pairs, and
single $c$ quarks are estimated using the {\sc alpgen/herwig} Monte
Carlo samples after a calibration to the parallel fractions in
inclusive jet data.  Then the contribution of each background type to
the data sample is determined by multiplying the corresponding fraction,
the event tagging efficiency for the particular configuration of $b$
and $c$ jets, and the number of $W$ + multijet events in the data
before $b$ tagging.

Another $W$ + multijet contribution comes from events where a light
flavor jet is misidentified as a heavy flavor jet.  Using jet data
events, a per-jet mistag rate is determined as a function of the
number of tracks, $\et$, $\eta$, and $\phi$ of the jet, and the scalar
sum of $\et$ for all jets with $\et>\gev{10}$ and $|\eta|<2.4$.  The
mistag rate is then applied to pretag data events in the signal region
to obtain the $W$ + light flavor contribution.

\subsubsection{Other backgrounds}

There are other minor contributions to the backgrounds: diboson
production ($WW$, $WZ$, and $ZZ$) associated with jets, and single top
production.  We use {\sc alpgen} Monte Carlo samples to estimate their
acceptance.  The NLO cross section
values~\cite{Campbell:1999ah,Harris:2002md} are used for
normalization.

\section{Jet Corrections and Systematics}
\label{sec:jetCorrSyst}

Jets of particles arising from quarks and gluons are the most
important reconstructed objects in the top quark mass measurement, but
are measured with poor energy resolution.  The jet measurements therefore
make the largest contribution to the resolution of the mass
reconstruction described in Section~\ref{sec:massRecon}.
Additionally, systematic uncertainties on the jet energy measurements
are the dominant source of systematic uncertainty on the top quark
mass.  We describe here the corrections applied to the measured jet
energies, as well as the systematic uncertainties on our modeling of
the jet production and detector response.  A more thorough treatment
of these topics is available elsewhere~\cite{r_NIMJES}.  Finally, we
introduce the jet energy scale quantity \jes, which is measured
\emph{in situ} using the $W$ boson mass resonance.

\subsection{Jet Corrections}
\label{ssec:jetCorr}

Matching reconstructed jets to quarks from the \ttbar decay has both
theoretical and experimental complications. A correspondence generally
can be assumed between measured jet quantities and the kinematics of
partons from the hard interaction and decay.  A series of corrections
are made to jet energies in order to best approximate the
corresponding quark energies.  Measured jet energies have a poor
resolution, and are treated as uncertain quantities in the mass
reconstruction.  The measured angles of the jets, in contrast, are
good approximations of the corresponding quark angles, so they are
used without corrections and are fixed in the mass reconstruction.

\subsubsection{Tower calibrations}

Before clustering into jets, the calorimeter tower energies are
calibrated as follows.  The overall electromagnetic scale is set using
the peak of the dielectron mass resonance resulting from decays of the
$Z$ boson.  The scale of the hadronic calorimeters is set using test
beam data, with changes over time monitored using radioactive sources
and the energy deposition of muons from $J/\psi$ decays, which are
minimum ionizing particles (MIPs) in the calorimeter.  Tower-to-tower
uniformity for the CEM is achieved by requiring the ratio of
electromagnetic energy to track momentum ($E/p$) of electrons to be
the same across the calorimeter.  In the CHA and WHA, the
$J/\psi\rightarrow\mu\mu$ MIPs are also used to equalize the response
of towers.  For the PEM and PHA, where tracks are not available, the
tower-to-tower calibrations use a laser calibration system and
$^{60}$Co sourcing.  The WHA calorimeter also has a sourcing system to
monitor changes in the tower gains.

\subsubsection{Process-independent corrections}
\label{sssec:genericJetCorr}

After clustering, jets are first corrected with a set of ``generic''
jet corrections, so called because they are intended to be independent
of the particular process under consideration. For these corrections,
the quark $\pt$ distribution is assumed to be flat. Since some of the
corrections are a function of jet $\pt$, and since the jet resolution
is non-negligible, this assumption has a considerable effect on the
derived correction.

These generic jet corrections scale the measured jet four-vector to
account for a set of well studied effects. First, a dijet balancing
procedure is used to determine and correct for variations in the
calorimeter response to jets as a function of $\eta$. These variations
are due to different detector technology, to differing amounts of
material in the tracking volume and the calorimeters, and to
uninstrumented regions.  In dijet balancing, events are selected with
two and only two jets, one in the well understood central region
($0.2<|\eta|<0.6$). A correction is determined such that the
transverse momentum of the other jet, called the probe jet, as a
function of its $\eta$, is equal on average to that of the central
jet. This relative correction ranges from about $+15\%$ to $-10\%$,
and can be seen in \fig{f:dijetBalance} in Section~\ref{ssec:jetSyst}.

After a small correction for the extra energy deposited by multiple
collisions in the same accelerator bunch crossing, a correction for
calorimeter non-linearity is applied so that the jet energies
correspond to the most probable in-cone hadronic energy assuming
a flat $\pt$ distribution.  First, the response of the calorimeter to
hadrons is measured using $E/p$ of single tracks in the data. Studies
of energy flow and jet shapes in the data also constrain the modeling
of jet fragmentation.  After tuning the simulation to model what we
observe in the data, the correction (+10\% to +30\%, depending on jet
$\pt$) is determined using a simulated sample of dijet events covering
a large \pt range.

\subsubsection{Process-specific corrections}

Jet corrections are then applied that have been derived specifically
for the \ttbar process. These corrections account for shifts in the
mean jet energy due to the shape of the $\pt$ distribution of quarks
from \ttbar decay, for the extra energy deposited by remnants of the
\ppbar collision not involved in the hard interaction (``underlying
event''), and for the energy falling outside the jet clustering cone.
Light-quark jets from $W$ boson decay ($W$ jets) and $b$ jets, which
have different $\pt$ distributions, fragmentation, and decay
properties, are corrected using different functions, but no separate
correction is attempted for $b$ jets with identified semi-leptonic
decays.  Each jet energy is also assigned an uncertainty arising from
the measurement resolution of the calorimeter.  Note that, since these
corrections depend on the flavor of the jet, they must be applied
after a hypothesis has been selected for the assignment of the
measured jets to quarks from the \ttbar decay chain.

The \ttbar-specific corrections are extracted from a large sample of
{\sc herwig} \ttbar events ($\mtop=\gevcc{178}$) in which the four
leading jets in $\et$ are matched within $\Delta R=0.4$ to the four
generator-level quarks from \ttbar decay.  The correction functions
are consistent with those extracted from a large {\sc pythia} sample.
The correction is defined as the most probable value (MPV) of the jet
response $(\ptquark-\ptjet)/\ptjet$, as a function of $\ptjet$ and
$\etajet$.  Since the $\etajet$ dependence is negligible for the
light-quark jets, their correction depends only on $\ptjet$.  The MPV
is chosen, rather than the mean of the asymmetric distribution, in
order to accurately correct as many jets as possible in the core of
the distribution. This increases the number of events for which the
correct jet-quark assignment is chosen by the fitter (see below),
resulting in a narrower core for the reconstructed mass distribution.
A corresponding resolution is found by taking the symmetric window
about the MPV of the jet response that includes 68\% of the total
area.  Figure~\ref{f_tscorrs} shows the corrections and resolutions as
a function of jet $\pt$ for several values of $|\eta|$.

\begin{cfigure}
\includegraphics{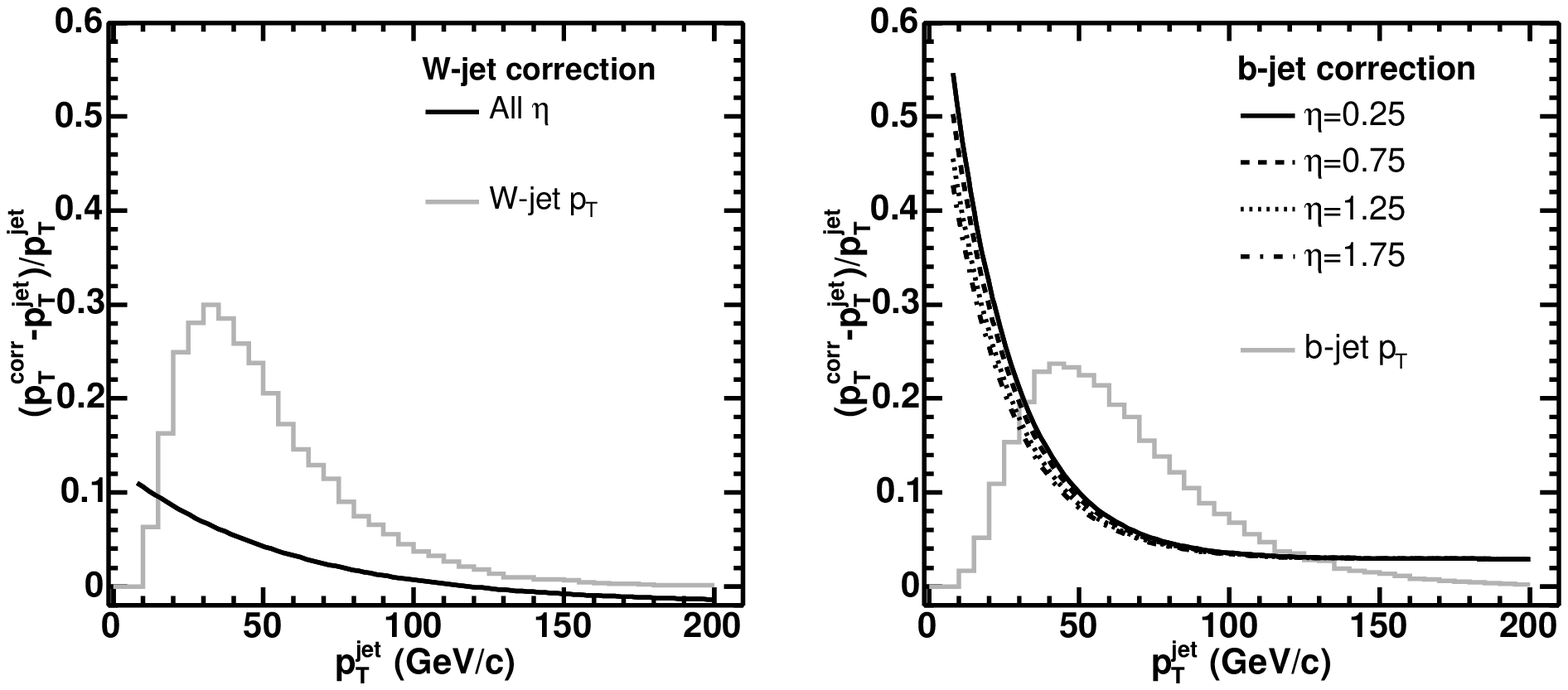}
\includegraphics{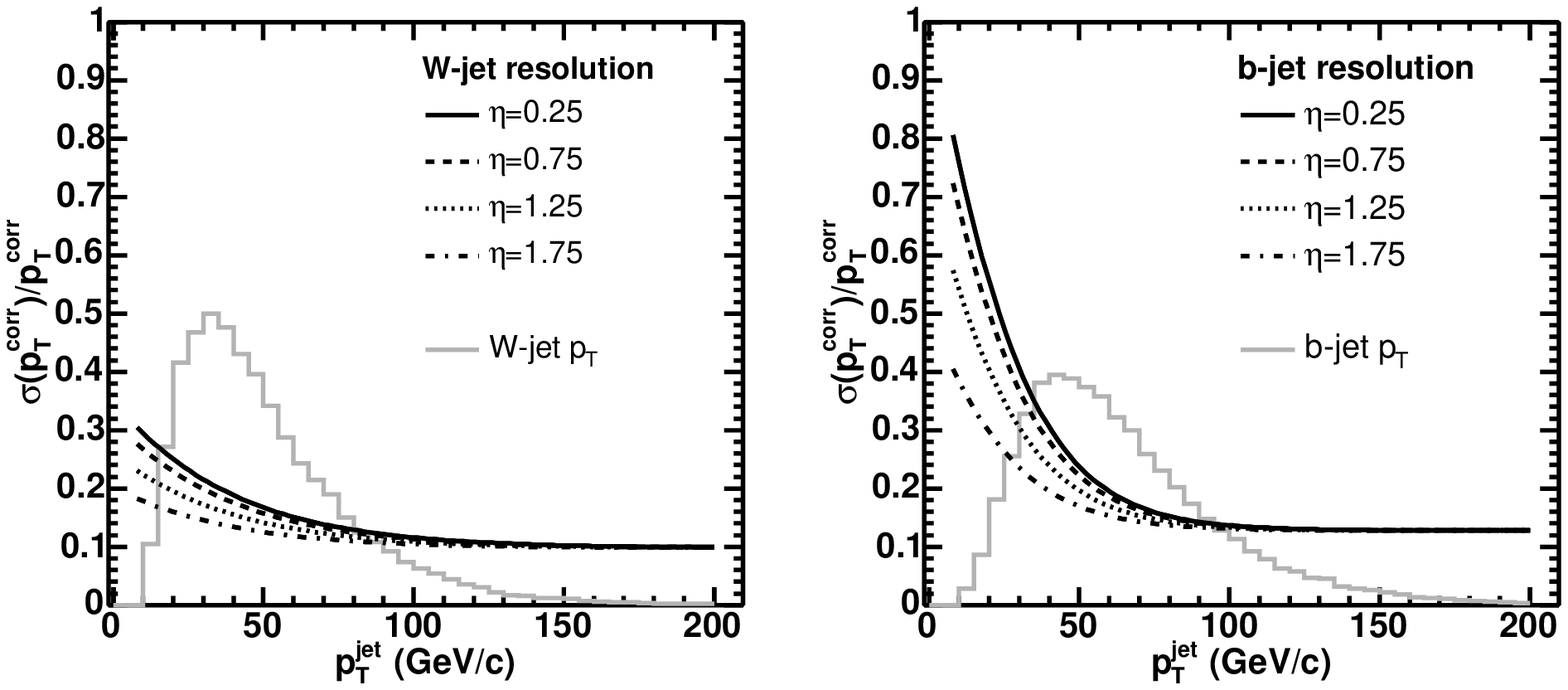}
\caption{The \ttbar-specific corrections are shown for $W$ jets (left)
and $b$ jets (right) as a function of jet \pt for several values of
$|\eta|$.  On the top is the correction factor, and on the bottom is
the fractional resolution passed to the fitter.  The histograms give
the distributions of jet \pt (arbitrarily normalized) from a signal
Monte Carlo sample with generated top quark mass of
\gevcc{178}.}
\label{f_tscorrs}
\end{cfigure}

As a final step in correcting the jet four-vector, the jet momentum is
held fixed while the jet energy is adjusted so that the jet has a mass
according to its flavor hypothesis.  A mass of $\gev{0.5}$ is used for
$W$ jets, and a mass of $\gev{5.0}$ is used for $b$ jets.  This is
done to match the generator-level quarks used to derive the
\ttbar-specific corrections.

\subsection{Systematics from Jet Energy Scale}
\label{ssec:jetSyst}

There are significant uncertainties on many aspects of the measurement
of jet energies. Some of these are in the form of uncertainties on the
energy measurements themselves; some are uncertainties on the detector
simulation, which is used to derive many corrections, and ultimately
to extract the top quark mass; still others are best understood as
theoretical uncertainties on jet production and fragmentation models
used in the generators.

\subsubsection{Calorimeter response relative to central}
The systematic uncertainties in the calorimeter response relative to
the central calorimeter range from 0.5\% to 2.5\% for jets used in
this analysis.  The uncertainties account for the residual $\eta$
dependence after dijet balancing, biases in the dijet balancing
procedure (especially near the uninstrumented regions) and the
variation of the plug calorimeter response with time.  Photon-jet
balancing is used to check the $\eta$ dependence after corrections in
data and simulated events, and the residual differences in this
comparison are also included in the systematic uncertainty.
\Fig{f:dijetBalance} shows the dijet balancing as a function of the
probe jet pseudorapidity, demonstrating that the simulation models
well the detector response for $|\eta|<2.0$. Since differing response
in neighboring regions of the detector is the primary source of biased
jet angle measurements, the plot also demonstrates that we can expect
angle biases to be well modeled in the simulated events.

\begin{cfigure}
\includegraphics{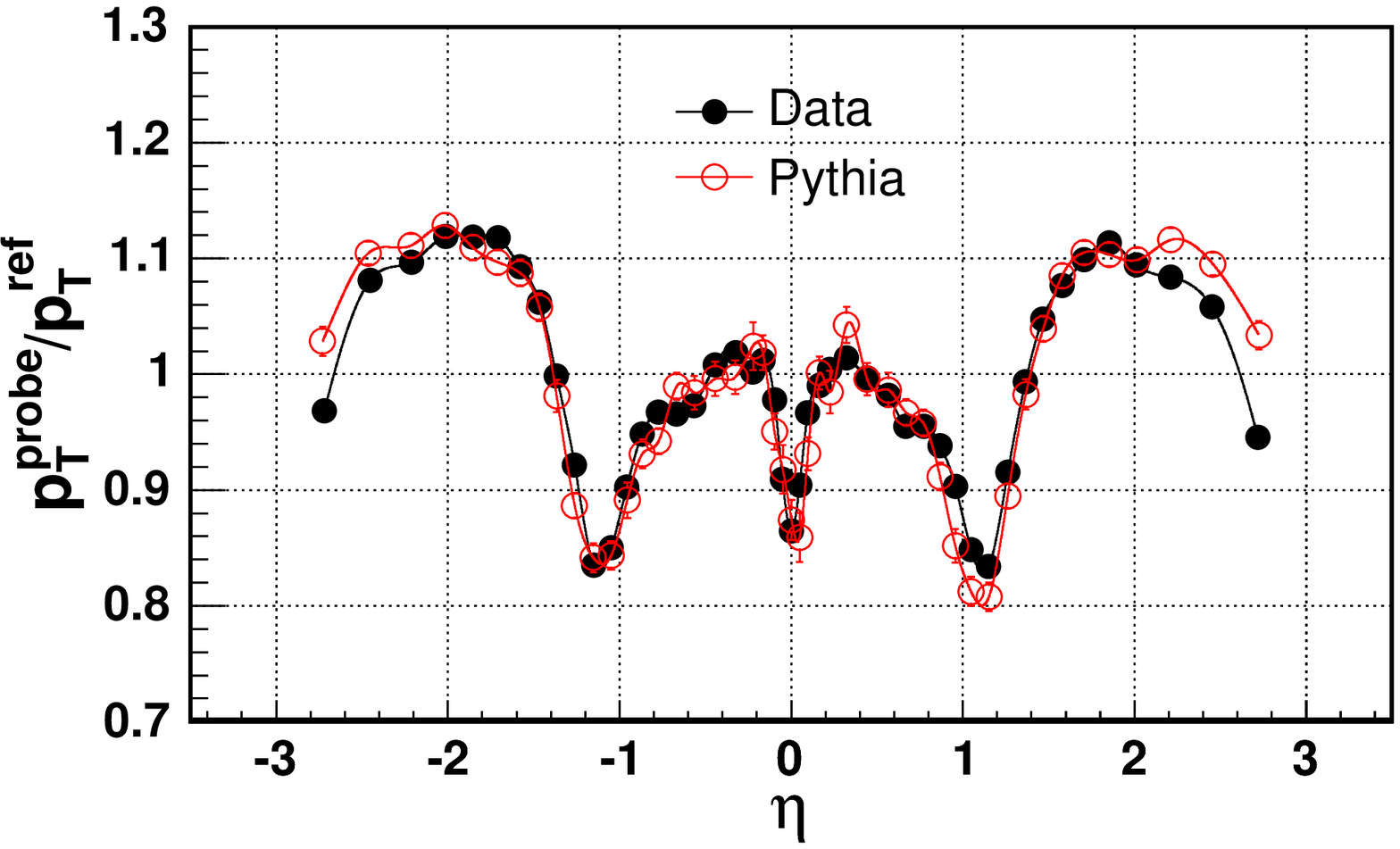}
\caption[Dijet balancing plot.]
{Results of the dijet balancing procedure are shown for data and
simulated dijet events with $\ptjet>\gev{20}$. Probe jets from
throughout the detector are compared with a reference jet in the
central region; the ratio of the \pt of the jets is plotted as a
function of the probe jet $\eta$.  The simulation models well the
detector response as a function of $\eta$.}
\label{f:dijetBalance}
\end{cfigure}

\subsubsection{Modeling of hadron jets}
The main systematic uncertainties at the hadronic level are obtained
by propagating the uncertainties on the single particle response and
the fragmentation, which are determined from studies on the data.
Smaller contributions are included from the comparison of data and
Monte Carlo simulation of the calorimeter response close to tower
boundaries in azimuth, and from the stability of the calorimeter
calibration with time.  There is also a small uncertainty on the
energy deposited by additional \ppbar interactions.  In all, this
uncertainty varies from $1.5\%$ to $3.0\%$, depending on jet \pt, and
only accounts for variations that affect the energy inside the jet
cone.

\subsubsection{Modeling of out-of-cone energy}
The uncertainty on the fraction of energy contained in the jet cone
(also primarily due to jet fragmentation modeling) is estimated in two
parts, one between $R=0.4$ and $R=1.3$ and the other for $R>1.3$. This
systematic, which is roughly 9\% at very low jet \pt but falls rapidly
to $<2\%$ for $\pt>\gev{70}$, is determined by comparing the energy
flow in jets from data and Monte Carlo for various event topologies.

\subsubsection{Modeling of underlying event}
The underlying event deposits energy uniformly in calorimeter towers
throughout the detector, some of which are clustered into jets.  Such
energy is subtracted from the jet energy in the corrections.  The
uncertainty on this correction decreases rapidly from 2\% at very low
$\pt$ to less than 0.5\% at about $\gev{35}$.

\subsubsection{Total uncertainty}
The systematic uncertainties on jet energies for jets in the reference
central region ($0.2<|\eta|<0.6$) are shown as a function of $\pt$ in
\fig{f:totalJetSyst}. For other $\eta$ regions, only the contribution
of the ``relative response'' uncertainty changes. The black line gives the
total uncertainty on the jet energy measurement, obtained by adding in
quadrature the contributions described above.

\begin{cfigure}
\includegraphics{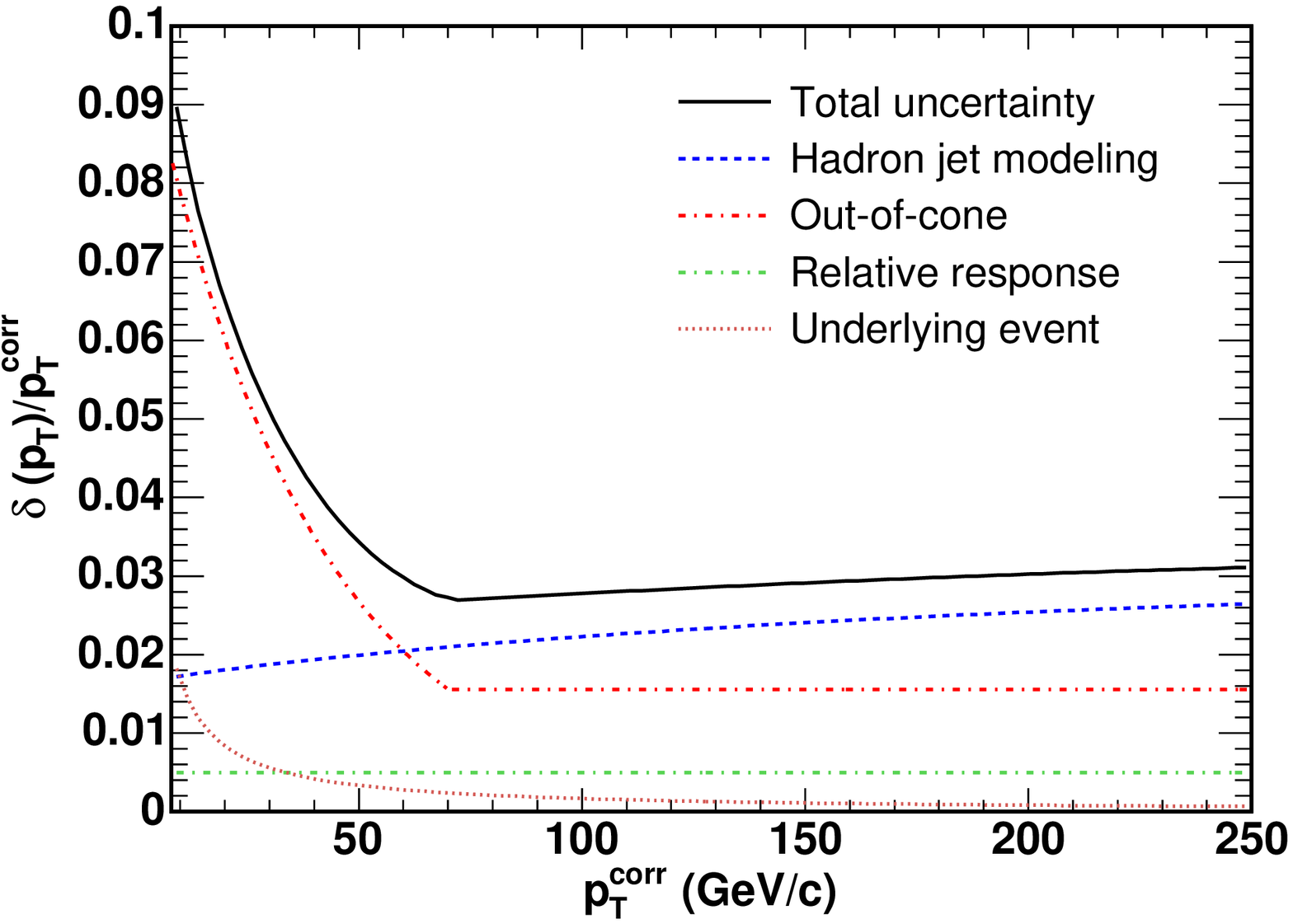}
\caption[Total jet systematic uncertainty.]
{The systematic uncertainties on jet energy are shown for jets in the
central calorimeter ($0.2<|\eta|<0.6$).  For non-central jets, the
total uncertainty has a different contribution from the eta-dependent
uncertainty.  In this plot the corrected jet transverse momentum
$\pt^{\text{corr}}$ is the process-independent estimate of the parton
$\pt$.  At low $\pt^{\text{corr}}$, the main contribution to the
systematic is from the uncertainty on the fraction of jet energy lost
outside the cone, while at high $\pt^{\text{corr}}$ it is from the
linearity corrections to obtain an absolute jet energy scale.}
\label{f:totalJetSyst}
\end{cfigure}

Events in which a jet recoils against a high energy photon are used to
check the absolute corrections.  We compare the corrected jet energy
to the photon energy, which is well calibrated using $\zee$ decays.
This $\gamma$-jet balancing is performed on data and Monte Carlo
samples, as a function of photon $\et$ and jet $\eta$, as a cross
check of the energy corrections and systematic uncertainties described
above.  \Fig{f:gammaJet} shows a comparison of the $\gamma$-jet
balancing in data and Monte Carlo after all jet corrections, along
with the $\pm\sigunit{1}$ range of the jet energy systematics.  The
agreement provides confidence that the systematic uncertainties are
reasonable.

\begin{cfigure}
\includegraphics[height=.5\textheight]{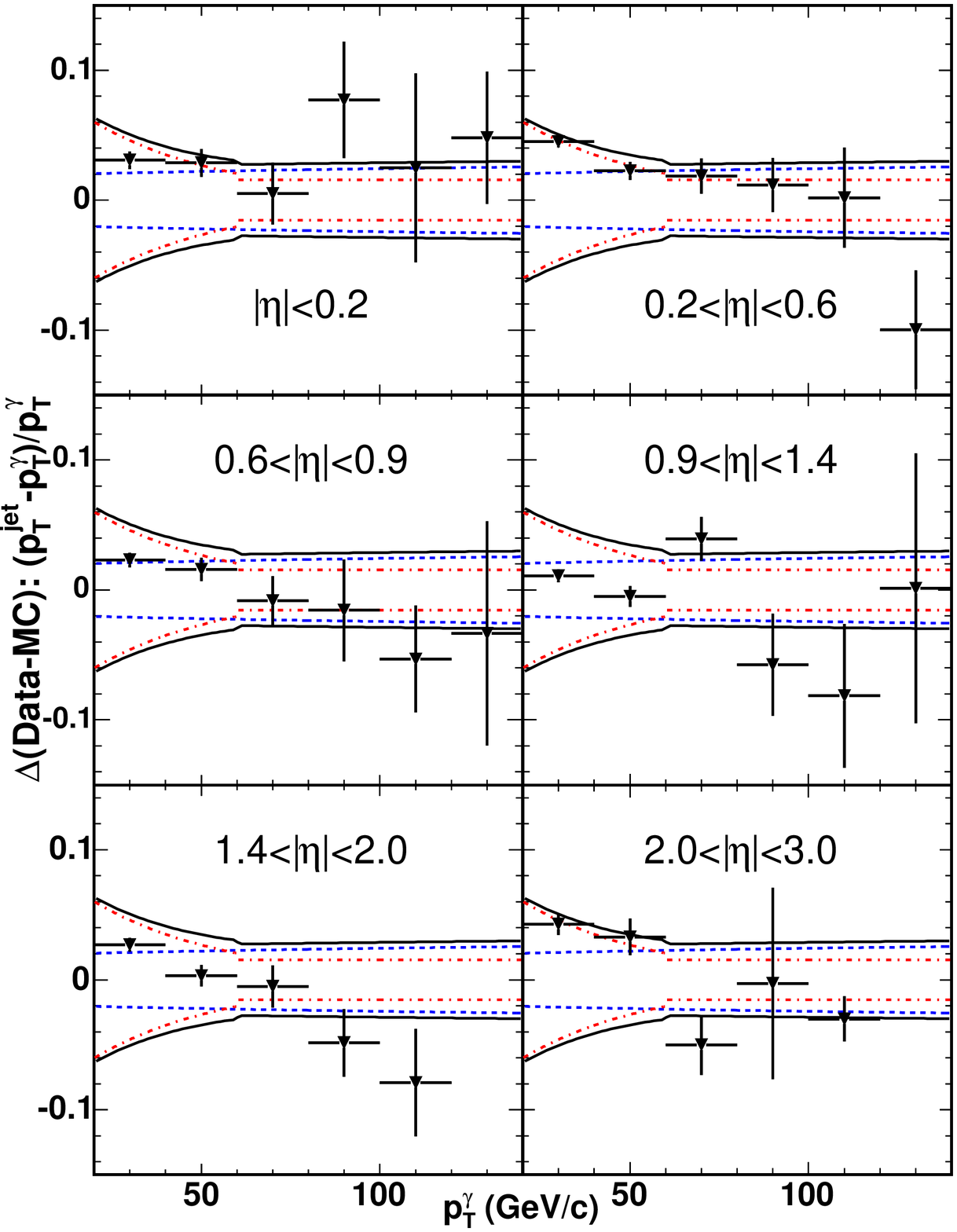}
\caption[$\gamma$-jet balancing summary.]
{For $\gamma$-jet events in both data and simulation, we find the fractional
difference in \pt between the jet and the photon after all jet corrections
are applied. Plotted here, for different ranges of jet $\eta$, is
the difference between this quantity in data and simulated events as a
function of photon $\pt$. The solid lines show the $\pm\sigunit{1}$
range given by the jet energy systematics. The other lines follow the same
definitions as in \fig{f:totalJetSyst}.}
\label{f:gammaJet}
\end{cfigure}

The systematic uncertainties on jet energies described here are
understood to apply to all jets. Clearly additional flavor-specific or
process-specific uncertainties could be present. In particular, any
systematics specific to the $b$ jets are extremely important in a
measurement of the top quark mass, and could arise from mismodeling of
$b$ quark fragmentation, semi-leptonic decays, or color connections
not present in the $W$ boson decay system. Uncertainties from these
sources have been studied and found to be relatively small; see
Section~\ref{ssec:jesSyst}.

\subsection{Jet Energy Scale}
\label{ssec:jes}

Since the jet energy systematics described in the previous section
generate the dominant systematic uncertainty on the top quark mass
measurement, a method has been developed to further constrain those
systematics using the $W$ boson mass resonance \emph{in situ}. In
particular, we measure a parameter \jes that represents a shift in the
jet energy scale from our default calibration.

Rather than defining \jes as a constant percentage shift of the jet
energies, we define it in units of the total nominal jet energy
scale uncertainty ($\sigma_c$), which is derived from the extrinsic
calibration procedures above. This $\sigma_c$ is the quantity depicted in
\fig{f:totalJetSyst} for central jets. Thus $\jes=\sigcunit{0}$
corresponds to our default jet energy scale; $\jes=\sigcunit{1}$
implies a shift in all jet energies by one standard deviation in the
uncertainty defined above; and so on. This choice has two
consequences. The first is that the effect of a shift in \jes is
different for jets with different \pt and $\eta$. For example, jets
with very low \pt have a larger fractional uncertainty, and therefore
have a larger fractional shift with a \sigcunit{1} change in \jes. The
second is that it is easy to incorporate the independent estimate of
the jet energy systematics (with its \pt and $\eta$ dependence) by
constraining \jes using a Gaussian centered at \sigcunit{0} and with a
width of \sigcunit{1}.

As described in Section~\ref{ssec:jetSyst}, the jet energy scale
uncertainty comprises many small effects, which have different
dependences on jet $\eta$ and \pt.  With more statistics, we would
choose to measure the various effects independently.  Currently,
however, we make the approximation of assuming that a single value of
the total jet energy scale parameter \jes applies to all jets in the
sample; that is, we measure a value of \jes that is averaged over jets
in the sample.  Additionally, by construction, our \jes measurement is
primarily sensitive to jets from the hadronic $W$ decay.  We estimate
the effect of this approximation as a systematic uncertainty on the
top quark mass measurement.

\section{Mass Reconstruction}
\label{sec:massRecon}

In this section, we describe the procedures for determining in each
event the reconstructed top quark mass \mreco and the dijet mass \mjj,
representing the mass of the
hadronically decaying $W$ boson.  We then discuss the results of
applying these reconstruction techniques.  Remember that by itself
\mreco is not an event-by-event measurement of the top quark mass;
rather it is a quantity whose distribution in the data will be
compared with simulated samples to extract the top quark mass (see
Section~\ref{sec:fitting}).  Similarly, the distribution of \mjj will
be used to constrain the calibration of the jet energy scale in the
reconstructed events.

Throughout the mass reconstruction, each event is assumed to be a
\ttbar event decaying in the lepton + jets channel, and the four
leading jets are assumed to correspond to the four quarks from the top
and $W$ decays. First, the measured four-vectors for the jets and
lepton in the event are corrected for known effects, and resolutions
are assigned where needed. Next, for the top quark mass
reconstruction, a \chisq fit is used to extract the reconstructed
mass, so that each event has a particular value of \mreco and a
corresponding \chisq value.  Some events are discarded from the event
sample when their minimized \chisq exceeds a cut value.  Meanwhile,
for the dijet mass reconstruction, the invariant mass \mjj is
calculated for each pair of jets without $b$ tags among the leading
four jets.

\subsection{Inputs to the mass reconstruction}
\label{ssec:inputs}

The \chisq fit takes as input the four-vectors of the jets and lepton
identified in the event.  All known corrections are applied to these
4-vectors, and Gaussian uncertainties are computed for the transverse
momenta, since they will be permitted to vary in the fit.  The
treatment of the neutrino four-vector is more complicated, since the
$\met$ is a derived quantity, and does not have an uncertainty
independent of the other measured values. The \chisq includes instead
information about a related fundamental quantity, the unclustered
energy, which is described below.

\subsubsection{Jet inputs}
\label{sssec:jetinputs}

The corrections made to the jet four-vectors are described in detail
in Section~\ref{ssec:jetCorr}. To summarize, a series of corrections
are applied to the jet energies in order to determine the energy of
the quark corresponding to each jet. The jet angles are relatively
well measured, and are fixed in the kinematic fit.  The final step of
the jet corrections is the \ttbar-specific correction that treats
separately $b$ jets and jets from the $W$ decay, and in addition
provides for the \pt of each jet a resolution that is used in the
\chisq expression.

\subsubsection{Lepton Inputs}
\label{sssec:lepinputs}

The electron four-vector has energy determined by its electromagnetic
calorimeter cluster, and angles defined by the associated track.  The
electron energy is corrected for differences in the calorimeter
response depending on where in the tower face the electron enters.
The electron mass is set to zero, and the angles are taken as
perfectly measured quantities.  The transverse momentum $(\pte = p
\sin \theta)$ of the electron is assigned an uncertainty of
\begin{equation}
\frac{\sigma_{\pte}}{\pte} = \sqrt{\left( \frac{0.135}
	{\sqrt{\pte[\gevcnoarg]}}\right)^{2} + (0.02)^{2}}.
\end{equation}

The muon four-vector uses the three-vector of the associated track,
also with a mass of zero.  Track curvature corrections due to chamber
misalignment are applied.  The angles and mass are given no
uncertainty; the transverse momentum has an uncertainty of
\begin{equation}
\frac{\sigma_{\ptmu}}{\ptmu} = 0.0011\cdot\ptmu[\gevcnoarg],
\end{equation}
The uncertainties on measured electron and muon transverse momenta are
obtained from studies of leptonic $Z^0$ decays.

\subsubsection{Neutrino Inputs: Unclustered Energy}
\label{sssec::neutrinoinputs}

The neutrino in a \ttbar event is not observed; its presence is
inferred by an imbalance in the observed transverse momentum.
Therefore, rather than treating the neutrino four-vector as an
independent input to the \chisq fit, the measured quantities, as
varied in the fit, are used to dynamically calculate the neutrino
transverse momentum.

All of the transverse energy in the calorimeter (towers with $|\eta| <
3.6$) that is not associated with the primary lepton or one of the
leading four jets is considered ``unclustered energy.'' For towers
clustered into a jet that has $\et>\gev{8}$ and $|\eta|<2.0$, but that
is not one of the leading four jets, the tower momenta are replaced
with the jet momentum after the generic jet corrections described in
Section~\ref{sssec:genericJetCorr}.  The rest of the tower momenta are
multiplied by a scale factor of 1.4, which is the estimated generic
correction factor for $\gev{8}$ jets.  Finally, the unclustered energy
includes the energy attributed to enter into the leading four jets
from the underlying event, and excludes the energy thought to fall
outside the jet cones of the leading four jets. This avoids
double-counting of energy that is included in the leading four jet
energies after all corrections.  Each transverse component of the
unclustered energy ($p_x^{UE}$, $p_y^{UE}$) is assigned an uncertainty
of $0.4 \sqrt{\sum\et^{\text{uncl}}}$, where $\sum\et^{\text{uncl}}$
is the scalar sum of the transverse energy excluding the primary
lepton and leading four jets.  The uncertainty comes from studies of
events with no real missing energy and no hard jet activity.

The unclustered energy is the observed quantity and the input to the
\chisq fit, but it is related to the missing energy through the other
measured physics objects in the event, since the $\ppbar$ system has
total transverse momentum close to 0.  The neutrino transverse momentum $\ptnu$ is calculated
at each step of the fit, using the fitted values of lepton, jet, and
unclustered transverse energies:
\begin{equation}
\vec{\ptnu} = -\left(\vec{\pt^{\ell}} + \sum\vec{\ptjet} +
				\vec{p_T^{UE}}\right)
\end{equation}
Note that this quantity, used in the mass fitting procedure, is
different from the missing energy described in
Section~\ref{ssec:selection} and used in event selection, where
simpler calorimeter energy corrections are used.

Although other treatments of the unclustered energy and missing energy
can be motivated, the $\met$ calculation does not have a large effect
on the results of the \chisq fit.  Various other approaches to
correcting the unclustered energy and assigning resolution were tried,
and no changes had any significant effect on the reconstructed top
quark mass resolution.

The mass of the neutrino is fixed at zero, and the longitudinal
momentum, $\pz^{\nu}$, is a free (unconstrained) parameter in the
fit. The initial value of $\pz^{\nu}$ is calculated using the initial
value of the lepton four-vector and the initial $\ptnu$, assuming that
they arise from a $W$ boson at the nominal pole mass.  Since these
conditions yield a quadratic equation, there are in general two
solutions for the $\pz^{\nu}$; a separate \chisq fit is done with each
solution used as the initial value of $\pz^{\nu}$.  When the solutions
are imaginary, the real part $\pm\:\gev{20}$ are the two values of
$\pz^{\nu}$ used to initialize the fit.

\subsection{Event \chisq fit}
\label{ssec:chisq}

Given the inputs described above, the event-by-event fit for the
reconstructed top quark mass proceeds as follows.  {\sc minuit} is
used to minimize a \chisq where \mreco is a free parameter.  For each
event, the \chisq is minimized once for each possible way of assigning
the leading four jets to the four quarks from the \ttbar decay. Since
the two $W$ daughter jets are indistinguishable in the \chisq
expression, the number of permutations is $\frac{4!}{2}=12$. In
addition, there are two solutions for the initial value of the
neutrino longitudinal momentum, so the minimization is performed a
total of 24 times for each event.  When $b$ tags are present,
permutations that assign a tagged jet to a light quark at parton level
are rejected. In the case of single-tagged events, the number of
allowed permutations is six, and for double-tagged events, it is
two. In the rare cases when an event has three $b$ tags, two of the
tagged jets must be assigned to $b$ quarks. We use the reconstructed
top quark mass from the permutation with the lowest \chisq after
minimization.

The \chisq expression has terms for the uncertainty on the
measurements of jet, lepton, and unclustered energies, as well as
terms for the kinematic constraints applied to the system:
\begin{eqnarray}
\label{eq_chi2}
\chisq &=&
\sum_{i = \ell, 4 jets} \frac{(p_T^{i,fit} - p_T^{i,meas})^2}{\sigma_i^2} \nonumber \\
&&+ \sum_{j = x,y} \frac{(p_j^{UE,fit} - p_j^{UE,meas})^2}{\sigma_{UE}^2} \nonumber \\
&&+ \frac{(M_{\ell\nu} - M_W)^2}{\Gamma_W^2}
+ \frac{(M_{jj} - M_W)^2}{\Gamma_W^2} \nonumber \\
&&+ \frac{(M_{b\ell\nu} - \mreco)^2}{\Gamma_t^2}
+ \frac{(M_{bjj} - \mreco)^2}{\Gamma_t^2}.
\end{eqnarray}

The first term constrains the \pt of the lepton and four leading jets
to their measured values within their assigned uncertainties; the
second term does the same for both transverse components of the
unclustered energy.  In the remaining four terms, the quantities
$M_{\ell\nu}$, $M_{jj}$, $M_{b\ell\nu}$, and $M_{bjj}$ refer to the
invariant mass of the sum of the four-vectors denoted in the
subscripts. For example, $M_{\ell\nu}$ is the invariant mass of the
sum of the lepton and neutrino four-vectors. $M_{W}$ is the pole mass
of the $W$ boson, $\gevcc{80.42}$~\cite{r_PDBook}, and \mreco is the
free parameter for the reconstructed top quark mass used in the
minimization. $M_{jj}$ is a quantity computed in the kinematic fit,
and should not be confused with \mjj, the measured dijet mass used to
constrain JES. The fit is initialized with $\mreco = \gevcc{175}$.
$\Gamma_W$ and $\Gamma_t$ are the total width of the $W$ boson and the
top quark. In order to use the \chisq formalism, the $W$ and top
Breit-Wigner lineshapes are modeled with Gaussian distributions, using
the Breit-Wigner full width at half maximum as the Gaussian sigma.
$\Gamma_{W}$ is $\gev{2.12}$~\cite{r_PDBook}, and $\Gamma_{t}$ is
$\gev{1.5}$~\cite{Oliveira:2001vw}.  Thus these terms provide
constraints such that the $W$ masses come out correctly, and the $t$
and $\bar{t}$ masses come out the same (modulo the Breit-Wigner
distribution, here modeled by a Gaussian, in both cases).

The jet-quark assignment (and $\pz^{\nu}$ solution) with the lowest
\chisq after minimization is selected for each event.  The \chisq of
this combination is denoted \chisqmin (or just \chisq when the context
is unambiguous), and the requirement $\chisqmin<9$ is imposed.  The
expected statistical uncertainty on the top quark mass does
not change much over a wide range of the value of the cut, even when
it is varied independently for the four event types.
The value of the cut chosen is close to the minimum of
expected top quark mass uncertainty.

\subsection{Dijet Mass and Jet Energy Scale}
\label{ssec:mdijetRecon}

We calculate the dijet masses used to constraint JES in the same data
sample used to reconstruct the observed top quark mass, with the
exception that there is no \chisq requirement on the jet-quark
assignments under consideration.  The imposition of the \chisq
requirement would impose a bias in the dijet masses being considered
and therefore reduce the sensitivity of the dijet mass distribution to
\jes. We calculate the dijet masses directly from the measured jet
four-vectors without the use of a kinematic fit, considering all
jet-quark assignments in each event for any of the leading 4 jets that
are not $b$-tagged. Monte Carlo studies have shown that the sensitivity
of the dijet mass distribution to the \jes parameter is maximized by
considering all dijet mass combinations that do not involve a $b$-tagged
jet in each event.  The number of possible assignments ranges from one
(for events with two $b$ tags) to six (for events with no $b$ tags).

\subsection{Mass reconstruction results}
\label{ssec:mrecoResults}

Typical reconstructed top quark mass distributions for signal Monte
Carlo ($\mtop=\gevcc{178}$) are shown for the four event categories as
the light histograms in \fig{f:sampletemp}. Each event in the sample
that passes both event selection and the $\chisq$ cut contributes
exactly one entry to these histograms. The distributions peak near the
generated mass of $\gevcc{178}$. But there is not an exact
correspondence between the generated mass and the mean or peak position of the
reconstructed mass. Differences can arise when ISR/FSR jets are
selected instead of the \ttbar decay products; even with the correct
jets, the fit may choose the wrong jet-quark assignment. In
particular, the broader shape, beneath the relatively sharp peak at
$\gevcc{178}$, comprises events where an incorrect permutation has
been chosen in the fit.  The dark histograms in the same figure show
the reconstructed mass distributions for events where the four leading
jets correspond to the four quarks from \ttbar decay, and where the
correct jet-quark assigment is chosen by the fit.  These histograms
have much smaller tails than the overall distributions, and account
for 47\% of the \twotag sample, 28\% of the \onetagt sample, 18\% of
the \onetagl, and 20\% of the \zerotag category.

\begin{cfigure}
\includegraphics{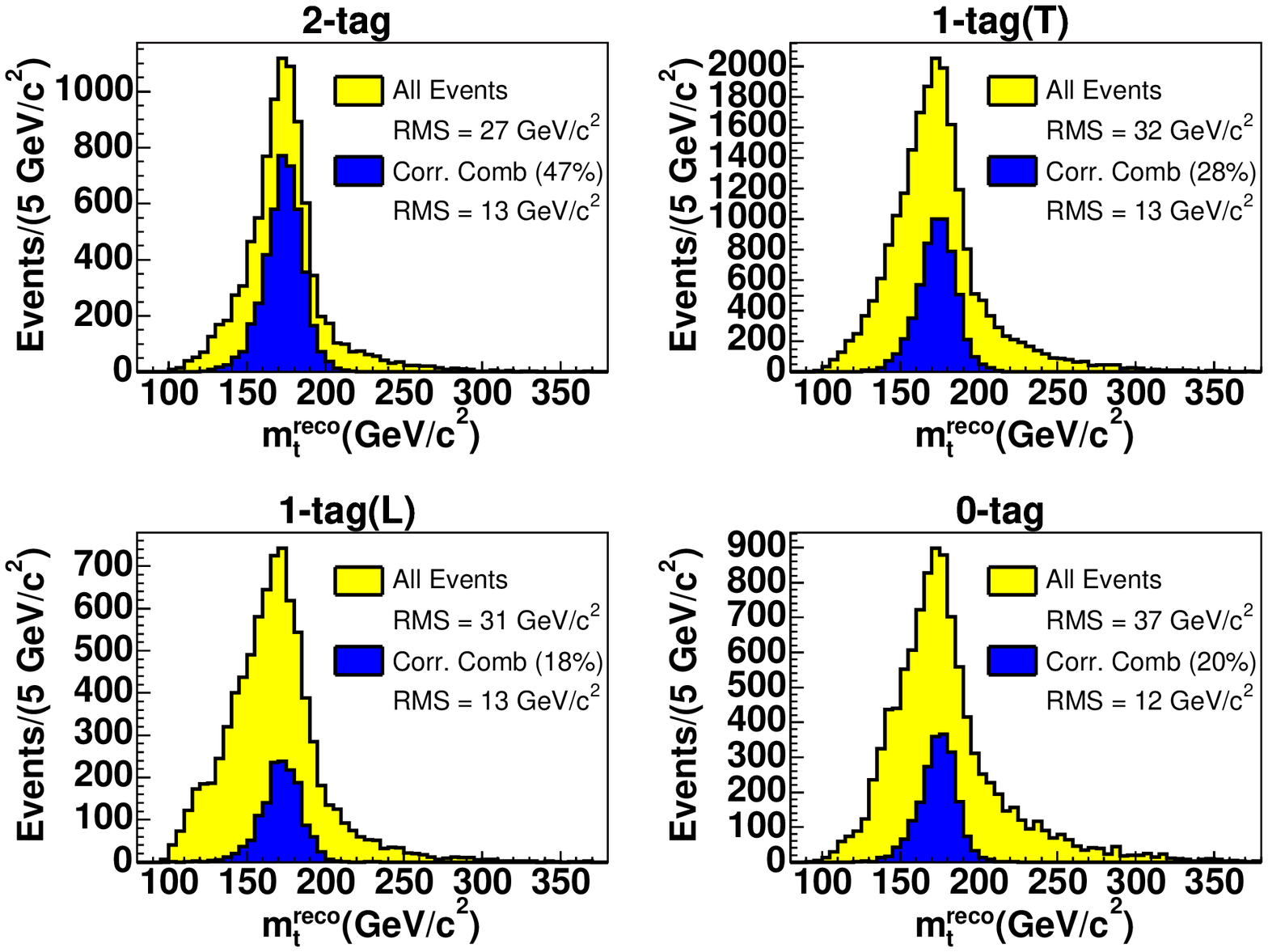}
\caption[\gevcc{178} reconstructed mass templates.]
{The light histograms show the reconstructed top quark mass
distribution for the $\gevcc{178}$ {\sc herwig} \ttbar sample at the
nominal jet energy scale.  Overlaid are darker histograms of the
reconstructed mass distributions using the subset of events for which
the leading four jets are matched (within $\Delta R=0.4$) to the four
quarks from the \ttbar decay and the correct jet-quark assignment has
the lowest \chisq.  Distributions are shown for \twotag (upper left),
\onetagt (upper right), \onetagl (lower left), and \zerotag (lower
right) events.}
\label{f:sampletemp}
\end{cfigure}

The corresponding dijet mass distributions for the $W$ boson
reconstruction are shown in \fig{f:samplemjjtemp} for the four
subsamples. Each event contributes 1, 3, or 6 entries to the
distributions, depending on the number of $b$ tags.  One sees a clear
$W$ boson mass signal, with a peak near the nominal $W$ boson mass of
\gevcc{80}.  The peak becomes more evident with increasing numbers of
$b$-tagged jets in the event, a consequence of the decreasing number
of combinations for $W$ boson jet daughters.

\begin{cfigure}
\includegraphics{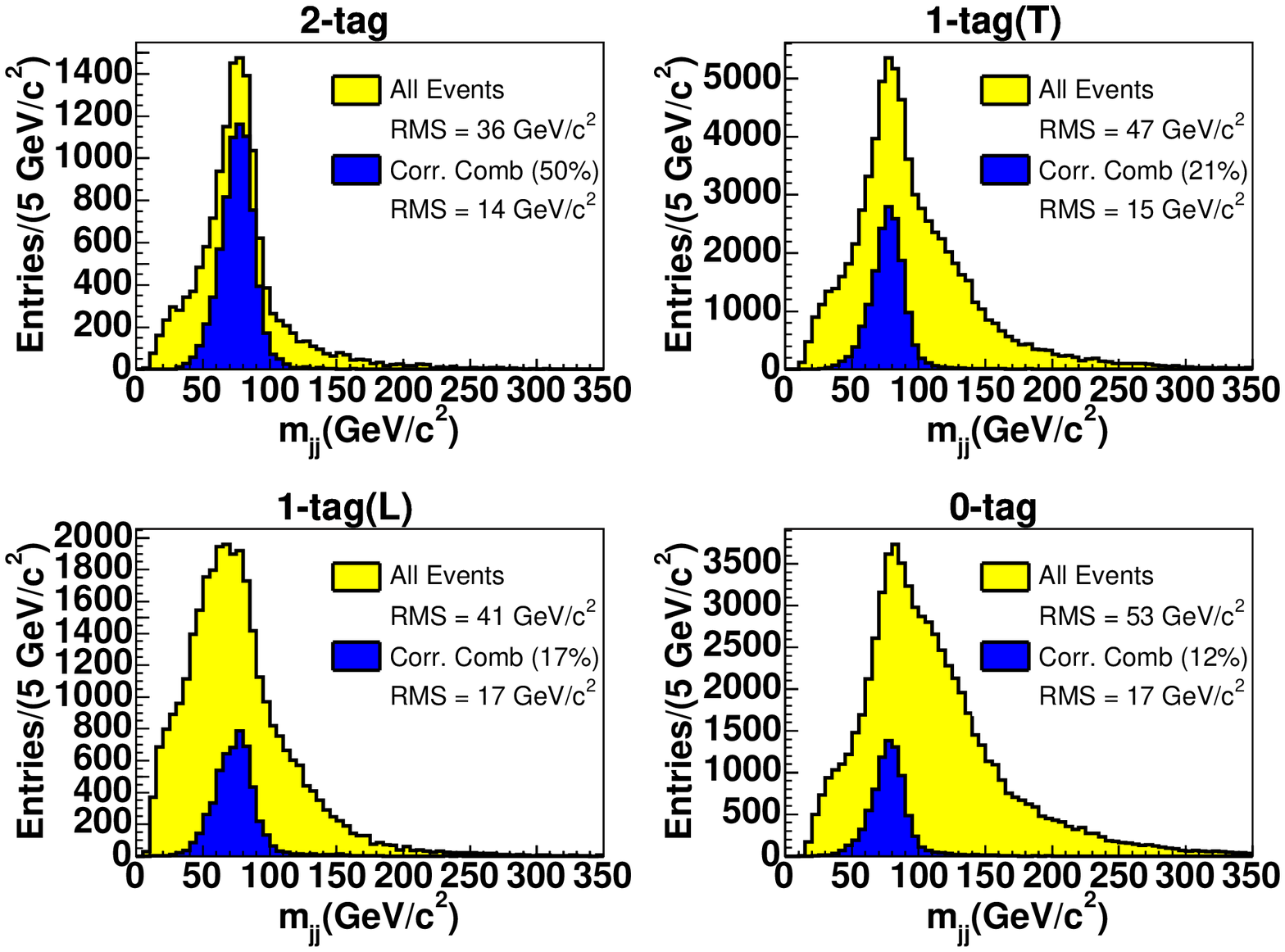}
\caption[\gevcc{178} dijet mass templates.]
{The reconstructed dijet mass distributions for the \gevcc{178} {\sc
herwig} \ttbar sample at the nominal jet energy scale. Overlaid are
darker histograms of the reconstructed mass distributions using the
subset of events for which the leading four jets include two jets
matched (within $\Delta R=0.4$) to the two quarks from the hadronic
$W$ decay, and plotting just the invariant mass of those two jets.
Distributions are shown for \twotag (upper left), \onetagt (upper
right), \onetagl (lower left), and \zerotag (lower right) events.}
\label{f:samplemjjtemp}
\end{cfigure}

Some results of the mass reconstruction on Monte Carlo \ttbar signal
($\mtop=\gevcc{178}$) and background samples are given in
\tab{t:massRecoMC}. The four subsamples have significantly different
\mreco and \mjj shapes for \ttbar signal and background, as
evidenced by their reconstructed mass mean and RMS values.  The \chisq
cut efficiency is lowest for \twotag events, especially for the
background processes, because there are fewer allowed jet-quark
assignments and thus fewer chances to pass the \chisq cut. The
efficiencies for signal events vary only weakly with the generated top
quark mass, and for the purposes of this analysis are assumed to be
constant.  The means of the background reconstructed mass
distributions are primarily driven by the jet cuts (see
\tab{t:eventTypes}).

\begin{table}
\caption[Result of mass reconstruction on Monte Carlo samples.]
{Monte Carlo samples of \ttbar signal and of background processes with
the expected relative weights are run through the \chisq mass
fitter. For signal and background in each of the four event
categories, the table shows the efficiency of the \chisq cut and the
mean and RMS of the resulting reconstructed mass distributions.  The
signal sample has $\mtop=\gevcc{178}$, and the nominal jet energy
scale is used for all events.}
\label{t:massRecoMC}
\begin{ruledtabular}
\begin{tabular}{lddddd}
Sample
& \multicolumn{1}{c}{\chisq cut}
& \multicolumn{2}{c}{\mreco (\gevccnoarg)}
& \multicolumn{2}{c}{\mjj (\gevccnoarg)} \\
Description
& \multicolumn{1}{c}{eff.}
& \multicolumn{1}{c}{Mean}
& \multicolumn{1}{c}{RMS}
& \multicolumn{1}{c}{Mean}
& \multicolumn{1}{c}{RMS} \\
\hline
\multicolumn{6}{l}{\textbf{Signal}} \\
\twotag   & 0.65 & 173.9 & 26.6 & 76.8 & 34.5 \\
\onetagt  & 0.85 & 174.0 & 31.8 & 98.1 & 49.1 \\
\onetagl  & 0.80 & 167.4 & 30.8 & 77.8 & 41.7 \\
\zerotag  & 0.91 & 179.3 & 36.9 & 112.6 & 57.9 \\
\hline
\multicolumn{6}{l}{\textbf{Background}} \\
\twotag   & 0.38 & 160.2 & 35.1 & 77.2 & 53.3 \\
\onetagt  & 0.73 & 166.4 & 42.2 & 95.8 & 59.7 \\
\onetagl  & 0.71 & 153.7 & 37.3 & 67.7 & 46.3 \\
\zerotag  & 0.83 & 182.6 & 46.5 & 114.0 & 69.3 \\
\end{tabular}
\end{ruledtabular}
\end{table}

The reconstructed top quark and dijet mass distributions for the
165~events found in the data can be seen in \fig{f:datamasses}. These
events consist of both \ttbar signal and background events.
\Fig{f:chi2} shows distributions of \chisq values from the top quark
mass reconstruction in data and simulated events, where the
distributions from simulation contain the expected mixtures of signal
and background events.  Kolmogorov-Smirnov tests, with probability
normalized using many trial distributions randomly selected from the
Monte Carlo predictions, show that the distributions agree well,
indicating that kinematic quantities and resolutions are correctly
simulated.

\begin{cfigure}
\includegraphics[height=.38\textheight]{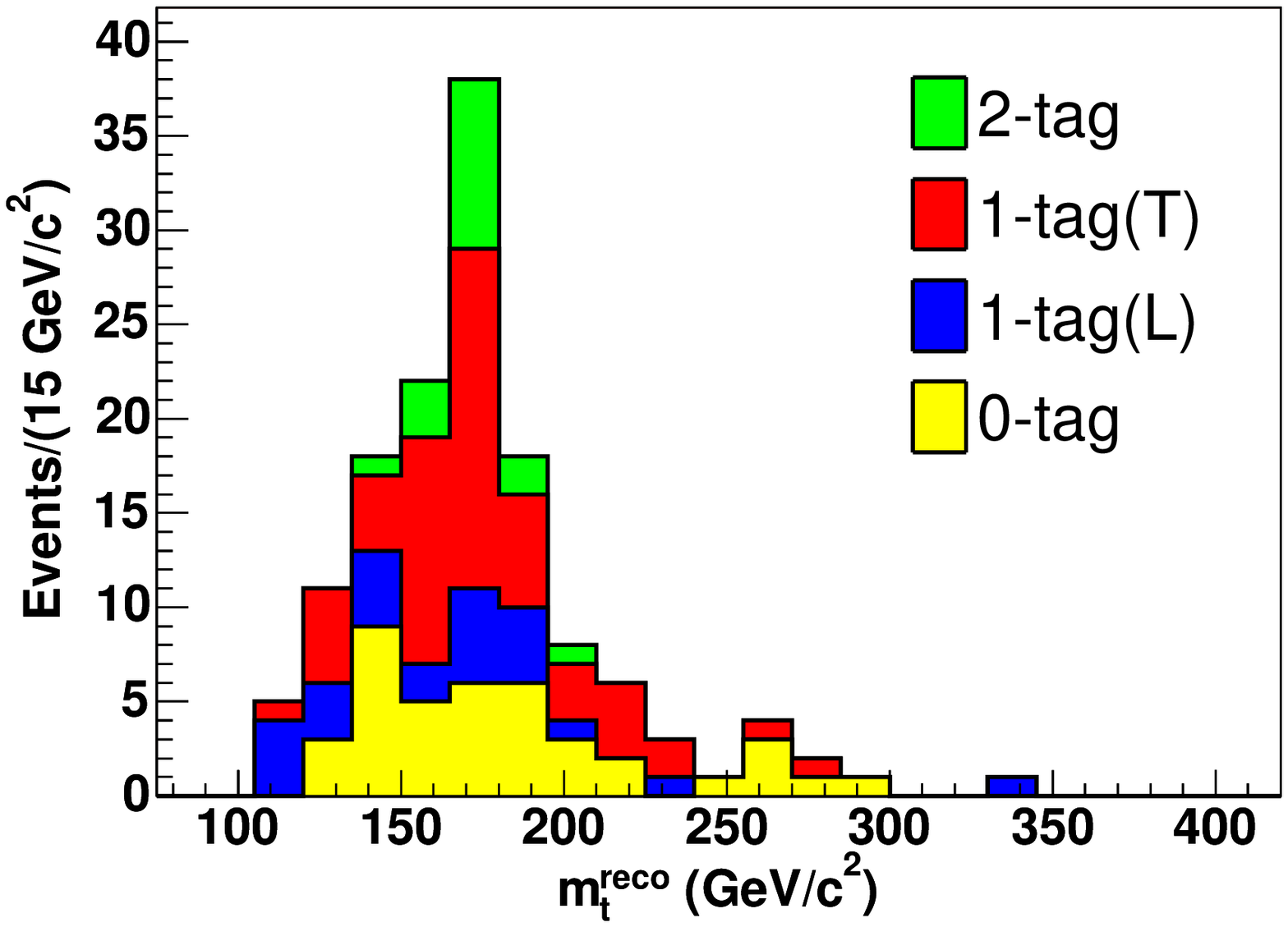}
\includegraphics[height=.38\textheight]{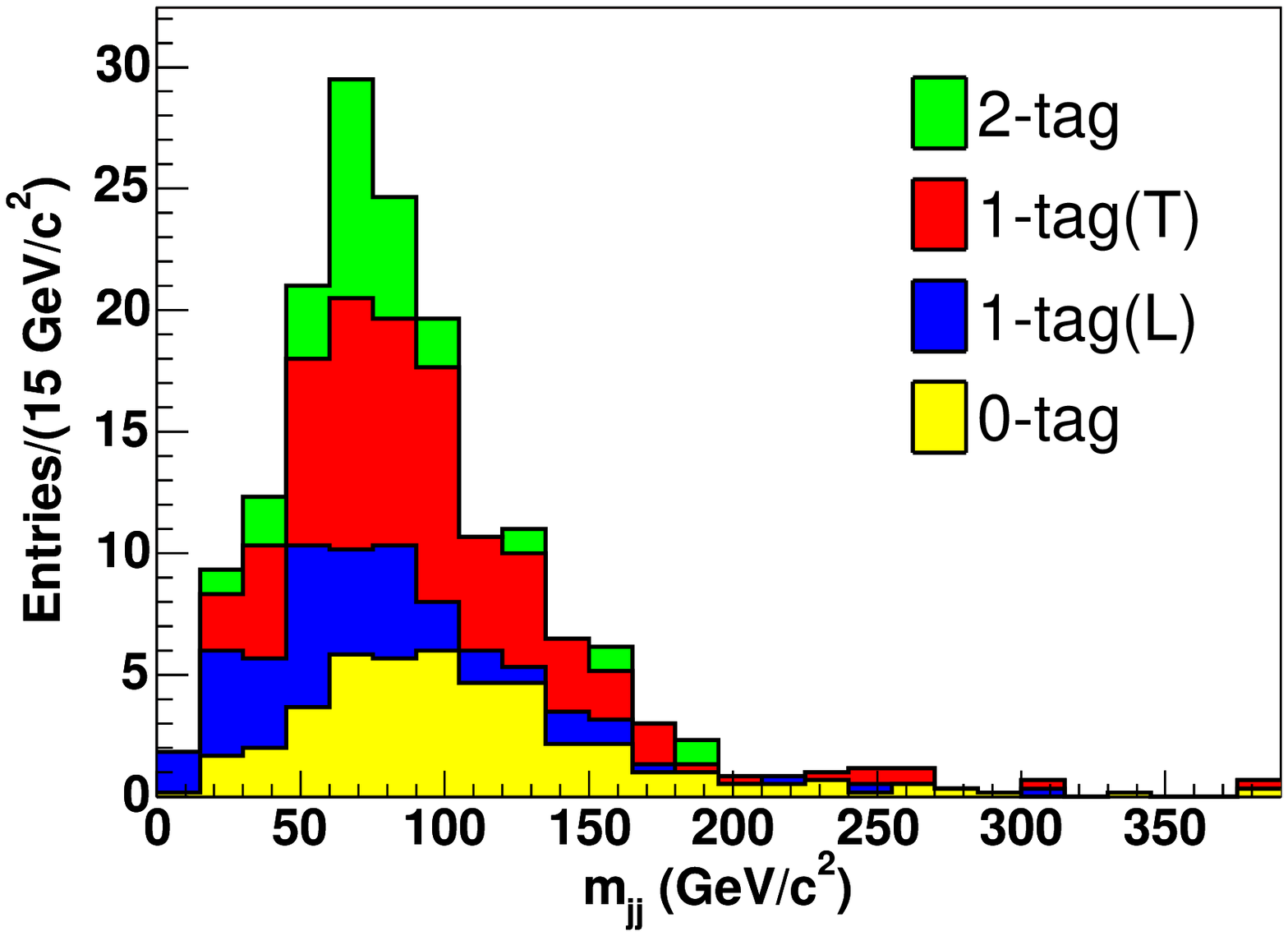}
\caption[Reconstructed masses of data events.]
{The top histogram shows the reconstructed top quark mass \mreco, and
the bottom histogram shows the dijet reconstructed mass \mjj, with
events from the four subsamples represented by separate stacked
histograms. In the \mjj plot, each event has a different number of jet
pairs, depending on the number of $b$ tags in the event, but the
entries are weighted so that the total contribution from each event is
one unit.  The highest \mjj bin contains overflow entries.}
\label{f:datamasses}
\end{cfigure}

\begin{cfigure}
\includegraphics{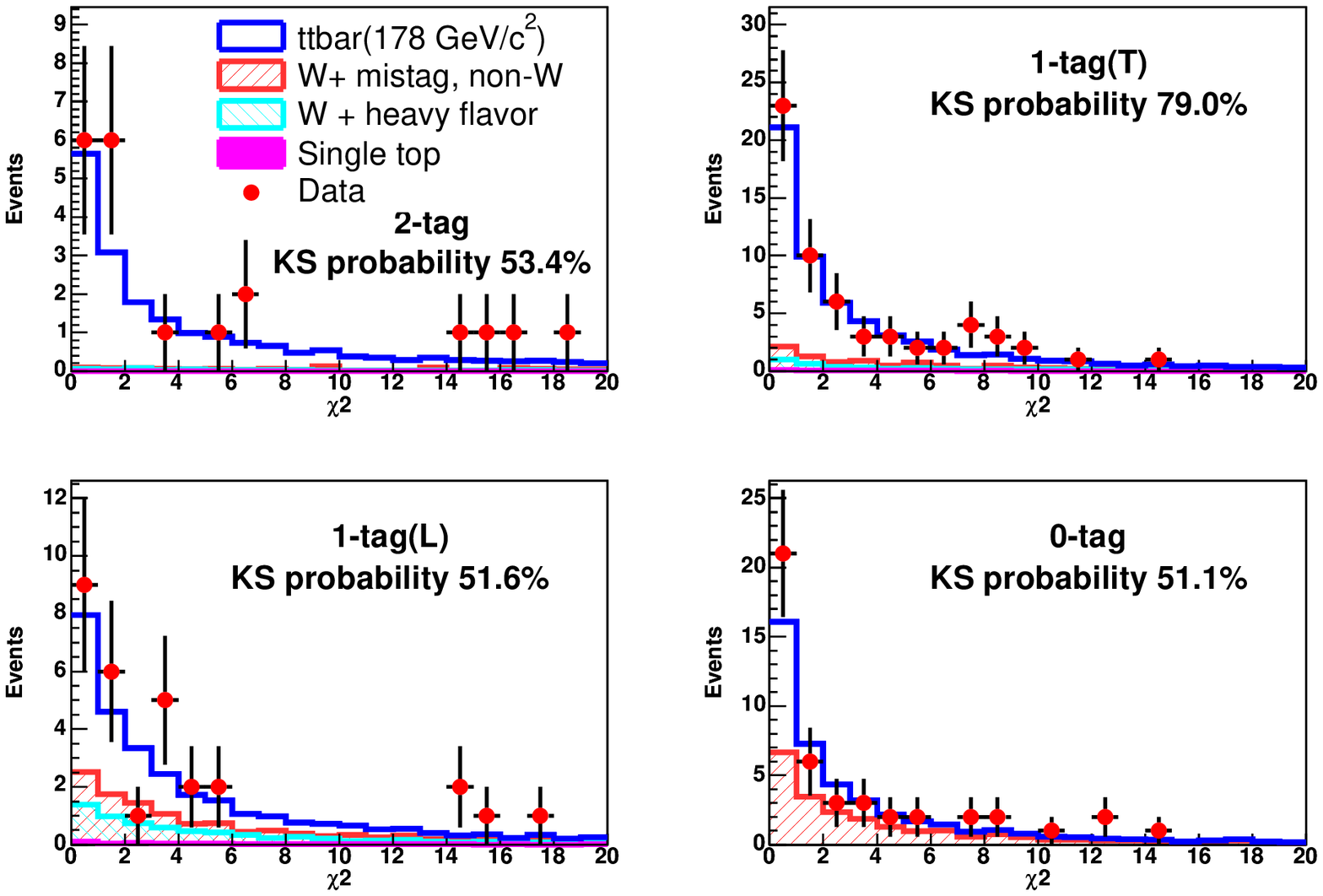}
\caption[Comparison of \chisq distribution in data and MC.]
{The \chisq distribution is shown for data events and for
signal and background simulated events in the expected ratio.
Distributions are shown for \twotag (upper left), \onetagt (upper right),
\onetagl (lower left), and \zerotag (lower right) events.}
\label{f:chi2}
\end{cfigure}

\section{Top Quark Mass Fitting}
\label{sec:fitting}

The distribution of reconstructed mass (either \mreco or \mjj) for a
particular top quark mass (or background process) and jet energy scale
is referred to as a template.  We compare the reconstructed top quark
mass distribution and the dijet mass distribution from data to the
Monte Carlo templates to measure simultaneously the top quark mass and
the jet energy scale.  First, probability density functions (p.d.f.'s)
for the reconstructed top quark and dijet masses are determined for
signal events and background events in each subsample by fitting a
functional form to the corresponding templates; the signal p.d.f.'s
depend on the top quark mass and jet energy scale.  The shift in
jet energy scale is given by \jes, which is the relative shift in
units of the nominal uncertainty in the jet energy scale derived from
the extrinsic calibration procedures (Section~\ref{ssec:jetSyst}).
Although the jet energy scale uncertainty varies with jet momentum and
pseudorapidity, a one unit shift in the \jes parameter is
approximately equivalent to a 3\% shift in the jet energy scale for
jets in \ttbar events.  We perform an unbinned likelihood fit to
determine the values of the top quark mass and jet energy scale that
best describe the data.  At the end of this section, we describe a
number of checks of the method using simulated events.

\subsection{Parameterization of Signal and Background Shapes}
\label{ssec:parameterization}

Since templates are available only at discrete values of top
quark mass and jet energy scale, the signal reconstructed mass
distributions in each subsample are parameterized by a flexible
functional form as a function of top quark mass and jet energy
scale in order to smooth the distributions and interpolate between the
templates.

For background events, the parameterization has no dependence on top
quark mass or jet energy scale; a single p.d.f.\ is used to describe
each background reconstructed mass shape in each subsample.  In
principle, a shift in the jet energy scale will change the shape of
the background templates.  However, we have determined from studies of
the background that the shape of the background templates are
insensitive to the jet energy scale.  Rather, the overall rate of
background events does show some sensitivity to the jet energy
response, and this uncertainty is incorporated into the uncertainty in
the rate of background events in the sample.

The same parameterizations are used for both \mreco and \mjj signal
p.d.f.'s, although of course the fitted parameters are different. In
the case of the background, different functional forms are required to
fit well the \mreco and \mjj templates.

\subsubsection{Signal shape parameterization}
\label{sssec:signalparam}

Signal templates are produced using sets of Monte Carlo samples with
the input top quark mass at 2.5--5 $\gevccnoarg$ intervals from
\gevcc{130} to \gevcc{230} and the jet energy scale varying from
$-3.0$ to $+3.0$ in steps of 0.5.  Examples of the template shapes
from each event category are given in \fig{f:sampletemp} (\mreco) and
\fig{f:samplemjjtemp} (\mjj). \Tab{t:sampletemps} shows the evolution
of the mean, most probable value, and RMS of the reconstructed top
quark mass templates as a function of top quark mass using
selected generated mass samples and the nominal jet energy scale.
\begin{table}
\caption[Signal template evolution.]
{The evolution of the \mreco template parameters is demonstrated using
selected signal Monte Carlo samples with generated top quark mass of
$\gevcc{145}$, $\gevcc{165}$, $\gevcc{185}$, and $\gevcc{205}$, with
the nominal jet energy scale. The mean, most probable value (MPV), and
RMS of the template are given for each subsample in each generated
mass. The mean and MPV of the templates are driven by the jet $\et$
cuts, and the widths are dominated by the fraction of events with
correct jet-quark assignments.}
\label{t:sampletemps}
\begin{ruledtabular}
\begin{tabular}{lcdddd}
  & $\mtop$ &
  \multicolumn{1}{c}{\twotag} &
  \multicolumn{1}{c}{\onetagt} &
  \multicolumn{1}{c}{\onetagl} &
  \multicolumn{1}{c}{\zerotag} \\
\hline
                & 145 & 151.5 & 155.1 & 147.1 & 163.9 \\
Mean            & 165 & 164.5 & 166.5 & 158.8 & 174.3 \\
$(\gevccnoarg)$ & 185 & 178.2 & 179.1 & 171.1 & 185.3 \\
                & 205 & 193.5 & 190.5 & 182.7 & 194.2 \\
\hline
                & 145 & 144.5 & 144.0 & 140.7 & 145.0 \\
MPV             & 165 & 163.8 & 159.5 & 156.5 & 159.3 \\
$(\gevccnoarg)$ & 185 & 179.9 & 178.1 & 171.7 & 179.5 \\
                & 205 & 198.5 & 194.7 & 185.4 & 193.9 \\
\hline
                & 145 & 25.1 & 31.7 & 28.5 & 39.2 \\
RMS             & 165 & 24.8 & 31.8 & 28.6 & 39.1 \\
$(\gevccnoarg)$ & 185 & 27.1 & 32.3 & 32.1 & 37.7 \\
                & 205 & 28.6 & 33.6 & 34.1 & 37.7 \\
\end{tabular}
\end{ruledtabular}
\end{table}

We derive from these distributions parametrized templates that are a
smoothly varying function of top quark mass and jet energy scale.
For any given top quark mass and jet energy scale, the
probability to observe a particular reconstructed mass is specified by
a function consisting of two Gaussians---intended to account for the
well reconstructed quantities---plus a gamma distribution---intended
to account for cases where the incorrect jets are used to reconstruct
the top quark or $W$ masses.  The 9 parameters necessary to specify
this combination of functions are themselves assumed to depend
linearly on the top quark mass and jet energy scale, so that the
full set of p.d.f.'s is specified by 27 parameters. This assumed
functional form works well in the limited range of top quark masses
and jet energy scales considered; as an example, letting the nine
parameters have quadratic dependence on \mtop or \jes does not improve
the fit. Thus the parameterization is as follows:
\begin{eqnarray}
\label{e:sigparam}
\lefteqn{P_{s}(m;\mtop,\jes)=} \nonumber \\
&&\alpha_{7} \cdot \frac{\alpha_{2}^{1 + \alpha_{1}}}
	{\Gamma(1 + \alpha_{1})}
	\cdot (m - \alpha_{0})^{\alpha_{1}} \cdot \exp\left(-\alpha_{2} (m - \alpha_{0})\right)
	\nonumber \\
&+& \alpha_{8} \cdot
	\frac{1} {\alpha_{4}\sqrt{2\pi}} \cdot
	\exp\left(\frac{-(m-\alpha_{3})^{2}} {2\alpha_{4}^{2}}\right) \nonumber \\
&+& (1 - \alpha_{7} - \alpha_{8}) \cdot
	\frac{1} {\alpha_{6}\sqrt{2\pi}} \cdot
	\exp\left(\frac{-(m-\alpha_{5})^{2}} {2\alpha_{6}^{2}}\right);
\end{eqnarray}
where
\begin{equation*}
\alpha_{i} = p_{i} + p_{i+9} \cdot (\mtop-175) + p_{i+18} \cdot (\jes).
\end{equation*}
The variable $m$ in $\gevccnoarg$ refers to the reconstructed top
quark or dijet mass, \mtop in $\gevccnoarg$ refers to the top
quark mass, and \jes refers to the shift in the jet energy scale
from that determined from our calibrations.  These template
parametrizations are normalized so that, for a given top quark
mass \mtop and jet energy scale \jes, the integral over all
reconstructed masses $m$ is unity.

A binned likelihood fit is used to determine the 27 parameter values
both for the \mreco templates and for the \mjj templates.  The \chisq
is calculated between the MC samples and the prediction from the fit,
after rebinning to ensure that each bin has at least five predicted
events.  The resulting \chisq values are given in \tab{t:parfits},
along with the number of degrees of freedom. Clearly the corresponding
probabilities are small, not surprising considering the limited
flexibility of the functional form and the large statistics of the
templates.  But the method check and calibration of
Section~\ref{ssec:methodCheck} show that the disagreement between the
templates and the parameterizations is not large enough to have a
significant effect on the measurement.

In \fig{f:fitvis}, four signal templates at varying generated masses
are shown overlaid with the fitted parameterization evaluated at each
mass. This figure exhibits the changing shape of the
reconstructed mass templates as a function of top mass. \Fig{f:fitmjjvis} shows the \mjj templates with varying jet
energy scale, overlaid with the fitted parameterization. One sees that
the location of the $W$ boson peak is sensitive to the jet energy
scale.

\begin{cfigure}
\includegraphics{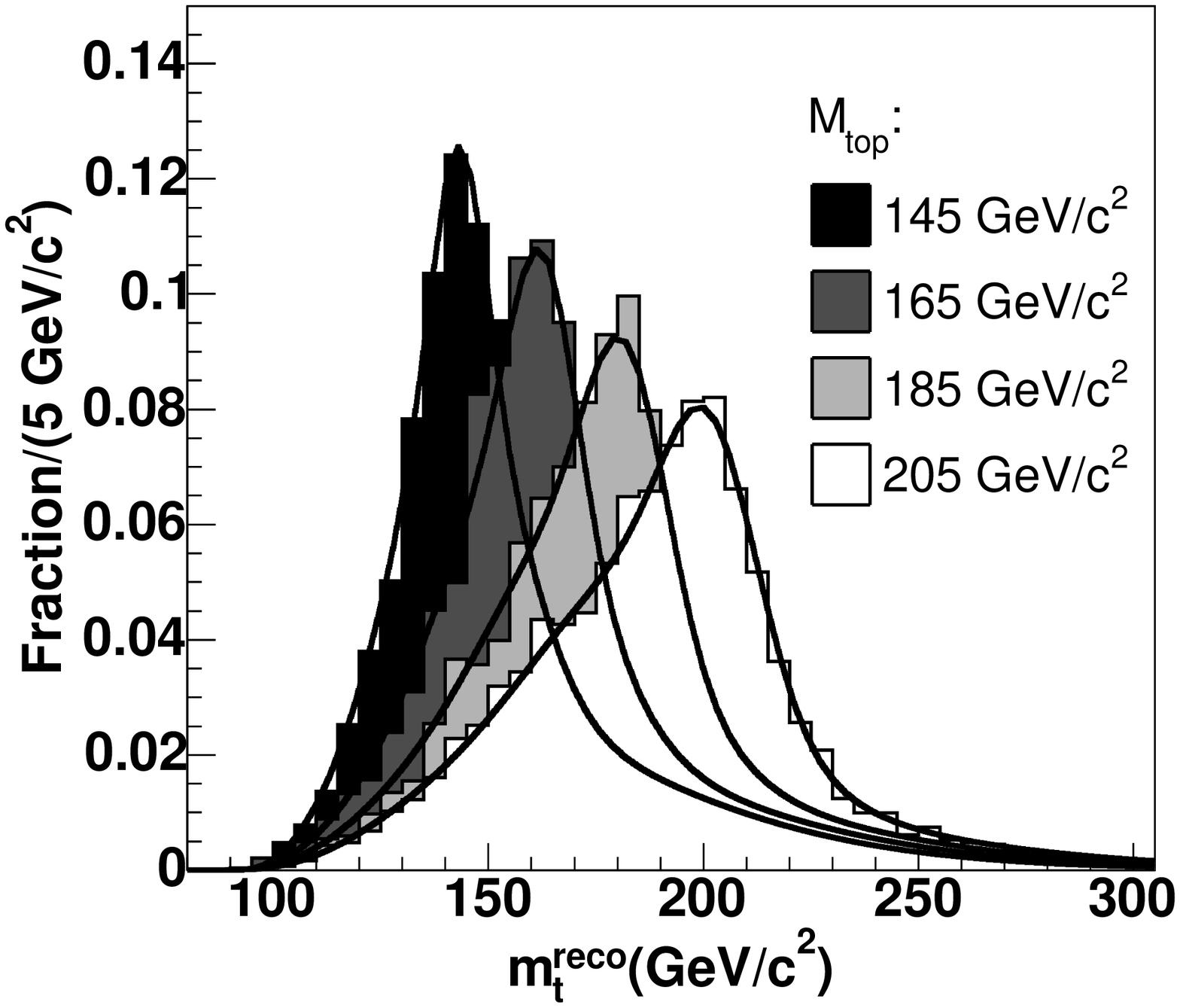}
\caption[Four \mreco templates at different masses, with fits overlaid.]
{Four \mreco signal templates for the \onetagt sample are shown, with
top quark masses ranging from \gevcc{145} to \gevcc{205} and with \jes
set to 0. Overlaid are the fitted parameterizations at each generated
mass, taken from the full parameterization given in
Eq.~\ref{e:sigparam}.}
\label{f:fitvis}
\end{cfigure}

\begin{cfigure}
\includegraphics{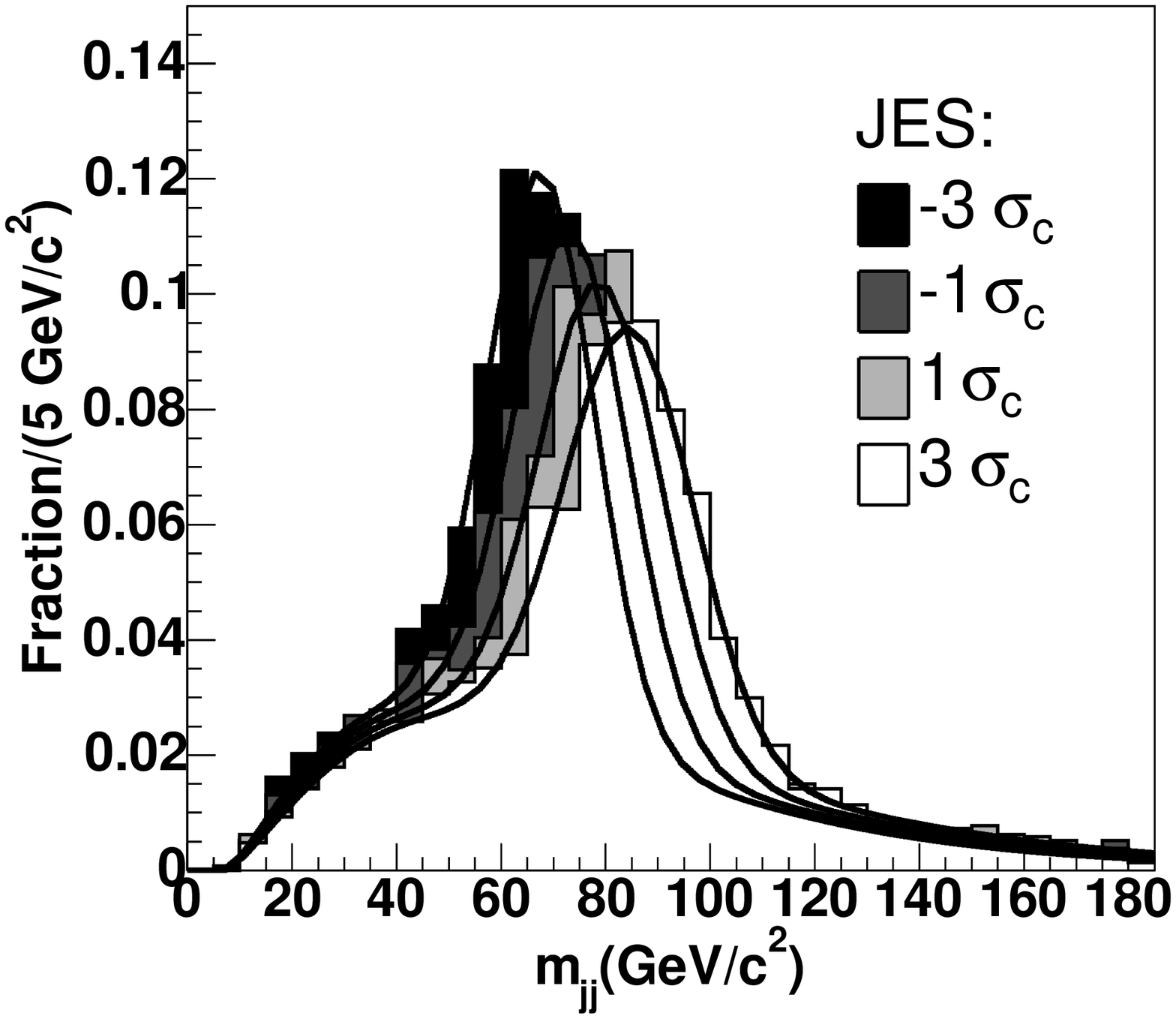}
\caption[Four \mjj templates at different jet energy scales, with fits overlaid.]
{Four \mjj signal templates for the \twotag sample are shown, with
\jes values ranging from \sigunit{-3} to \sigunit{3} and with the 
top quark mass set to \gevcc{180}.  Overlaid are the fitted
parameterizations at each value of jet energy scale, taken from the
full parameterization given in Eq.~\ref{e:sigparam}.}
\label{f:fitmjjvis}
\end{cfigure}

\subsubsection{Background shape parameterization}
\label{sssec:bkgdparam}

Monte Carlo simulations of the various processes listed in
Section~\ref{ssec:backgrounds} are used to model the reconstructed top
quark mass shape and dijet mass shape for background processes. When
possible, a single large-statistics sample is used to represent
several background processes that have similar template shapes.

For the tagged backgrounds, the $W$ + heavy flavor processes
($W\bbbar$, $W\ccbar$, $Wc$) all have similar reconstructed mass
shapes, as shown in \fig{f:bkgdcomp} for \mreco in the \onetagt
sample, and thus are all modeled with a high-statistics $W\bbbar$
simulated sample.  $WW$ and $WZ$ events, a negligible contribution to
the total expected background, are also included in this category.
The shapes for the three subsamples with tagged events are found by
reconstructing the simulated events exactly as is done for the data
and signal Monte Carlo. Similarly, the simulated s- and t-channel
single top quark events are used to obtain corresponding mass
templates.

\begin{cfigure}
\includegraphics{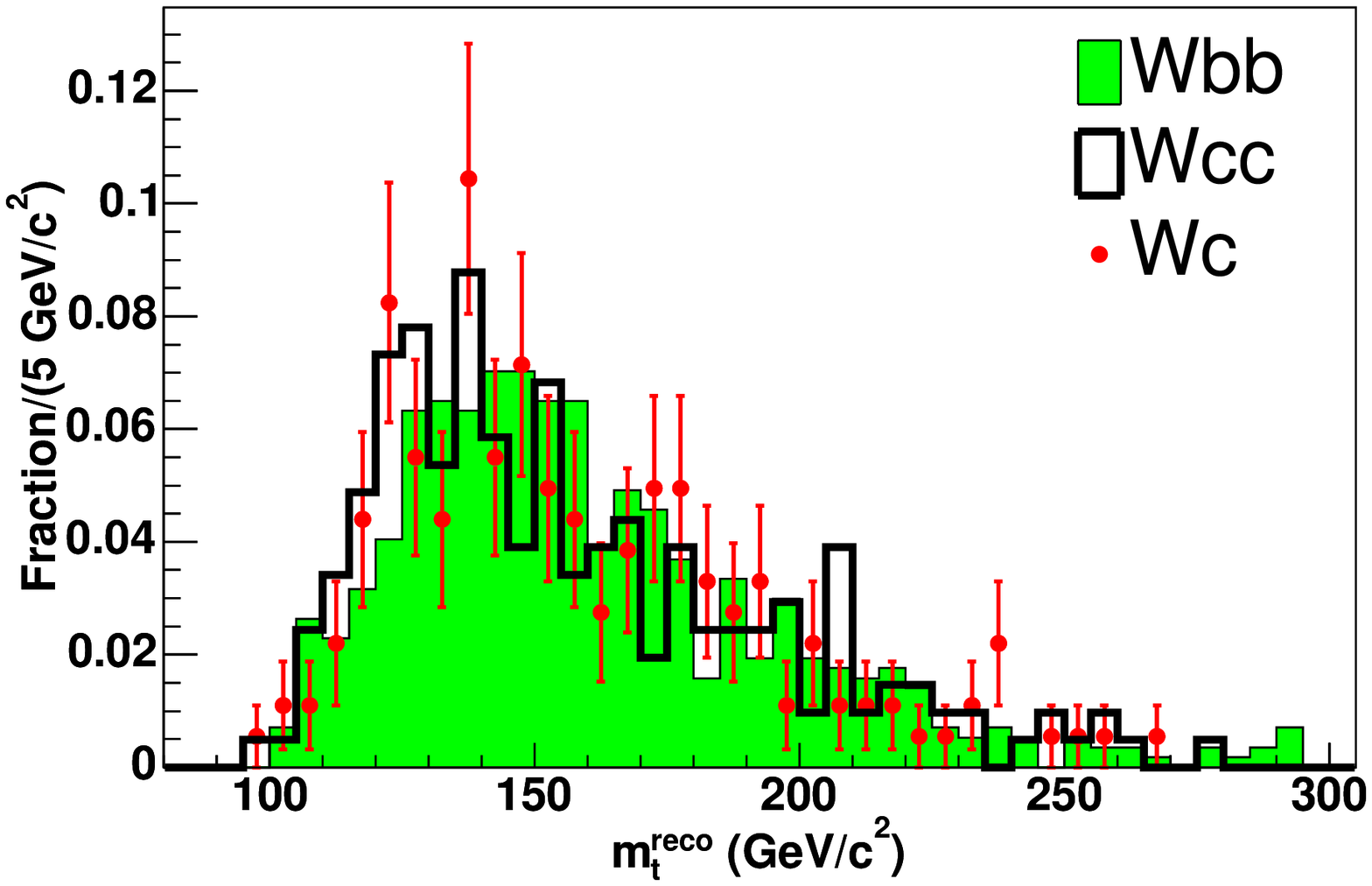}
\caption[Plot comparing Wbb, Wcc, Wc templates.]
{The templates for $W\ccbar$ and $Wc$ background processes are
compared to the high-statistics $W\bbbar$ template for all tagged
events. The agreement is good in both cases, so the $W\bbbar$ template
is used to represent all $W$ + h.f.\ processes.}
\label{f:bkgdcomp}
\end{cfigure}

The mass templates for the \wjets backgrounds in the tagged
subsamples, i.e.\ ``mistags,'' are not obtained using the Monte Carlo
$b$ tagging, which is not expected to model well the rate or kinematic
dependences of fake tags. Instead, a mistag matrix, derived from the
data, is used to give the probability for a jet to be falsely tagged
as a function of its $\et$, $\phi$, $\eta$, number of tracks, and the
$\Sigma\et$ for all jets in the event.  Then for each simulated event
($W$ + 4 partons, generated by {\sc alpgen} and showered by {\sc
herwig}), every possible tag configuration on the leading four jets is
considered. For every tag configuration, the fit with lowest \chisq
among the jet-quark assignments consistent with the assumed tags is
selected, and the appropriate mass template is filled with a weight
corresponding to the probability of observing that set of tags. The
result is a weighted template for the mistag backgrounds.

The backgrounds that are least amenable to Monte Carlo modeling arise
from QCD background events, i.e.\ events with no real $W$ to produce
the isolated lepton and $\met$. These events are difficult to
simulate, but can be studied by selecting events in the data with
non-isolated leptons, which are enriched in this type of background,
but kinematically similar to events chosen in the default selection.

The mass reconstruction described in Section~\ref{sec:massRecon} is
expected to produce similar results for QCD background events and
\wjets background events. This is because the leptonic $W$ system, in
which these types of events differ, does not have a strong effect on
the mass reconstruction, since \met is poorly measured and since the
$W$ mass is constrained in the \chisq expression. In the kinematic
properties of the jets, to which the mass reconstruction is very
sensitive, these two types of events are similar since in both cases
the jets arise from hard QCD radiation.

Indeed, within the limited statistics available, the reconstructed
mass distributions of the QCD-enriched data events are consistent with
those of simulated \wjets events.  Given these similarities, the $W$ +
jets reconstructed mass templates are used also for the expected
contributions from QCD for both the reconstructed top quark mass and
the dijet invariant mass.  An additional check, treating the subset of
QCD events where the primary lepton is a jet misidentified as an
electron, is performed using a large QCD-dominated dataset with at
least five jets. One jet with a large fraction of its energy in the
electromagnetic calorimeter is assigned to ``fake'' an electron, and
the mass reconstruction proceeds under that hypothesis. Very good
agreement is found between the reconstructed mass distributions of
these events and those of simulated \wjets events.  A systematic
uncertainty on the background modeling (see
Section~\ref{ssec:otherSyst}) is assigned using the differences
between the templates obtained from these three samples: \wjets
events, events with non-isolated leptons, and events with one jet
assigned to ``fake'' an electron.

The background for the \zerotag subsample is treated separately from
the others. The dominant process is $W$ + jets, with a smaller
($\sim20\%$) contribution expected from non-$W$ (QCD) events.  Since
we model the reconstructed mass of QCD events using \wjets events, the
entire \zerotag background shape comes from $W$ + 4 parton Monte Carlo
events, simulated by {\sc alpgen} and showered by {\sc herwig}.

We do not allow the normalization of each background contribution to
vary independently in the final likelihood fit. Instead, for each
subsample, the templates from all background processes are combined in
their expected ratios according to \tab{t:backgroundnumber}. A single
function is fitted to the combined background for each subsample and
is used to describe the background shape in the final likelihood fit
(Section~\ref{ssec:likelihood}). The overall background normalization
for each subsample is then permitted to vary, within its constraint
where applicable.

We determine the p.d.f.'s for the background reconstructed top quark
mass templates using a parameterization similar in spirit to that of
the signal, but simpler in form. First, there is no dependence on top
quark mass or jet energy scale.
Second, no narrow Gaussian peak is expected, so the full shape is
modeled by the integrand of the gamma distribution. Specifically,
\begin{eqnarray}
\label{e_bgparam}
P_{b}(\mreco) &=& \frac{p_{2}^{1 + p_{1}}} {\Gamma(1 + p_{1})}
   \cdot (\mreco - p_{0})^{p_{1}} \nonumber \\
&& \cdot \exp\left(-p_{2} (\mreco - p_{0})\right).
\end{eqnarray}
In the case of the \zerotag background events, a slightly more
sophisticated function is used to achieve a good fit:
\begin{eqnarray}
\label{e_bgparamzerotag}
P_{b}(\mreco) &=& p_{6} \frac{p_{2}^{1 + p_{1}}} {\Gamma(1 + p_{1})}
    \cdot (\mreco - p_{0})^{p_{1}} \nonumber \\
&&  \cdot \exp\left(-p_{2} (\mreco - p_{0})\right) \nonumber \\
&+& (1-p_{6}) \frac{p_{5}^{1 + p_{4}}} {\Gamma(1 + p_{4})}
    \cdot (\mreco - p_{3})^{p_{4}} \nonumber \\
&&  \cdot \exp\left(-p_{5} (\mreco - p_{3})\right).
\end{eqnarray}

In the background events, the following parameterizations are used to
fit the reconstructed dijet mass templates. For the background samples
with $b$ tags,
\begin{eqnarray}
\label{e_bgparamtag_wjj}
P_{b}(\mjj) &=& \alpha_{5} \cdot \frac{\alpha_{2}^{1 + \alpha_{1}}}
    {\Gamma(1 + \alpha_{1})} \cdot (\mjj - \alpha_{0})^{\alpha_{1}}
    \nonumber \\
&&  \cdot \exp\left(-\alpha_{2} (\mjj - \alpha_{0})\right) \nonumber \\
&+& (1 - \alpha_{5}) \cdot \frac{1} {\alpha_{4}\sqrt{2\pi}} \nonumber \\
&&  \cdot \exp\left(\frac{-(\mjj-\alpha_{3})^{2}} {2\alpha_{4}^{2}}\right),
\end{eqnarray}
and for the \zerotag sample,
\begin{eqnarray}
\label{e_bgparamnotag_wjj}
P_{b}(\mjj) &=& \alpha_{7} \cdot \frac{\alpha_{2}^{1 + \alpha_{1}}}
    {\Gamma(1 + \alpha_{1})} \cdot (\mjj - \alpha_{0})^{\alpha_{1}}
    \nonumber \\
&&  \cdot \exp\left(-\alpha_{2} (\mjj - \alpha_{0})\right) \nonumber \\
&+& \alpha_{8} \cdot \frac{1} {\alpha_{4}\sqrt{2\pi}} \nonumber \\
&&  \cdot \exp\left(\frac{-(\mjj-\alpha_{3})^{2}} {2\alpha_{4}^{2}}\right)
    \nonumber \\
&+& (1 - \alpha_{7} - \alpha_{8}) \cdot \frac{1} {\alpha_{6}\sqrt{2\pi}}
    \nonumber \\
&&  \cdot \exp\left(\frac{-(\mjj-\alpha_{5})^{2}} {2\alpha_{6}^{2}}\right).
\end{eqnarray}

The final background templates for the reconstructed top quark mass
and dijet mass for the four subsamples are shown in \fig{f:bkgdtemps}
and \fig{f:bkgdmjjtemps}, respectively, overlaid with the fitted
parameterization.

\begin{table}
\caption[Parameterization fit information.]
{The \chisq and number of degrees of freedom are given for the signal
parameterization fits in each of the four subsamples.}
\label{t:parfits}
\begin{ruledtabular}
\begin{tabular}{cccc}
\twotag & \onetagt & \onetagl & \zerotag \\
\hline
\multicolumn{4}{c}{\chisq/n.d.o.f.\ for reconstructed top quark mass
(Eq.~\ref{e:sigparam})} \\
$13977/13081$ & $18736/17052$ & $14209/13444$ & $18791/16752$ \\
\hline
\multicolumn{4}{c}{\chisq/n.d.o.f.\ for dijet mass
(Eq.~\ref{e:sigparam})} \\
$19541/17000$ & $29410/25732$ & $23506/19827$ & $38315/30510$ \\
\end{tabular}
\end{ruledtabular}
\end{table}

\begin{cfigure}
\includegraphics{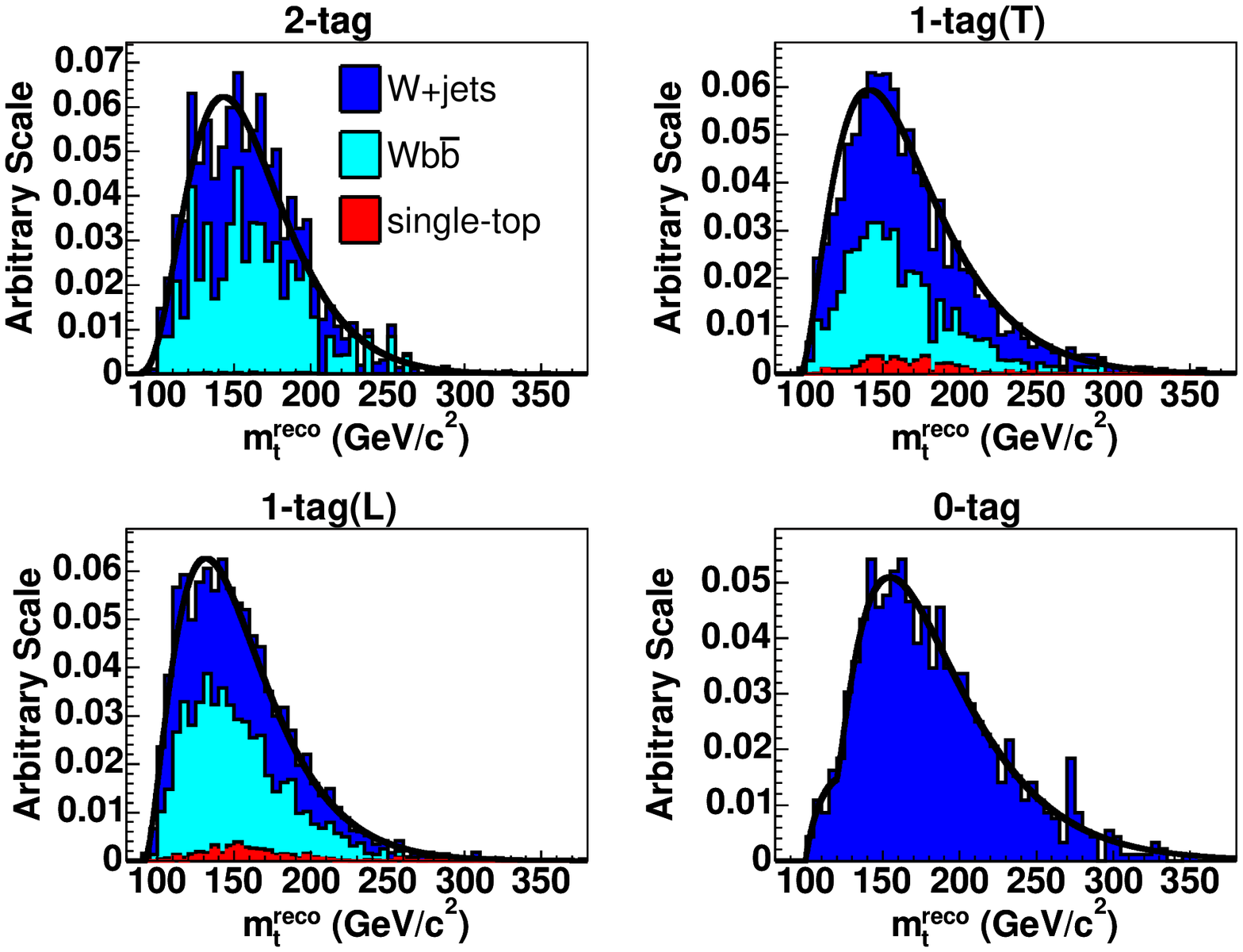}
\caption[Reconstructed top quark mass distributions for background.]
{Reconstructed top quark mass distributions of the combined
backgrounds in each subsample.  The contributions from different
background templates are shown stacked; overlaid are the fitted curves
(see Eq.~\ref{e_bgparam} and Eq.~\ref{e_bgparamzerotag}).}
\label{f:bkgdtemps}
\end{cfigure}

\begin{cfigure}
\includegraphics{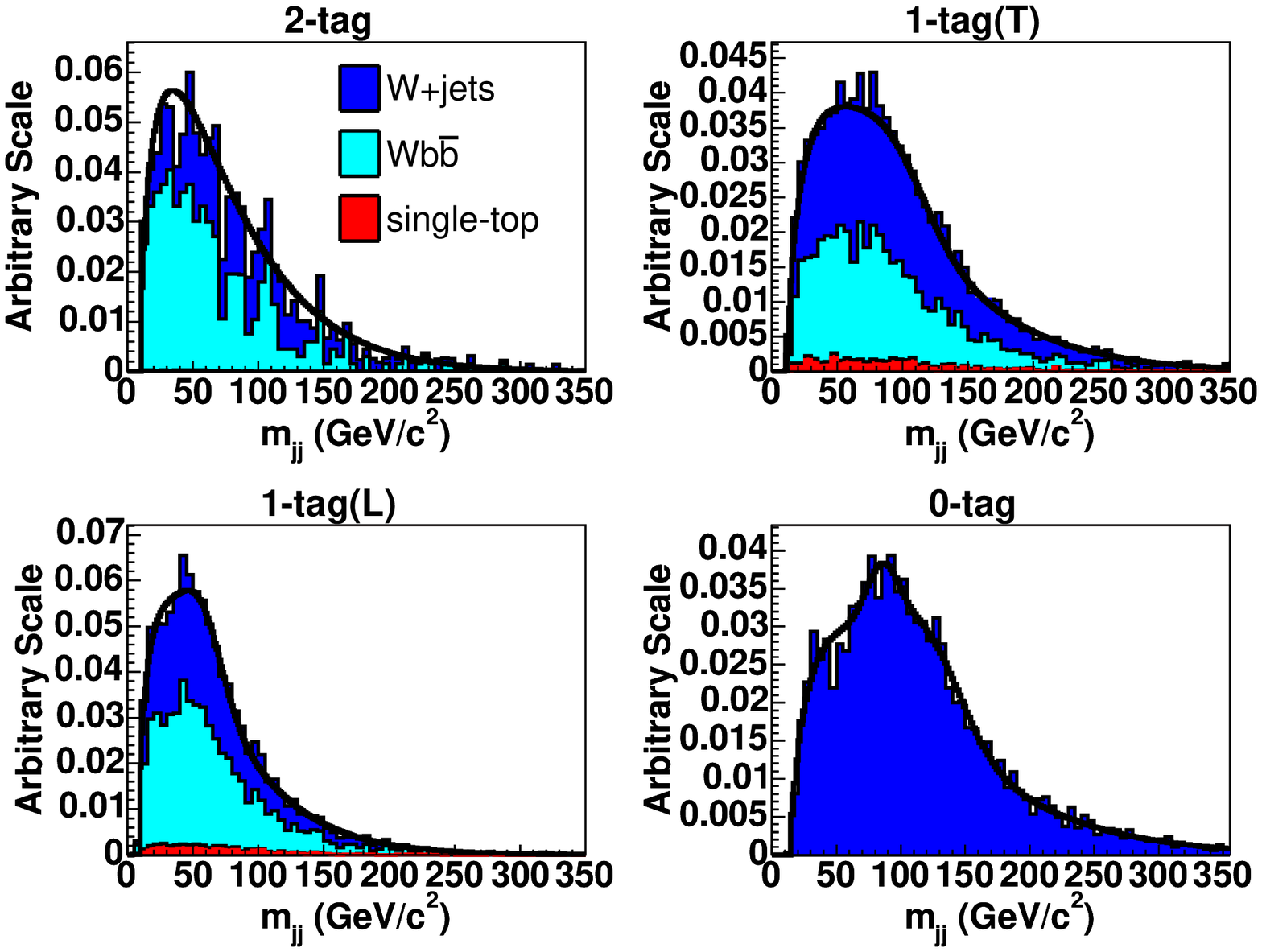}
\caption[Reconstructed dijet mass distributions for background.]
{Reconstructed dijet mass distributions of the combined backgrounds in
each subsample.  The contributions from different background templates
are shown stacked; overlaid are the fitted curves (see
Eq.~\ref{e_bgparamtag_wjj} and Eq.~\ref{e_bgparamnotag_wjj}).}
\label{f:bkgdmjjtemps}
\end{cfigure}

\subsection{Likelihood Fit for Top Quark Mass}
\label{ssec:likelihood}

The reconstructed mass distributions from data are simultaneously
compared to the templates from signal and background sources using an
unbinned extended likelihood fit.  The likelihood includes free
parameters for the number of expected signal events
$n_s$ and background events $n_b$ in each subsample, and for the top
quark pole mass \mtop and the jet energy scale \jes.  For each
subsample, the likelihood is given by:
\begin{equation}
{\mathcal L_{\text{sample}}} =
	\lshapemreco \times \lshapemjj \times \lnev \times \lbg,
\end{equation}
where
\begin{eqnarray}
 \lshapemreco &=& \prod_{k=1}^{N^{\chisq}} \frac{ \begin{array}{l}
 [ \epsilon_s n_{s} P_{s}(\mrecok;\mtop,\jes) \\ \;\;\;\; \;\;\;\;
 \;\;\;\; \;\;\;\; \;\;\;\; \;\;\;\; \;\;\;\; \;\;\;\; + \epsilon_b
 n_{b} P_{b}(\mrecok) ] \end{array} } {\epsilon_s n_{s} + \epsilon_b
 n_{b}}; \nonumber
\\
 \lshapemjj &=& \prod_{k=1}^{N\cdot C_i} \frac{n_{s}
 P_{s}(\mjjk;\mtop,\jes) + n_{b} P_{b}(\mjjk)} {n_{s}+n_{b}};
 \nonumber
\\
 \lnev &=& \sum_{N_s+N_b=N} \text{Pois}(N_s;n_s) \times
 \text{Pois}(N_b;n_b) \times \nonumber
\\
 & & \left [ \sum_{N_s^{\chisq} + N_b^{\chisq} =
 N^{\chisq}}^{N_{s,b}^{\chisq} \le N_{s,b}} \!\!\!\!\!\!\!\! 
 B(N_s^{\chisq}; N_s, \epsilon_s) B(N_b^{\chisq}; N_b, \epsilon_b)
 \right ]; \nonumber
\\
 \lbg &=& \exp\left( -\frac{(n_{b}-n_{b}^{0})^2}
 {2\sigma_{n_{b}}^{2}} \right).
\end{eqnarray}
The values $\epsilon_s$ and $\epsilon_b$ represent the efficiency of
the \chisq cut for signal and background events, respectively, and are
given in \tab{t:massRecoMC}. $N$ and $N^{\chisq}$ are the number of
events observed in the data before and after the \chisq cut. All other
symbols are explained below.

The most important information on the top quark mass is provided
by the products in \lshapemreco, the $k$th term of which gives the
probability of observing the $k$th data event with reconstructed mass
\mrecok, given the background reconstructed top quark mass template,
$P_{b}(\mrecok)$, and the signal reconstructed top quark mass template
with a top quark mass of \mtop and energy scale shift \jes,
$P_{s}(\mrecok;\mtop,\jes)$.

The second term, \lshapemjj, is sensitive primarily to the value of
\jes. It reflects the product of probabilities for each of $C_{i}$
ways of assigning the two $W$ daughter jets in each event, where
$C_{i}=1,3,6$ for 2,1,0 tags, respectively.  The probabilities are
analogous to those in the first term, but are defined using the dijet
mass template parameterizations.

The third term in the likelihood, \lnev, represents the information
arising from the number of signal and background events in the top
quark mass and dijet mass samples, which are correlated.  Since the
number of expected signal and background events in the $W\rightarrow
jj$ sample are $n_s$ and $n_b$, the expected numbers of signal and
background events in the \mreco sample are given by $\epsilon_s n_s$
and $\epsilon_b n_b$, respectively.  This term expresses the
likelihood associated with observing $N$ and $N^{\chisq}$ events in
the two samples given the expected number of events as defined above.
We introduce the variables $N_s$ and $N_b$, the (unknown) number of
signal and background events actually in our sample.  We sum over the
possible values of these two variables that are consistent with the
total number of observed events.  Thus the first two factors express
the Poisson probability to observe $N_s$ signal and $N_b$ background
events given Poisson means of $n_s$ and $n_b$, respectively.  For the
final factor we introduce and sum over the possible values of
$N_s^{\chisq}$ and $N_b^{\chisq}$, variables for the (unknown) actual
number of signal and background events remaining after the \chisq
cut. The summand is the binomial probability to observe $N_s^{\chisq}$
signal events and $N_b^{\chisq}$ background events in the \mreco
sample given the assumed numbers of observed events $N_s$ and $N_b$ in
the \mjj sample and the \chisq cut efficiencies.

Finally, in the fourth term, \lbg, the background normalization is
constrained in the likelihood fit by a Gaussian probability distribution
centered at $n_{b}^{0}$ and with width $\sigma_{n_{b}}$. The background
normalizations are constrained for the \twotag, \onetagt, and \onetagl
samples.  For the \zerotag subsample, no background
normalization estimate is available, so no background constraint is
used.  Both $n_s$ and $n_b$ are required to be greater than zero.

As described in Sections~\ref{ssec:jetSyst} and~\ref{ssec:jes},
independent detector calibrations and studies of other processes allow
us to independently determine the jet energy scale \jes, and this
information is used in the reconstruction of the data events and in
the determination of the signal templates.  We include in the
likelihood fit the knowledge of this independent jet energy
calibration through an additional term in the overall likelihood:
\begin{eqnarray}
\label{e:JESConstraint}
 \ljes &=& \exp\left(-\frac{(\jes-\jes^{0})^2}
 {2\sigma_{\jes}^{2}}\right) \nonumber
\\
 &=& \exp\left(-\frac{\jes^2}{2}\right),
\end{eqnarray}
where the simplification arises because, by our definition of \jes,
the calibrated shift in energy scale is $\jes^{0}=0$ and the
uncertainty is $\sigma_{\jes}=1.0$.

The total likelihood is given by the product of the likelihoods for
the four subsamples and the jet energy scale constraint:
\begin{equation}
\mathcal{L} =
\mathcal{L}_{\text{\twotag}}  \times
\mathcal{L}_{\text{\onetagt}} \times
\mathcal{L}_{\text{\onetagl}} \times
\mathcal{L}_{\text{\zerotag}} \times
\ljes.
\end{equation}
The top quark pole mass \mtop and the jet energy scale \jes are shared
between the four likelihoods. The likelihood is maximized with respect
to all ten parameters ($n_s$ and $n_b$ for four subsamples, \jes, and
\mtop) using the {\sc minuit} package. A likelihood curve as a
function of \mtop is found by maximizing the likelihood with respect
to all other parameters for a series of fixed \mtop. The statistical
uncertainty from the fit procedure is taken from the points $\mtop^+$
and $\mtop^-$ where the log-likelihood changes by $-1/2$ unit
from its maximum. The positive and negative uncertainties are then
scaled to achieve unit pull widths as described in the next
Section~\ref{ssec:methodCheck}.

\subsection{Method Check}
\label{ssec:methodCheck}

The method described above is checked for any possible systematic
biases by running large numbers of ``pseudo-experiments,'' where we
create, using Monte Carlo simulation, samples of signal and background
events with an assumed value of the top quark mass and jet energy
scale and with the same statistical properties as our observed sample.
We then perform likelihood fits to each pseudo-experiment and
characterize the accuracy of the technique in determining the correct
values.

For each pseudo-experiment, the procedure is first to determine the
number of signal and background events, then to generate a reconstructed
top quark mass and dijet mass for each event, and finally to fit the
resulting pseudo-data using our standard machinery.
The number of background events in
each subsample is Poisson fluctuated around the central value given in
\tab{t:backgroundnumber}. The central value for the \zerotag
background is estimated by subtracting the estimated number of
\zerotag signal events from the observed number of \zerotag events in
the data.  The number of signal events is Poisson fluctuated around
the number observed in the data, minus the central value for the
background expectation, for each subsample.  For each event,
reconstructed masses \mreco and \mjj are selected at random from the
templates corresponding to signal or background processes. Some of the
events are eliminated from the \mreco sample, according to the \chisq
cut efficiencies given in \tab{t:massRecoMC}.  The resulting list of
reconstructed masses is fit using exactly the same machinery used on
the data, described in the previous Section~\ref{ssec:likelihood}.
Although this default procedure does not model correlations among the
\mreco and \mjj values in each event, a separate check showed a
complete modeling of the correlations to have a negligible effect on
the checks described here.

For each pseudo-experiment, the likelihood fit provides a measured top
quark mass \mtop and jet energy scale \jes, as well as positive and
negative errors (\errp and \errm) for each from the $\Delta \ln L=-1/2$
procedure.  We check the pull distribution for \mtop, defined using a
symmetrized uncertainty on the top quark mass as the distribution of
$(\mtop - \mtop^{\text{input}}) / 0.5(\mterrp + \mterrm)$, where
$\mtop^{\text{input}}$ is the generated top quark mass.  A pull
distribution is generated for each of 12 input values for the top
quark mass, keeping \jes fixed to zero, where 2500 pseudo-experiments
are generated for each input mass value, and each pull distribution is
fitted using a Gaussian function.  We determine a similar set of pull
distributions for various values of the \jes parameter, keeping the
top quark mass fixed to \gevcc{180}, although the results in this
case are correlated since they all use the same Monte Carlo event
sample.  The mean and sigma of the fitted functions are shown in
\fig{f:pulls}. Defining the pull distributions using the reported
asymmetric uncertainties does not systematically change the results.

\begin{cfigure}
\includegraphics[width=.48\columnwidth]{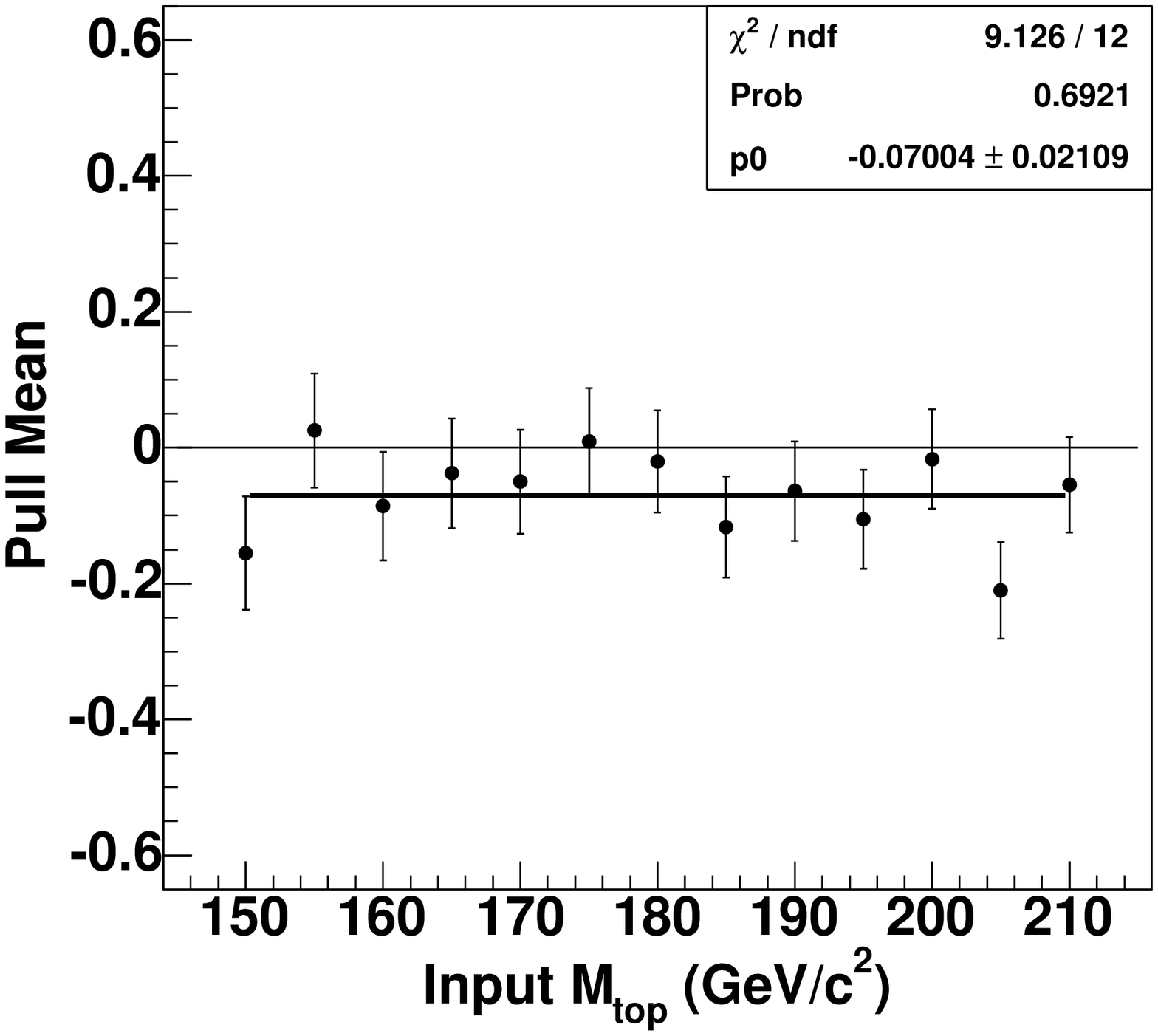}
\includegraphics[width=.48\columnwidth]{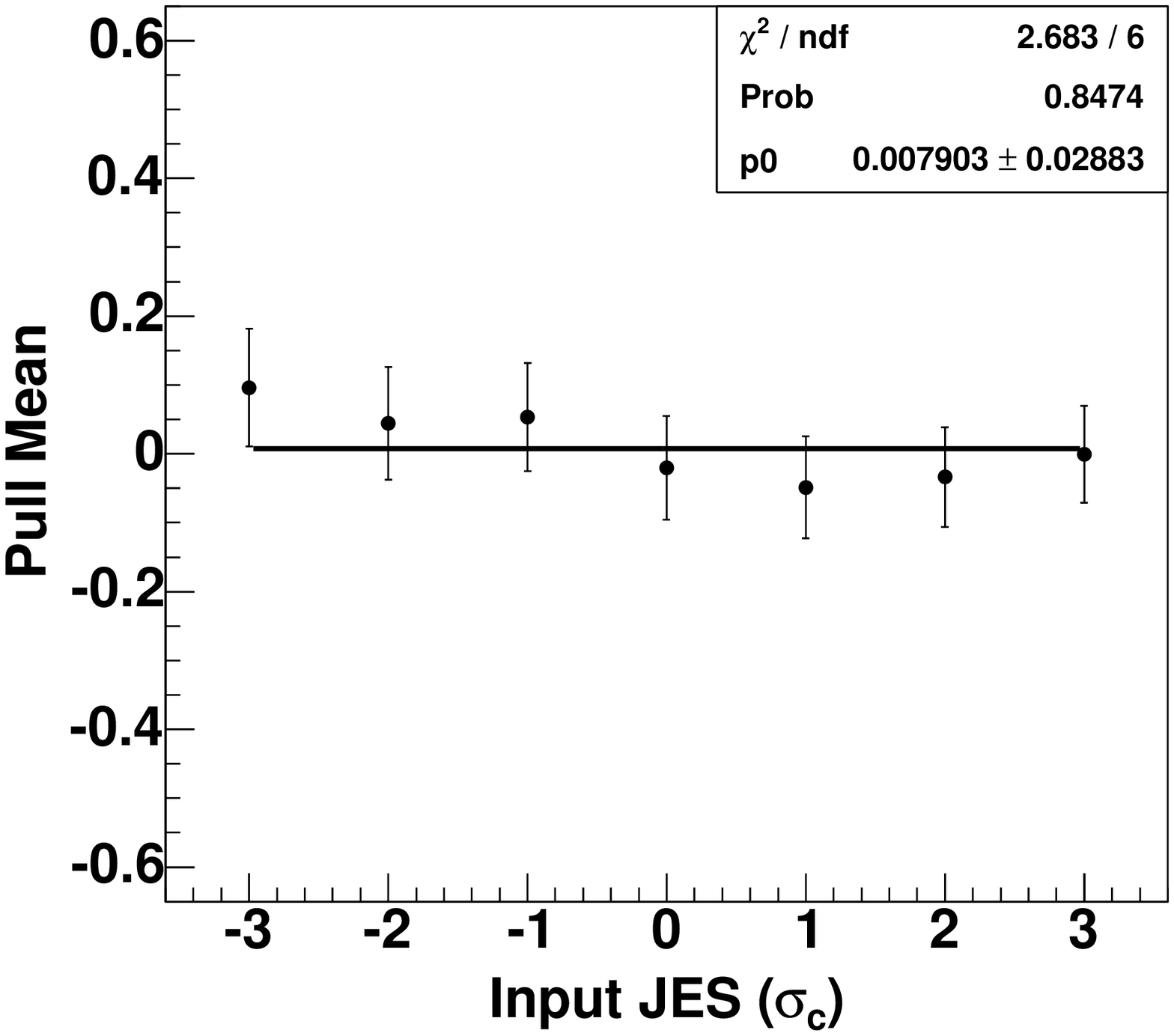}
\includegraphics[width=.48\columnwidth]{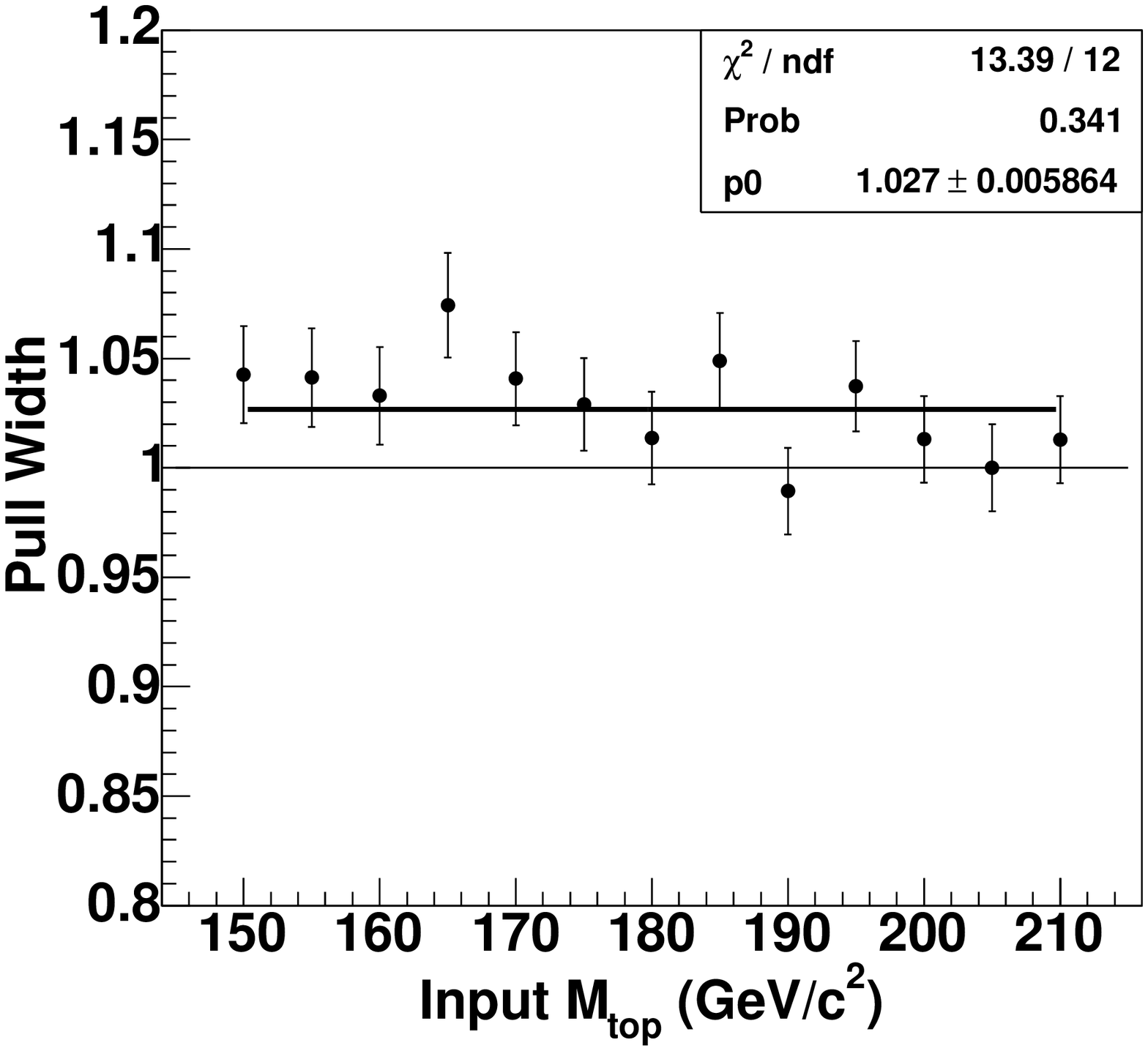}
\includegraphics[width=.48\columnwidth]{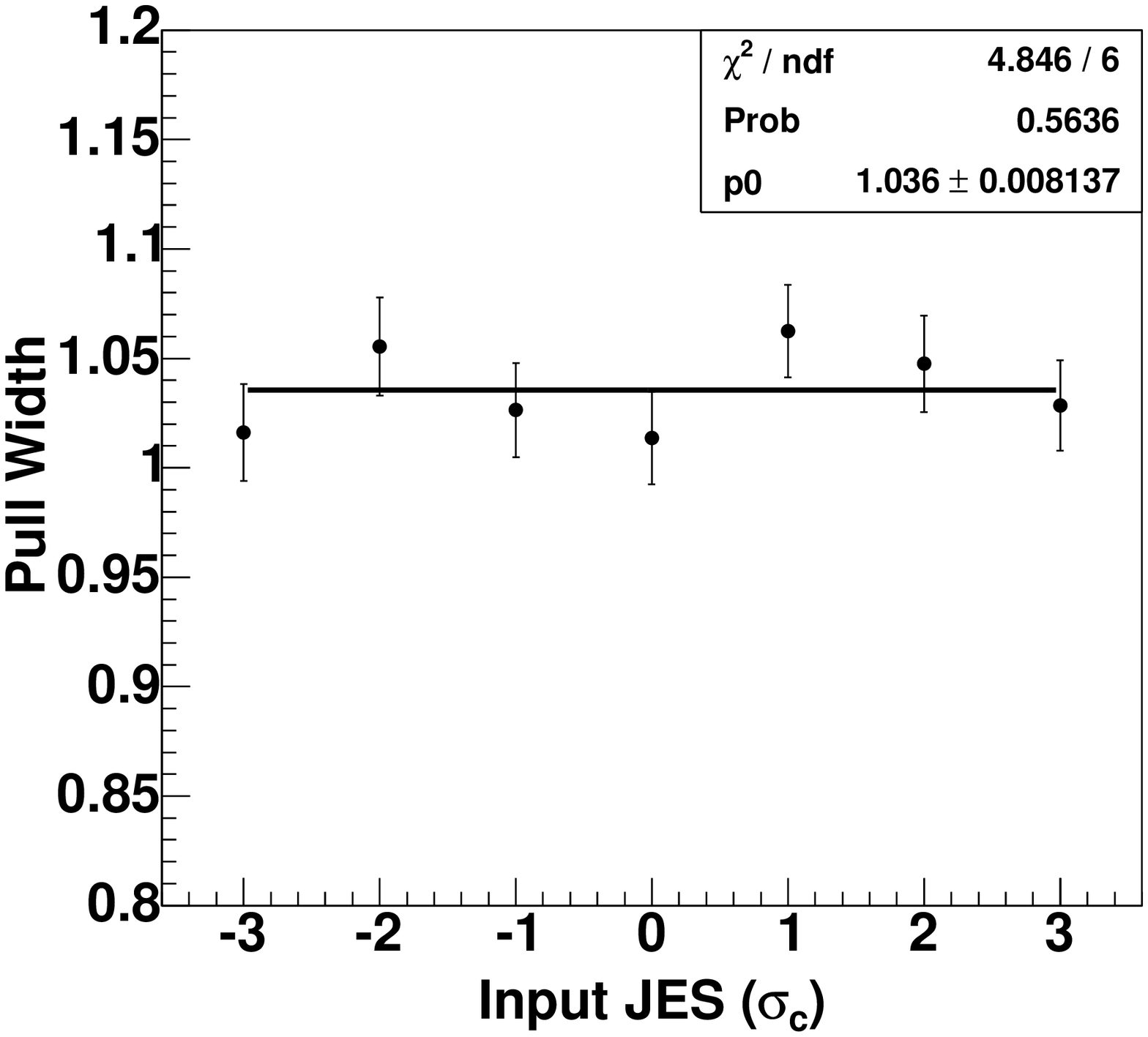}
\caption[Pull distribution means and widths for measured \mtop vs
\mtop and \jes.]
{The mean (top) and width (bottom) of pull distributions from sets of
2500 pseudo-experiments are shown. On the left, the jet energy scale
is fixed at its nominal value, and the generated top quark mass is
varied from \gevcc{150} to \gevcc{210}. On the right, the top quark
mass is fixed at \gevcc{180}, and the input jet energy scale is varied
from \sigunit{-3} to \sigunit{+3}.  The error bars come mostly from
the limited statistics of the Monte Carlo samples from which the
pseudodata is taken.}
\label{f:pulls}
\end{cfigure}

In the pull distributions as a function of top quark mass, the pull
means show a small offset for this particular slice of the \mtop-\jes
plane. Since the value of this offset varies with location in the
\mtop-\jes space, instead of directly correcting \mtop we take the
average offset of \gevcc{0.3} as a systematic uncertainty on the
measurement.

In addition the pull widths are slightly larger than one due to the
modest statistics of the event sample. For the expected number of
events with current luminosity, and for templates such as the ones
described in Section~\ref{ssec:parameterization}, the resulting
likelihood curve is typically non-Gaussian, and in fact, typically
shallower than Gaussian. The pull distributions become more Gaussian
(with width one) as pseudo-experiments with more events are
performed. With ten times the statistics, the pull widths are
consistent with unity.

For the current data sample, the quoted measurement with uncertainties
is designed to have pull width equal to one by scaling the errors
taken from $\Delta\ln L=-1/2$. The scale factor is the pull width from
the lower left plot of \fig{f:pulls}, averaged over the values of 
top quark mass, giving 1.027.

\section{Results on the Data}
\label{sec:results}

We fit the events in the data using the procedure described in
Section~\ref{sec:fitting}. After detailing the results, we present
several additional results performed as cross-checks on the primary
measurement.

\subsection{Subsample Likelihood Curves}
\label{ssec:resultsSub}

The likelihood fit is first performed on each subsample separately.
In these fits, each subsample likelihood contains the \jes constraint
term given in Eq.~\ref{e:JESConstraint}.  For a series of
top quark masses and \jes values, the mass and \jes parameters are
fixed, while the likelihood is maximized with respect to the remaining
parameters ($n_s$ and $n_b$) using {\sc minuit}. The resulting
likelihood contours in the \mtop-\jes plane are shown in
\fig{f:lnlsubs}.

\begin{cfigure}
\includegraphics[width=.47\columnwidth]{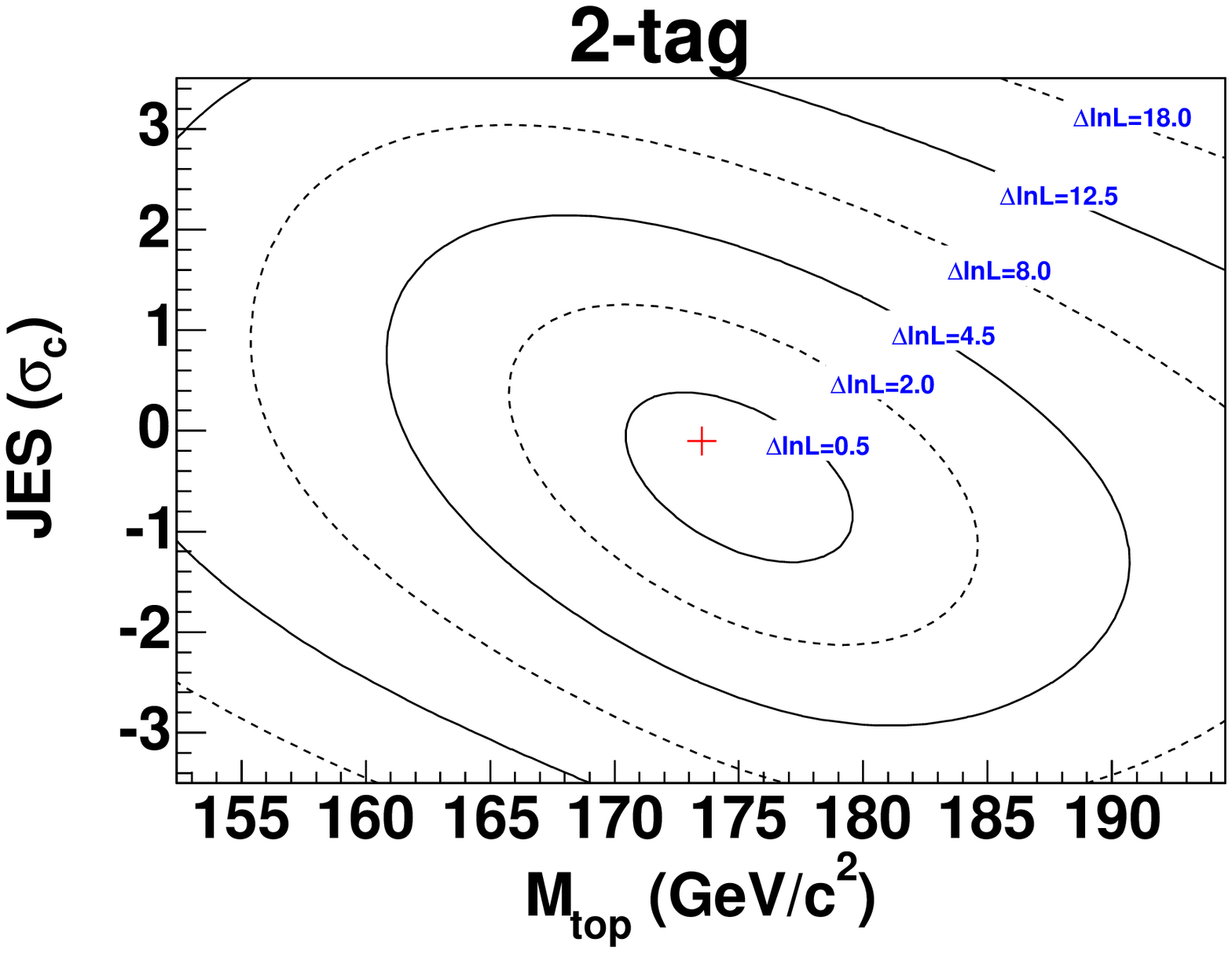}
\includegraphics[width=.47\columnwidth]{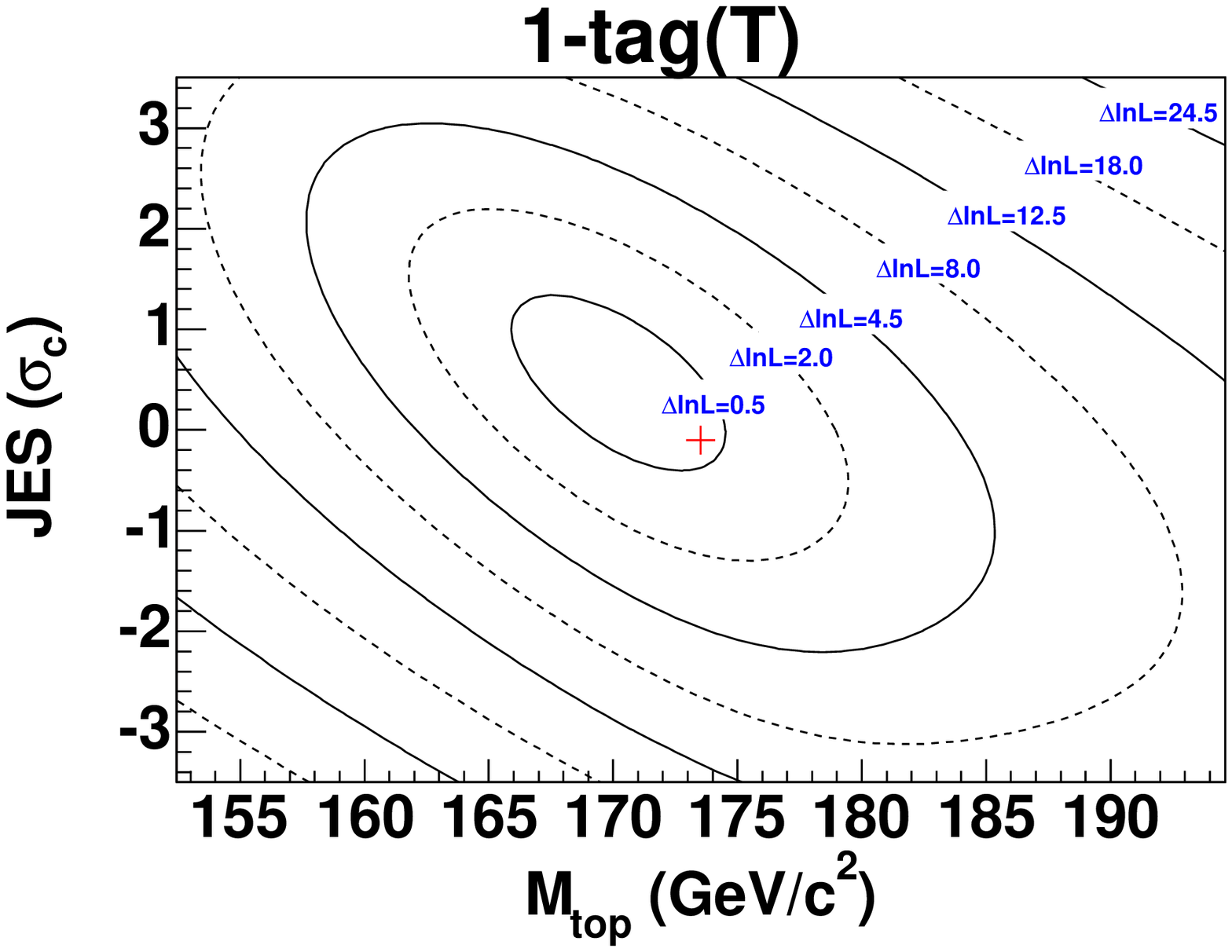}
\includegraphics[width=.47\columnwidth]{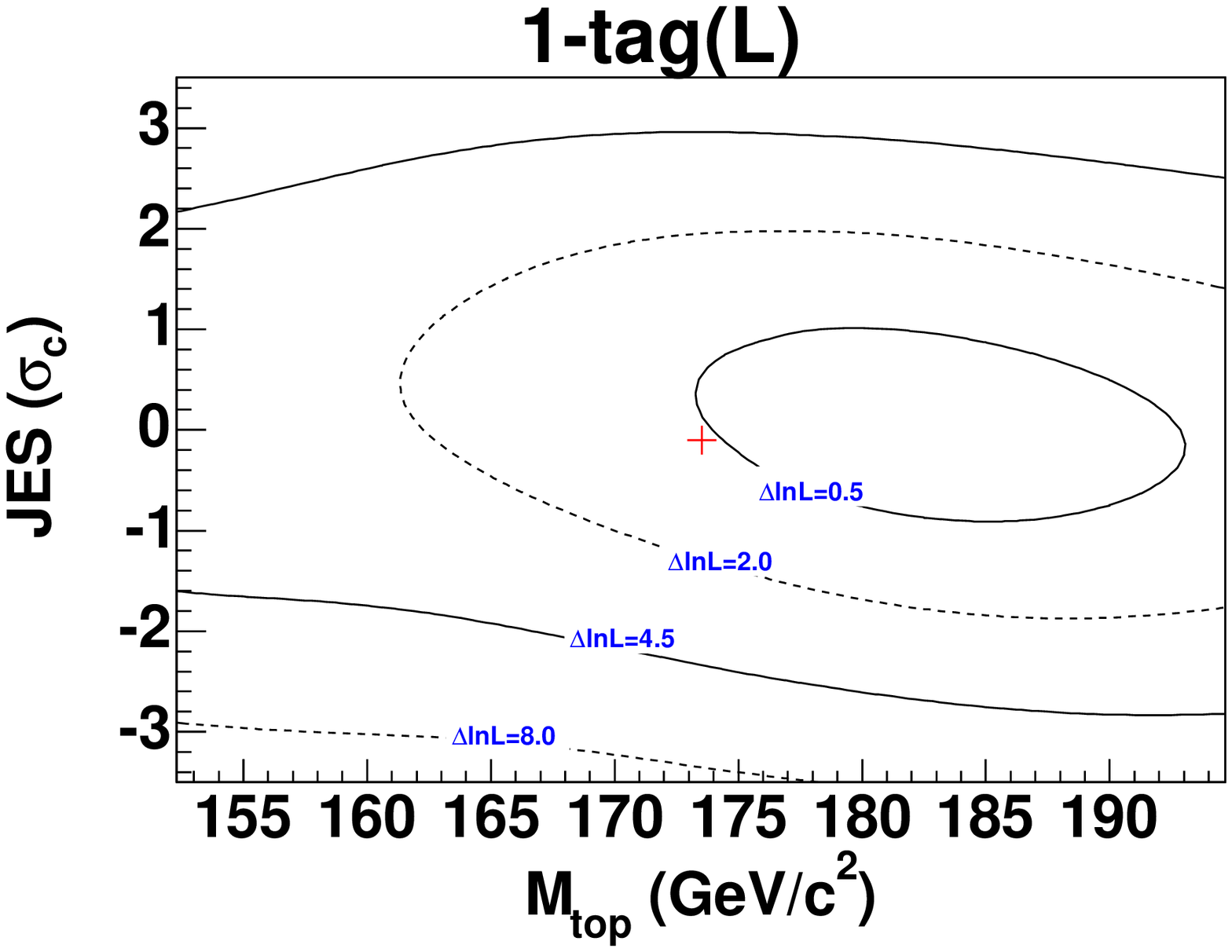}
\includegraphics[width=.47\columnwidth]{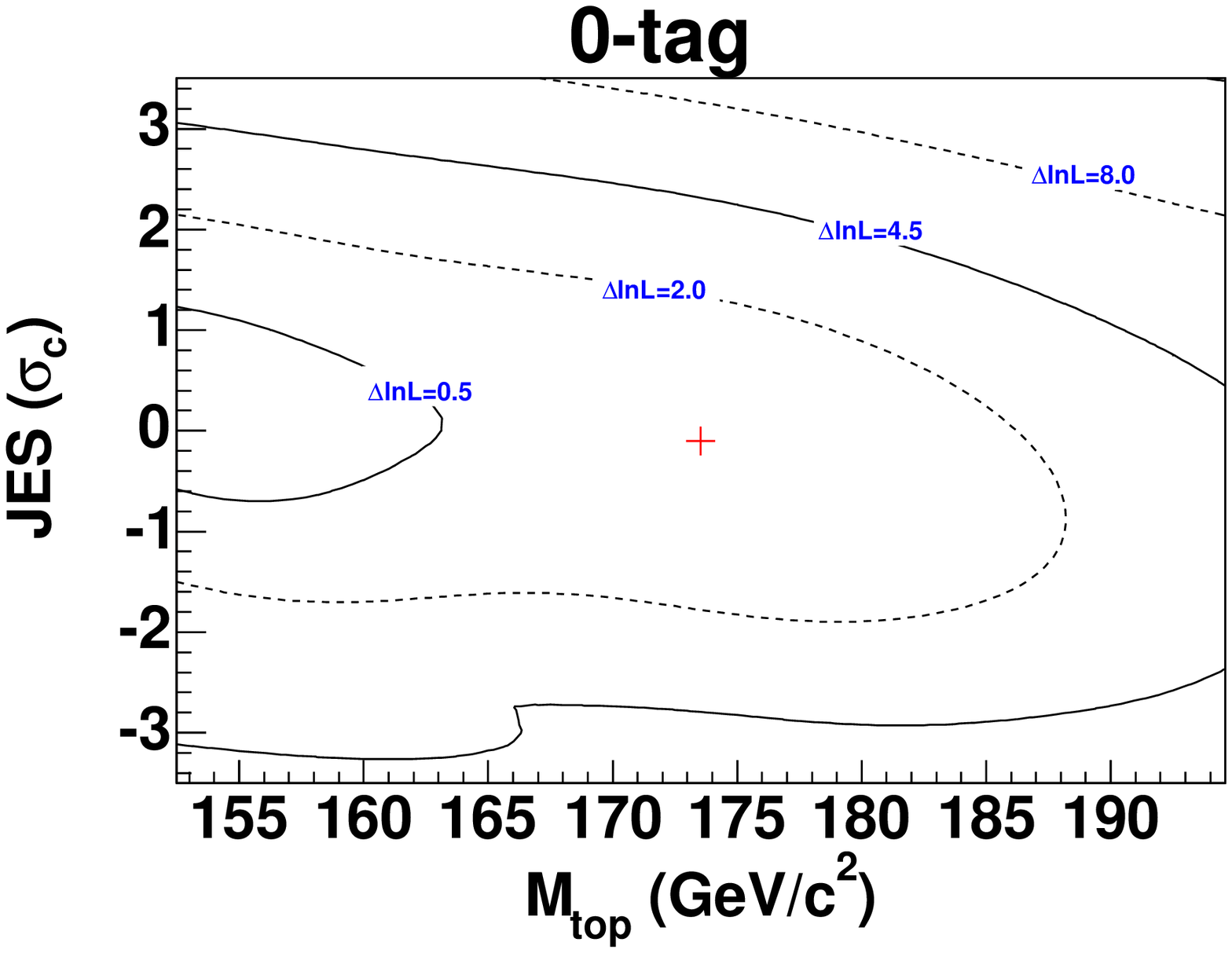}
\caption[Subsample fits likelihood curves.]
{The contours of the likelihood in the \mtop-\jes plane for the
independent fit to each subsample in the data.  At each point in the
plane, the likelihood is maximized with respect to the other free
parameters.  A crosshair shows the maximum likelihood point from the
combined fit, and contours are given at regular intervals in
$\Delta\ln L$, the change in log-likelihood from its maximum.  Upper
left: \twotag events; upper right: \onetagt events; lower left:
\onetagl events; lower right: \zerotag events.}
\label{f:lnlsubs}
\end{cfigure}

\subsection{Results of Combined Likelihood}
\label{ssec:resultsCombined}

Finally, the likelihood is maximized with respect to all parameters
using all four subsamples. The result, after scaling the $\Delta(\ln
L)=-1/2$ errors as described in Section~\ref{ssec:methodCheck}, is a
top quark mass of \gevcc{\measAStatJES{173.5}{3.7}{3.6}}.  The
simultaneous measurement of the jet energy scale is
\sigcunit{\measAErr{-0.10}{0.78}{0.80}}. The correlation between the
top quark mass and JES fits is -0.676. The combined likelihood as a
function of top quark mass and \jes is shown in \fig{f:lnlcomb}.  For
each value of the top quark mass and \jes, the likelihood is maximized
with respect to all other parameters.  This likelihood is not the
simple product of the four likelihoods shown in \fig{f:lnlsubs}
because the \jes constraint term \ljes is included in each of the
subsample fits, but of course only once in the combined fit.

\begin{cfigure}
\includegraphics{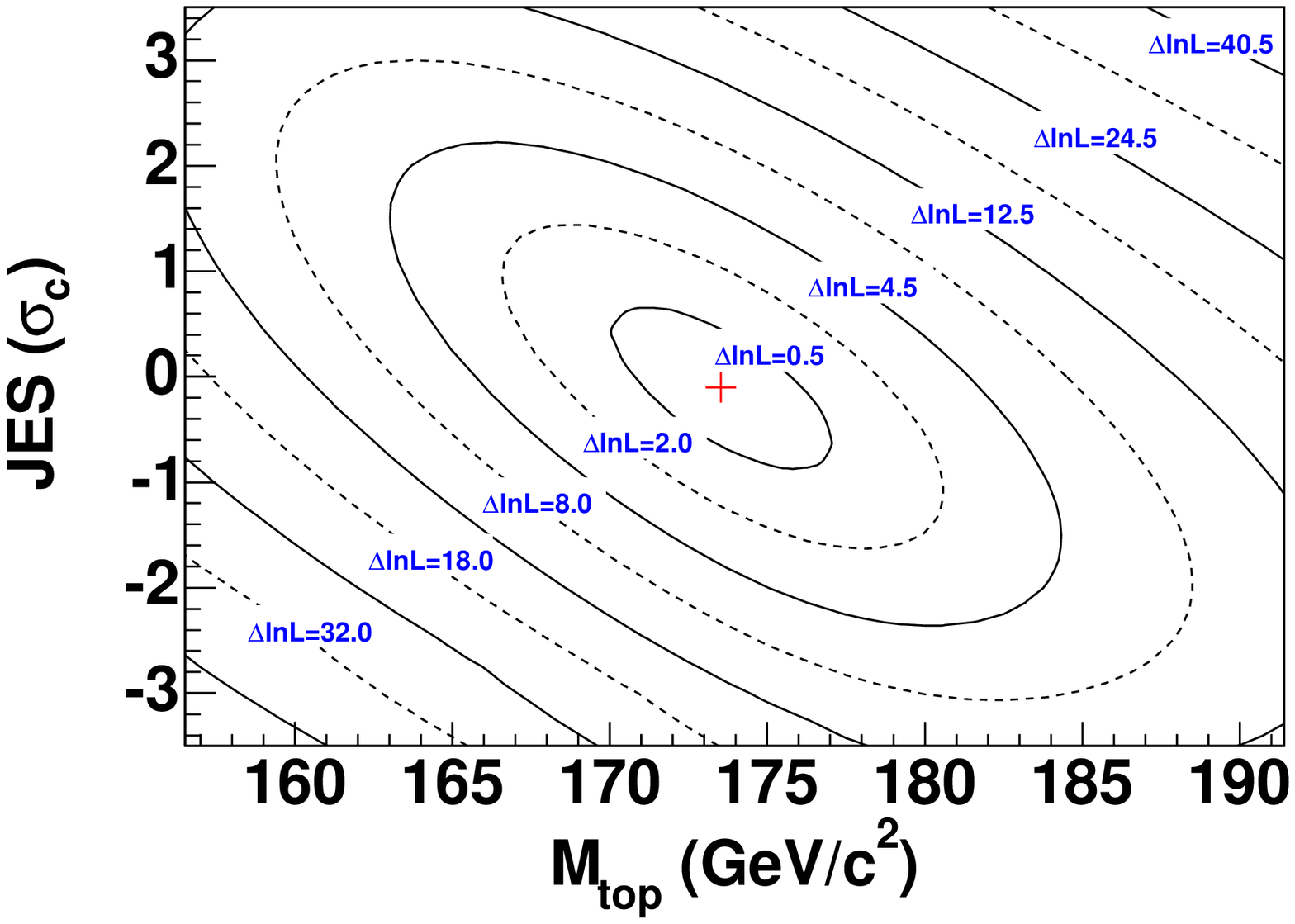}
\caption[Combined fit likelihood contours.]
{The contours of the likelihood in the \mtop-\jes plane for the
combined fit to all four subsamples. At each point in the plane, the
likelihood is maximized with respect to the other free parameters.
The crosshair shows the best fit point, and contours are given at at
regular intervals in $\Delta\ln L$, the change in log-likelihood from
its maximum.}
\label{f:lnlcomb}
\end{cfigure}

The uncertainty on \mtop from the likelihood fit is a combination of
the statistical uncertainty in extracting a measurement of \mtop and
the systematic uncertainty due to allowed variations of \jes. It is
possible to get an idea of the size of each contribution. Fixing \jes
to its fitted value of \sigcunit{-0.10} and fitting for \mtop alone
yields a top quark mass measurement of
\gevcc{\measAStat{173.5}{2.7}{2.6}}, corresponding to the ``pure
statistical'' uncertainty. Subtracting this uncertainty in quadrature
from the full uncertainty gives an \mtop uncertainty due to the jet
energy scale of \gevcc{\pm2.5}.

The input constraints and fit results for the combined fit are given
in \tab{t:fitResults}. \Fig{f:fitConsistencyMreco} shows the
consistency of the reconstructed top quark mass distribution in each
subsample with the combined fit results, while
\fig{f:fitConsistencyMjj} shows the same for the reconstructed dijet
mass.

\begin{table}
\caption[Combined fit results.]
{The input constraints and fitted values are given for all free
parameters in the combined likelihood fit.}
\label{t:fitResults}
\begin{ruledtabular}
\begin{tabular}{llcccc}
\multicolumn{2}{l}{Category} & \twotag & \onetagt & \onetagl & \zerotag \\
\hline
\mtop & constr. & \multicolumn{4}{c}{None} \\
      & fit & \multicolumn{4}{c}{\mathversion{bold}\gevcc{\measAStatJESBold{173.5}{3.7}{3.6}}}\\
      &     & \multicolumn{4}{c}{(\gevcc{\measAStatSepJES{173.5}{2.7}{2.6}{2.5}})} \\
\hline
\jes  & constr. & \multicolumn{4}{c}{\sigcunit{\measErr{0.0}{1.0}}} \\
      & fit & \multicolumn{4}{c}{\sigcunit{\measAErr{-0.10}{0.78}{0.80}}} \\
$n_s^W$ & constr. & \multicolumn{4}{c}{None} \\
      & fit & \measErr{23.5}{5.0} & \measErr{53.9}{7.9} &
              \measErr{14.3}{5.2} & \measErr{28.3}{8.3} \\
\hline
$n_b^W$ & constr. & \measErr{1.89}{0.52} & \measErr{10.4}{1.72} &
                     \measErr{14.3}{2.45} & None \\
      & fit & \measErr{1.8}{0.5} & \measErr{10.1}{1.7} &
              \measErr{15.5}{2.2} & \measAErr{15.7}{8.0}{7.1} \\
\end{tabular}
\end{ruledtabular}
\end{table}

\begin{cfigure}
\includegraphics{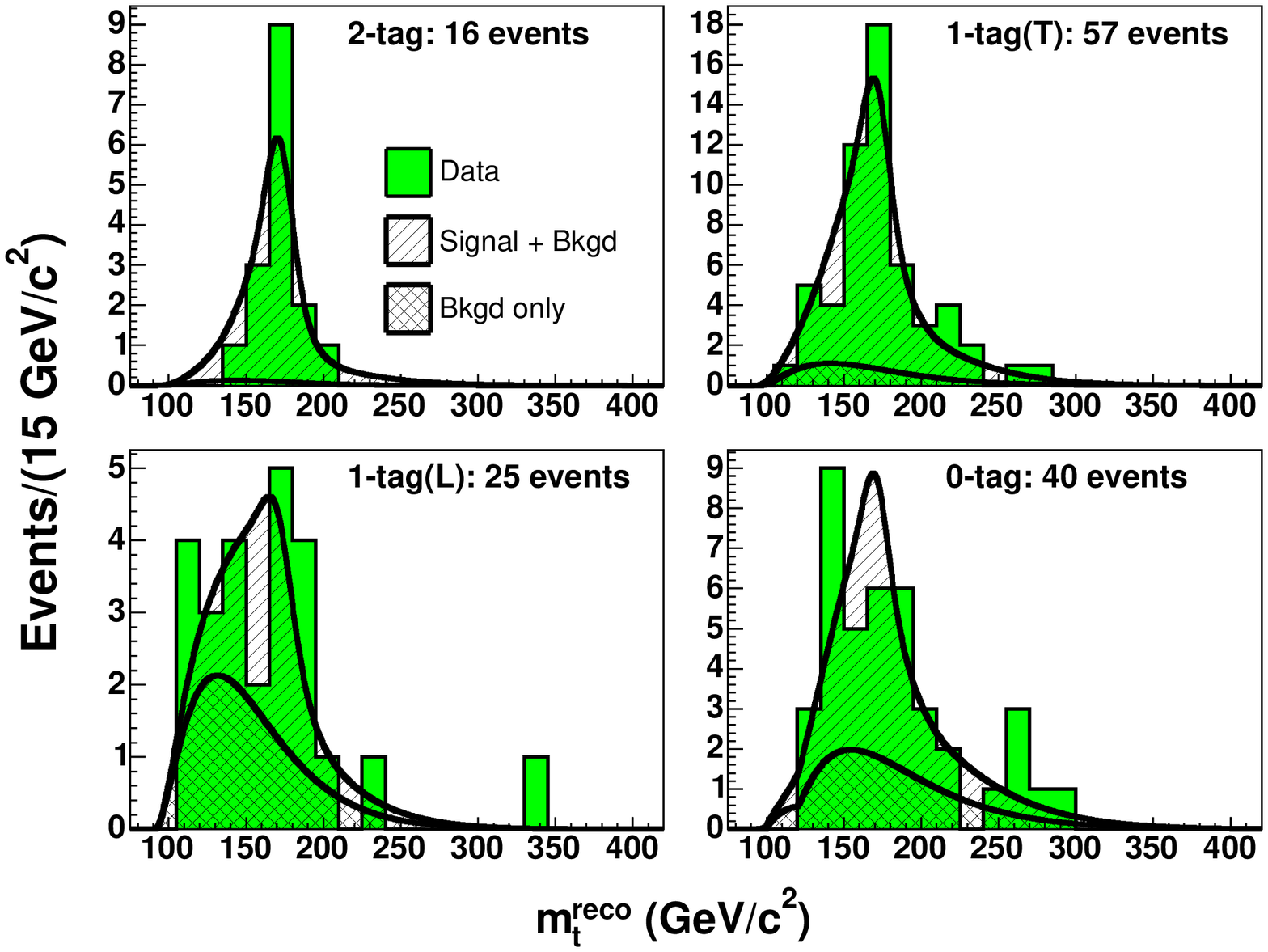}
\caption[Subsample fits consistency checks: \mreco.]
{The reconstructed top quark mass distribution for each subsample is
shown overlaid with the expected distribution using the top quark
mass, jet energy scale, signal normalization, and background
normalization from the combined fit.}
\label{f:fitConsistencyMreco}
\end{cfigure}

\begin{cfigure}
\includegraphics{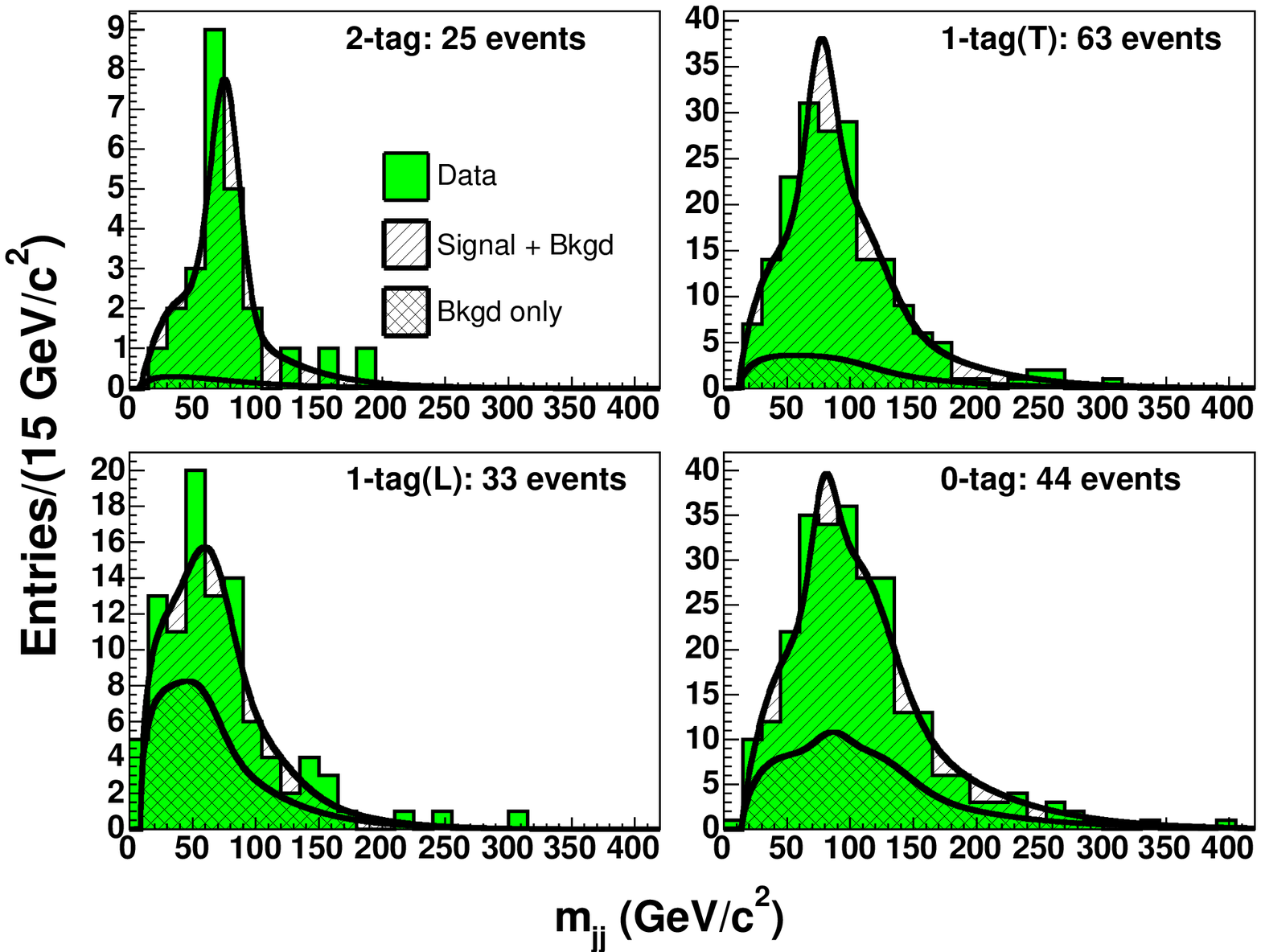}
\caption[Subsample fits consistency checks: \mjj.]
{The reconstructed dijet mass distribution for each subsample is shown
overlaid with the expected distribution using the top quark mass, jet
energy scale, signal normalization, and background normalization from
the combined fit.}
\label{f:fitConsistencyMjj}
\end{cfigure}

A set of pseudo-experiments is generated with a top quark mass of
\gevcc{172.5} (close to the central value from the fit), the nominal
jet energy scale, and with the number of events in each subsample
equal to the number observed in our data (\tab{t:eventTypes}). In
\fig{f:expStatErr}, the positive and negative uncertainties from the
likelihood fits are plotted. Arrows indicate the uncertainties from
the fit to the data. Although smaller than the median uncertainties
from the pseudo-experiments, the uncertainties on the data are
reasonable---9.2\% of the pseudo-experiments have smaller
uncertainties than those returned by the fit to the data.  The
distributions do not change significantly if a top quark mass
value of \gevcc{165} or \gevcc{180} is used. The better-than-expected
uncertainties are consistent with the sharpness of the reconstructed
top quark mass peaks in the \twotag and \onetagt subsamples, as shown
in \fig{f:fitConsistencyMreco}.

\begin{cfigure}
\includegraphics{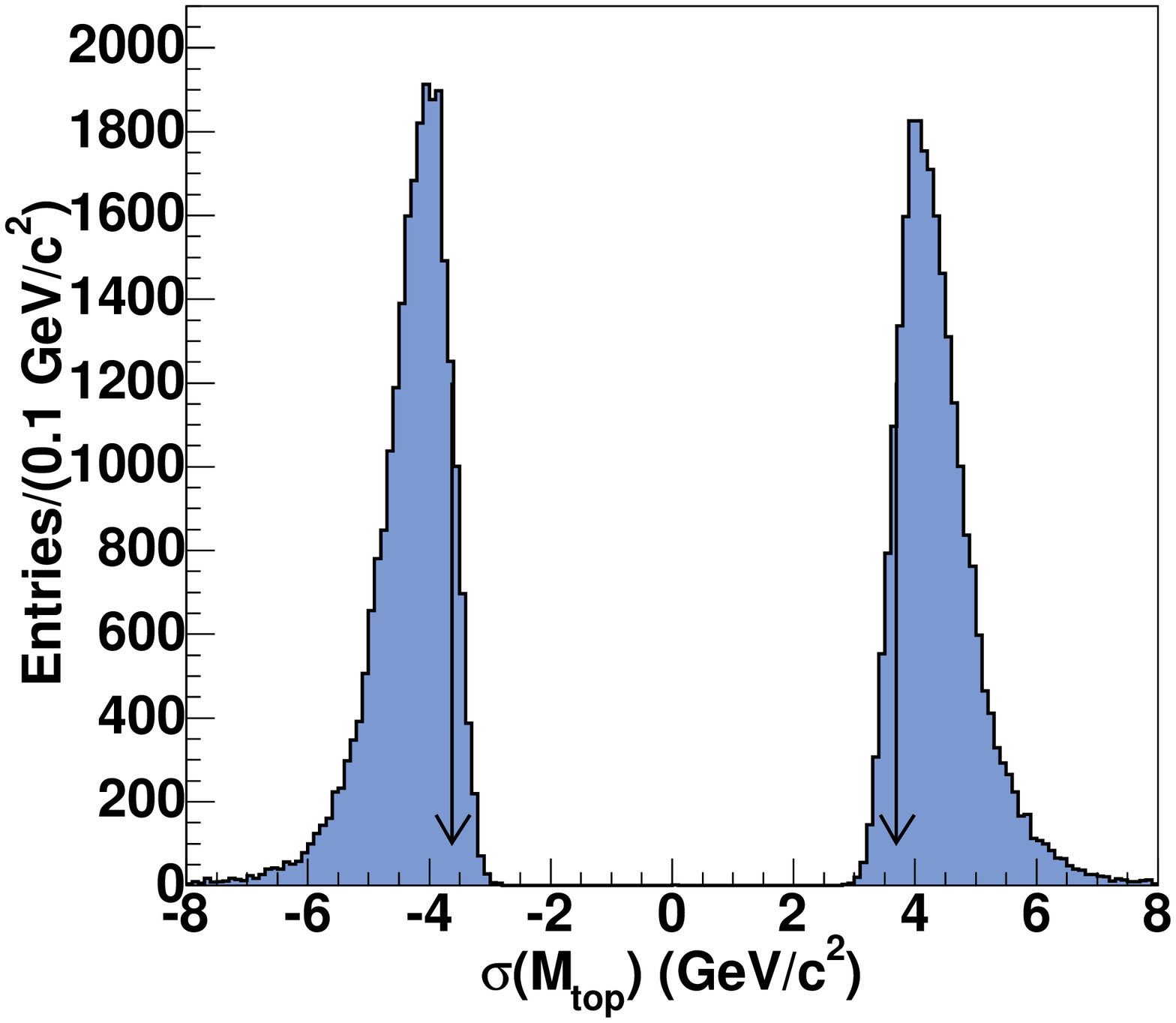}
\caption[Likelihood uncertainty distributions for data-like fit.]
{The distributions of positive and negative uncertainties from the
likelihood fit are shown, for pseudo-experiments generated with a 
top quark mass of \gevcc{172.5}, the nominal jet energy scale, and the
number of events in each subsample as observed in the data.  Arrows
indicate the positive and negative uncertainties from the likelihood
fit to the data; 9.2\% of the pseudo-experiments have smaller
uncertainties.}
\label{f:expStatErr}
\end{cfigure}

\subsection{Alternate Fits}
\label{ssec:otherfits}

In addition to the primary result described above, a number of
additional fits are performed as cross-checks and to investigate the
effect of certain assumptions on our measurement. The differences
between the primary fit and the alternate fits are briefly described
below, along with the resulting top quark mass
measurements. \Tab{t:alternateFits} summarizes the results. The
results from the alternate methods are quite similar, though the
methods are highly correlated.

\subsubsection{Fit without \jes constraint}
\label{sssec:altFitNoJES}

In the primary fit, the \jes measurement is treated as an update to
the extrinsic calibration by including the \jes Gaussian constraint
\ljes in the likelihood. Here the \ljes term is removed, so that
\emph{all} the jet energy scale information comes from the \emph{in
situ} calibration to the resonance of the hadronically decaying $W$
boson. The resulting top quark mass measurement is
\gevcc{\measStatJES{174.0}{4.5}}, and the simultaneous fit for \jes
gives \sigcunit{\measStat{-0.25}{1.22}}. Although the systematic
uncertainties are not explicitly evaluated for this approach, they are
not expected to be significantly different from those of the primary
analysis.

\subsubsection{Traditional \mtop-only fit}
\label{sssec:altFit1D}

For this alternate result, the traditional fit for a single variable,
\mtop, is performed using a single reconstructed quantity, \mreco.
This fit is virtually identical to the analysis performed in
\runi~\cite{Affolder:2000vy}. The event selection and \mreco
reconstruction are exactly as described earlier. With only one
reconstructed quantity and one measured quantity, the template
parameterizations are simpler.  Signal and background p.d.f.'s for
\mreco are fitted without any \jes dependence, but otherwise identical
in form to those described above. The form of the likelihood used is
also much simpler, since the term \lshapemjj is absent and \lnev is
greatly simplified with only the sample of events after the \chisq cut
used. For each subsample, the likelihood is given by
\begin{eqnarray}
 {\mathcal L} &=& \text{Pois}(N; n_s + n_b) \times \nonumber
\\
 & & \prod_{k=1}^{N} \frac{n_{s} P_{s}(\mrecok; \mtop) + n_{b}
 P_{b}(\mrecok)}{n_{s}+n_{b}} \times \nonumber
\\
 & & \exp\left ( -\frac{(n_{b} - n_{b}^{0})^2} {2\sigma_{n_{b}}^{2}}
 \right );
\end{eqnarray}
where in this context $n_{s}$, $n_{b}$, and $N$ refer to the number of
events expected and observed in the sample \emph{after} the \chisq
cut.  The combined likelihood is simply the product of the subsample
likelihoods.

The fitted value of the top quark mass using this method is
\gevcc{\measAStat{173.2}{2.9}{2.8}}, with a central value very
close to the result from the primary measurement.  For this result, of
course, since the jet energy scale systematic uncertainty is not
accounted for in the likelihood fit, its effect on the top quark mass
uncertainty must be estimated separately and added in quadrature.  The
negative log-likelihood curves for the \mtop-only fit in each
subsample are shown in \fig{f:lnlsubs1d}. As can be seen in the
lower-right panel, the \zerotag subsample contributes very little to
the overall measurement. This is because the fit prefers a small signal
contribution in this sample with no background constraint,
which results in very little sensitivity to \mtop.

\begin{cfigure}
\includegraphics[width=.47\columnwidth]{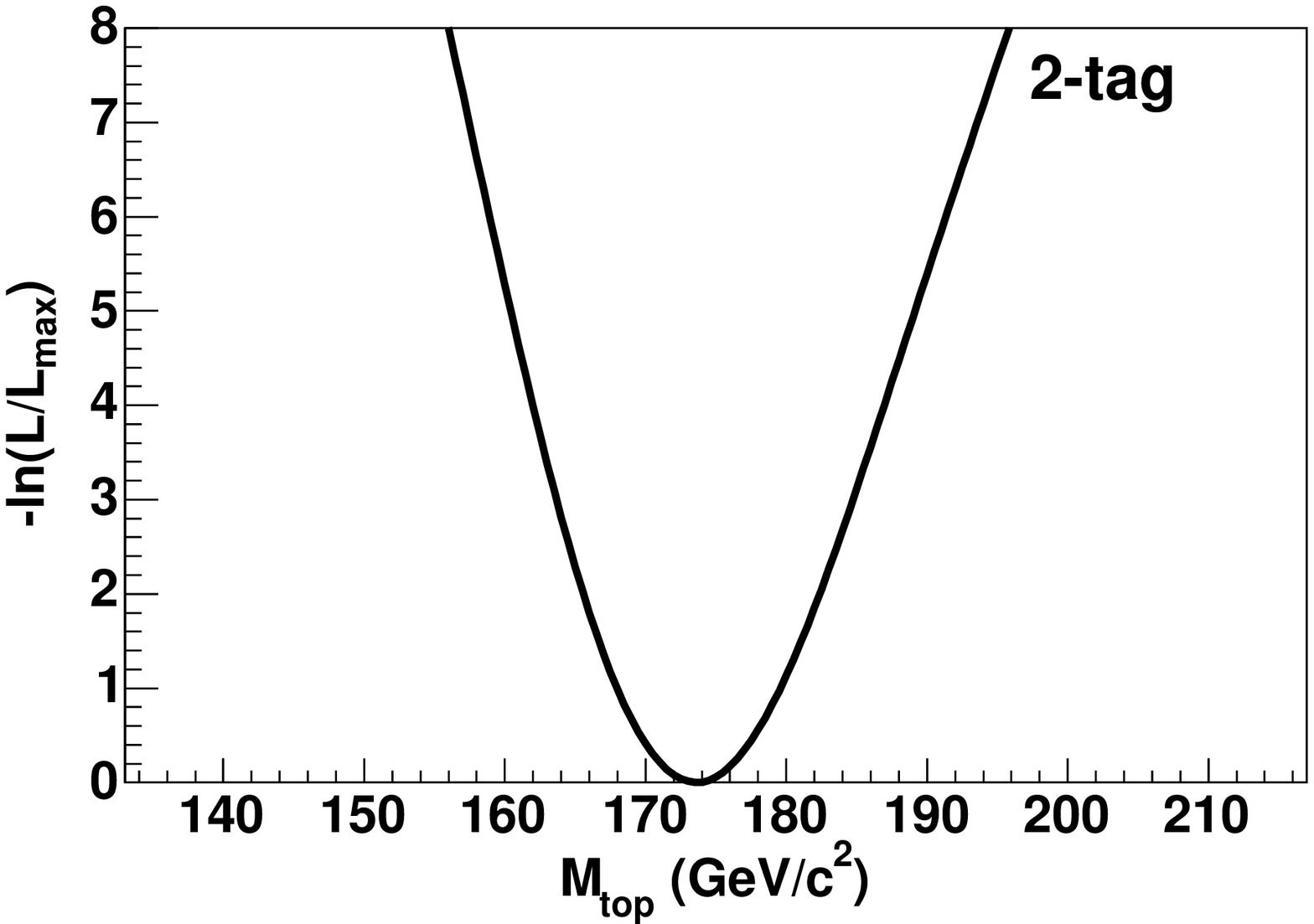}
\includegraphics[width=.47\columnwidth]{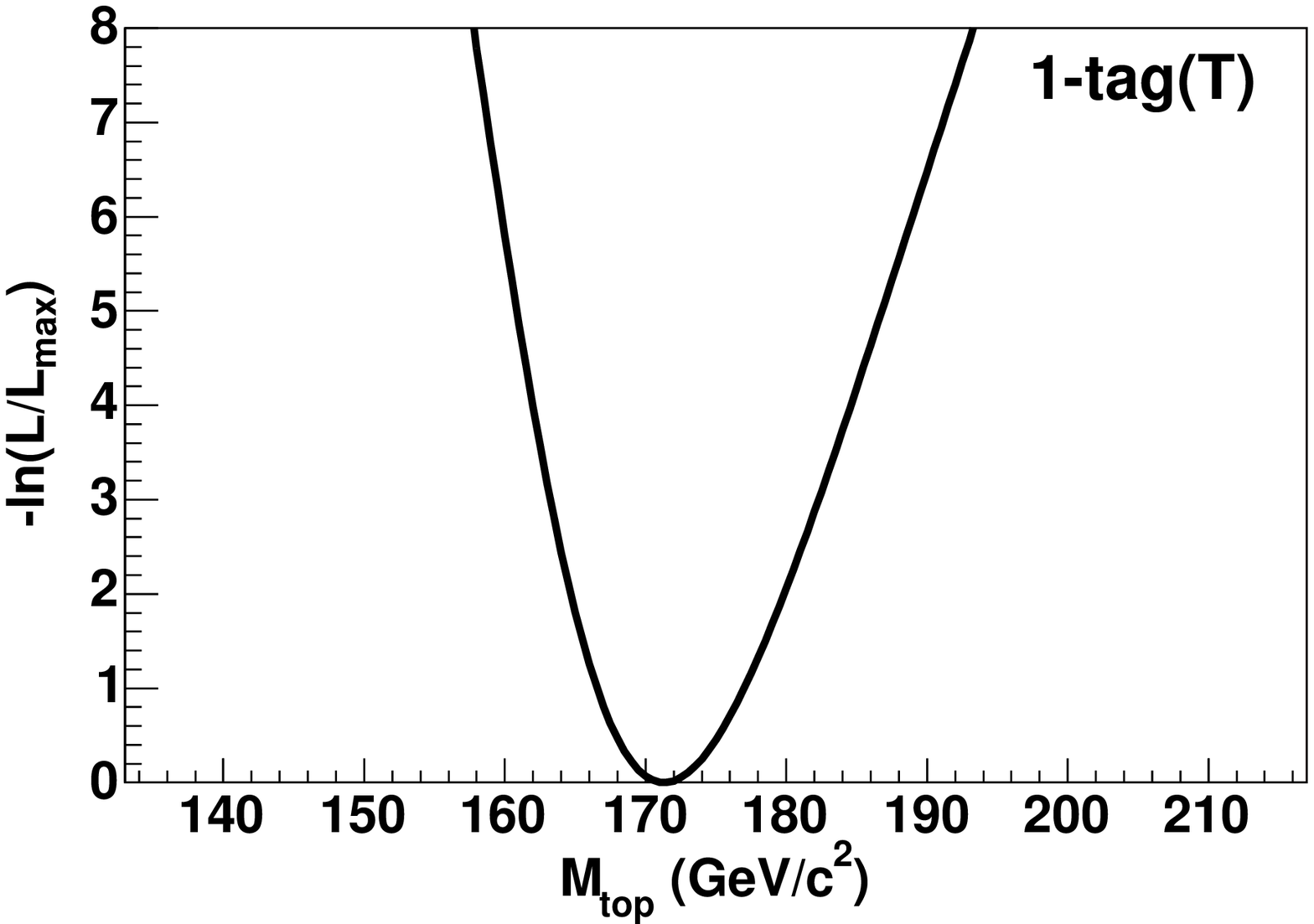}
\includegraphics[width=.47\columnwidth]{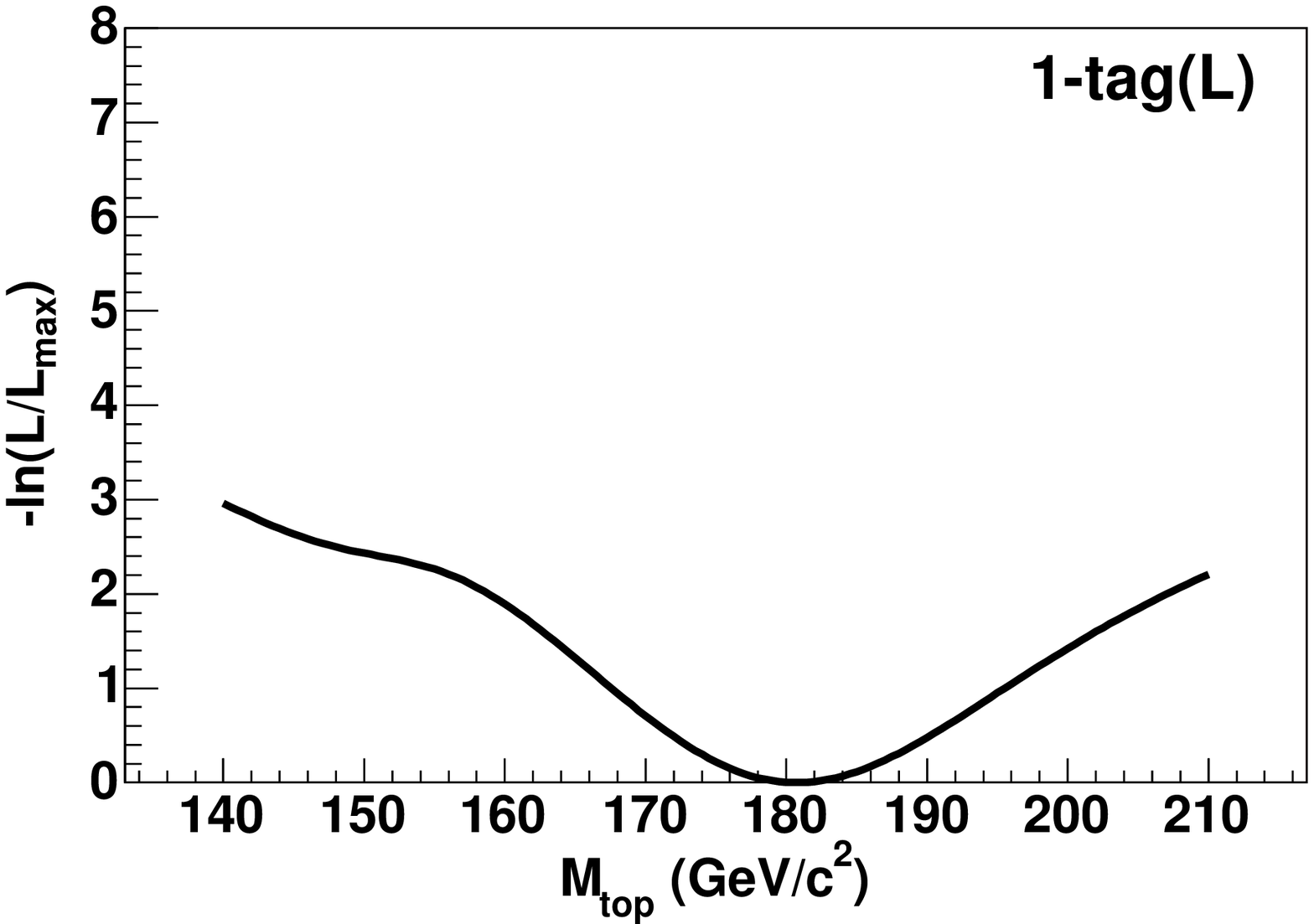}
\includegraphics[width=.47\columnwidth]{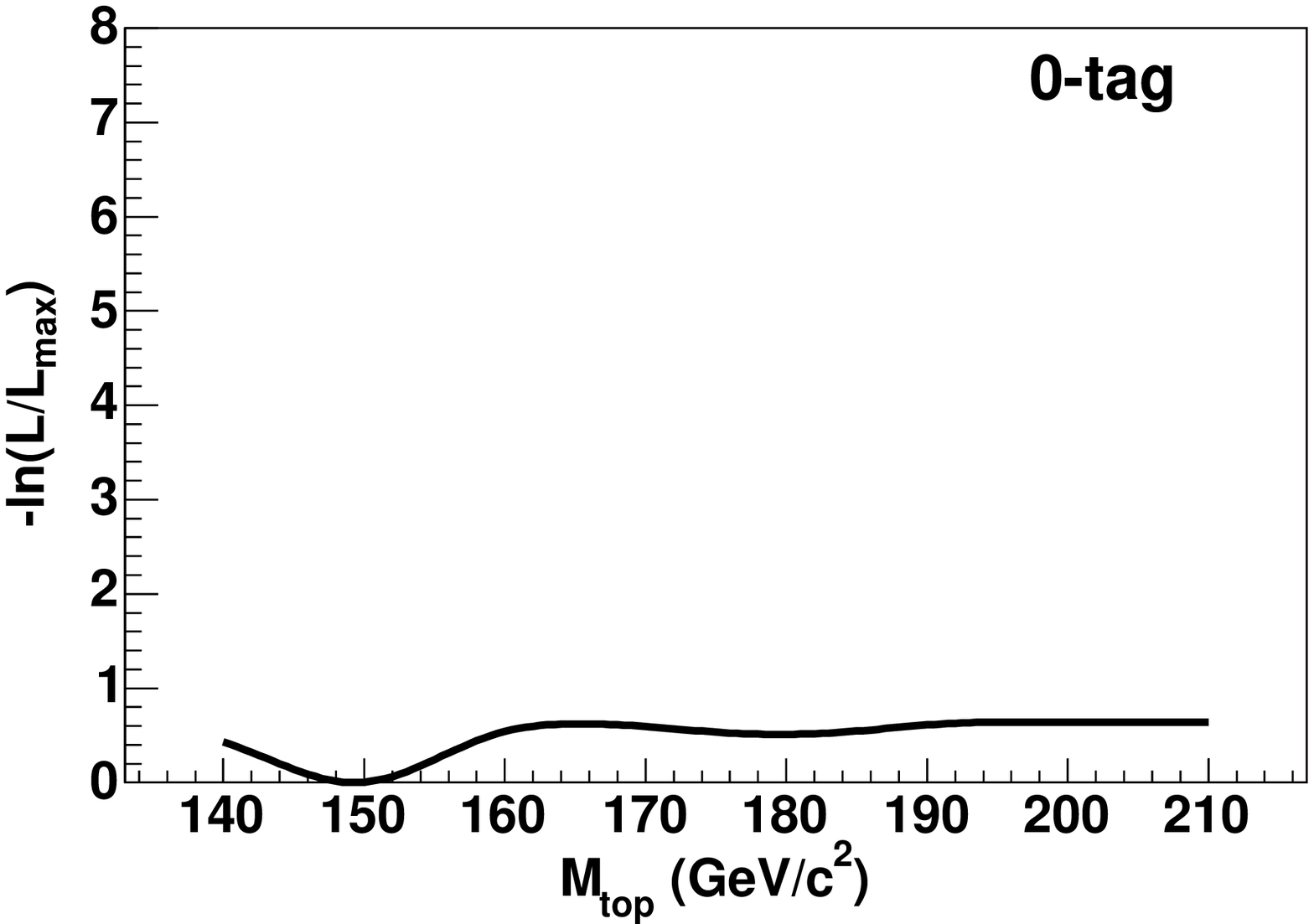}
\caption[Subsample fits likelihood curves, \mtop-only fit.]
{The negative log-likelihood curves as a function of the 
top quark mass are shown for
the \mtop-only fit to each subsample in the data. Upper left:
\twotag events; upper right: \onetagt events; lower left: \onetagl events;
lower right: \zerotag events.}
\label{f:lnlsubs1d}
\end{cfigure}

\subsubsection{Traditional \mtop-only fit with additional tag category}
\label{sssec:altFitJetProb}

Events with two $b$ tags carry the most information about the top
quark mass because of their high purity and narrow reconstructed mass
templates.  In this alternate analysis, we increase the number of
events with two $b$ tags by allowing one of the tags to come from the
Jet Probability tagger.  We establish a new category of events with
exactly one secondary vertex tag and an additional JPB tag; the former
requirement ensures that these events do not overlap with the \twotag
subsample.  The events in this category are then removed from the
\onetagt and \onetagl samples so that all the subsamples remain
disjoint.  Eighteen events are found in this category in the data
sample: 4 out of 18 events were newly categorized from \onetagl and 14
events from \onetagt in the default configuration.

The expected backgrounds in the new event category are estimated to be
$0.52 \pm 0.26$ events from $W\bbbar$, $Wc$, and $W\ccbar$ processes,
$0.15 \pm 0.08$ events from non-$W$ background,
$0.38 \pm 0.19$ events from mistagged \wjets events,
$0.08 \pm 0.04$ events from single top, and
$0.05 \pm 0.03$ events from the diboson processes $WW$ and $WZ$.
The total number of background events is thus estimated to be 
$1.2 \pm 0.6$ for the new subsample. The background estimates for the
exclusive one-tag subsamples change to account for the reduced acceptances.

The likelihood used to extract the top quark mass from this data is
that described above in Section~\ref{sssec:altFit1D}, i.e.\ using an
\mtop-only fit. \Fig{f:jetProbResult} shows the reconstructed top
quark mass distribution for the events with one secondary vertex tag
and one JPB tag, along with the expected distribution using parameters
taken from the fit to only this set of events.  The inset shows the
negative log-likelihood curve for this subsample alone.  Using only
the 18 events in this subsample, the measured top quark mass is
\gevcc{\measAStat{173.3}{6.1}{6.5}}.

\begin{cfigure}
\includegraphics{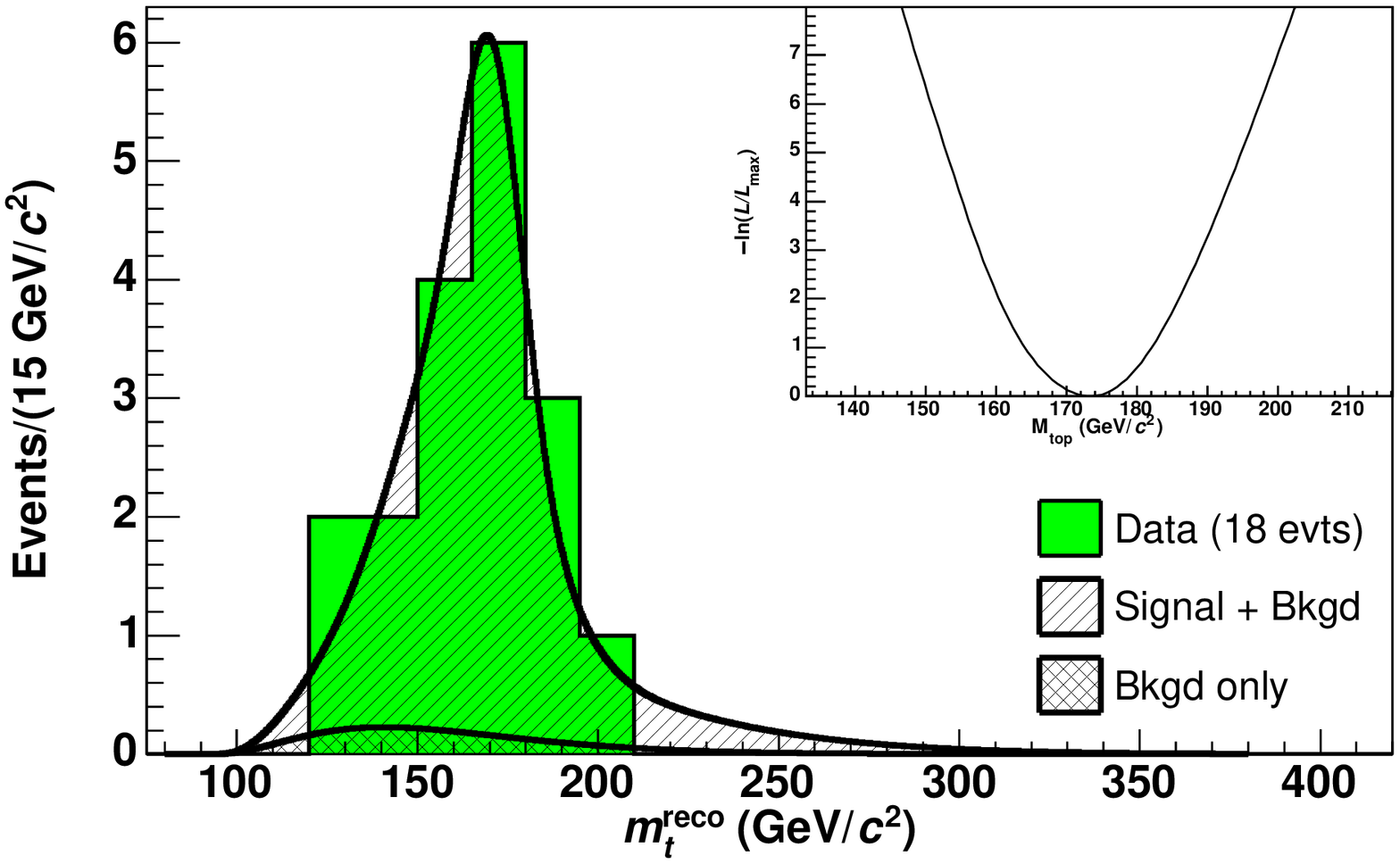}
\caption[Secondary vertex tag + JPB tag sample consistency check.]
{The reconstructed top quark mass distribution for the 18 events with
one secondary vertex tag and one JPB tag, overlaid with the expected
distribution from the fit to this subsample.  The inset shows the
shape of $- \Delta$ log ${\cal L}$ for the fit to these events as a
function of the top quark mass.}
\label{f:jetProbResult}
\end{cfigure}

This result using this data sample can be combined with the other four
categories of events. A sensitivity study shows that the combined
likelihood including the new class of events improves the expected
statistical uncertainty by 2.6\%. In addition to the statistical
improvement, increasing the number of double-tagged events improves
the jet energy systematic uncertainty.  The resulting combined top
quark mass measurement on the five subsamples is
\gevcc{\measAStat{173.0}{2.9}{2.8}}.

\begin{table}
\caption[Alternate fit results.]
{The results of alternate fits are summarized. For the cases that do
not include the jet energy scale systematic effect in the likelihood
fit result, the independently determined systematic is given for
comparison (see Section~\ref{ssec:jetSystOnMtop} for more details).}
\label{t:alternateFits}
\begin{ruledtabular}
\begin{tabular}{lcc}
Method & \mtop fit result & \jes fit result \\
       & [\gevccnoarg]    & [$\sigma_c$] \\
\hline
Default & \measAStatJES{173.5}{3.7}{3.6} & \measAErr{-0.10}{0.78}{0.80} \\
No \jes constr. & \measStatJES{174.0}{4.5} & \measErr{-0.25}{1.22} \\
\mtop-only & $\measAStat{173.2}{2.9}{2.8}\oplus3.1~(\jes)$ & N/A \\
\hspace{15pt}+ JPB & $\measAStat{173.0}{2.9}{2.8}\oplus3.0~(\jes)$ & N/A \\
\end{tabular}
\end{ruledtabular}
\end{table}

\subsection{Cross-checks on the results}
\label{ssec:xchecks}

The measurement of the top quark mass is checked by performing the
analysis in various subsamples and with different jet corrections and
background normalization constraints in order to ensure the robustness
of the result. In these cross-checks, we use the traditional
\mtop-only fit described in Section~\ref{sssec:altFit1D} for
simplicity.

\Fig{f:xcheck1} shows the resulting top quark mass measurement for
various modifications to the method. Any inconsistencies would most
likely indicate problems with the detector or the analysis method.
First the measurement using top-specific corrections (the default) and
using generic out-of-cone jet corrections is compared.  Next fits
using the background constraints (the default) and without the
constraints are shown. For the remaining comparisons, the dataset is
divided into two subsamples. First results are shown from two
different run periods, then events with a primary electron vs those
with a primary muon, and finally positive-charge primary leptons vs
negative-charge primary leptons.  Except in the case of the generic
jet corrections, the default reconstructed mass templates are
used. All the results are consistent with each other and with the
primary measurement.

\begin{cfigure}
\includegraphics{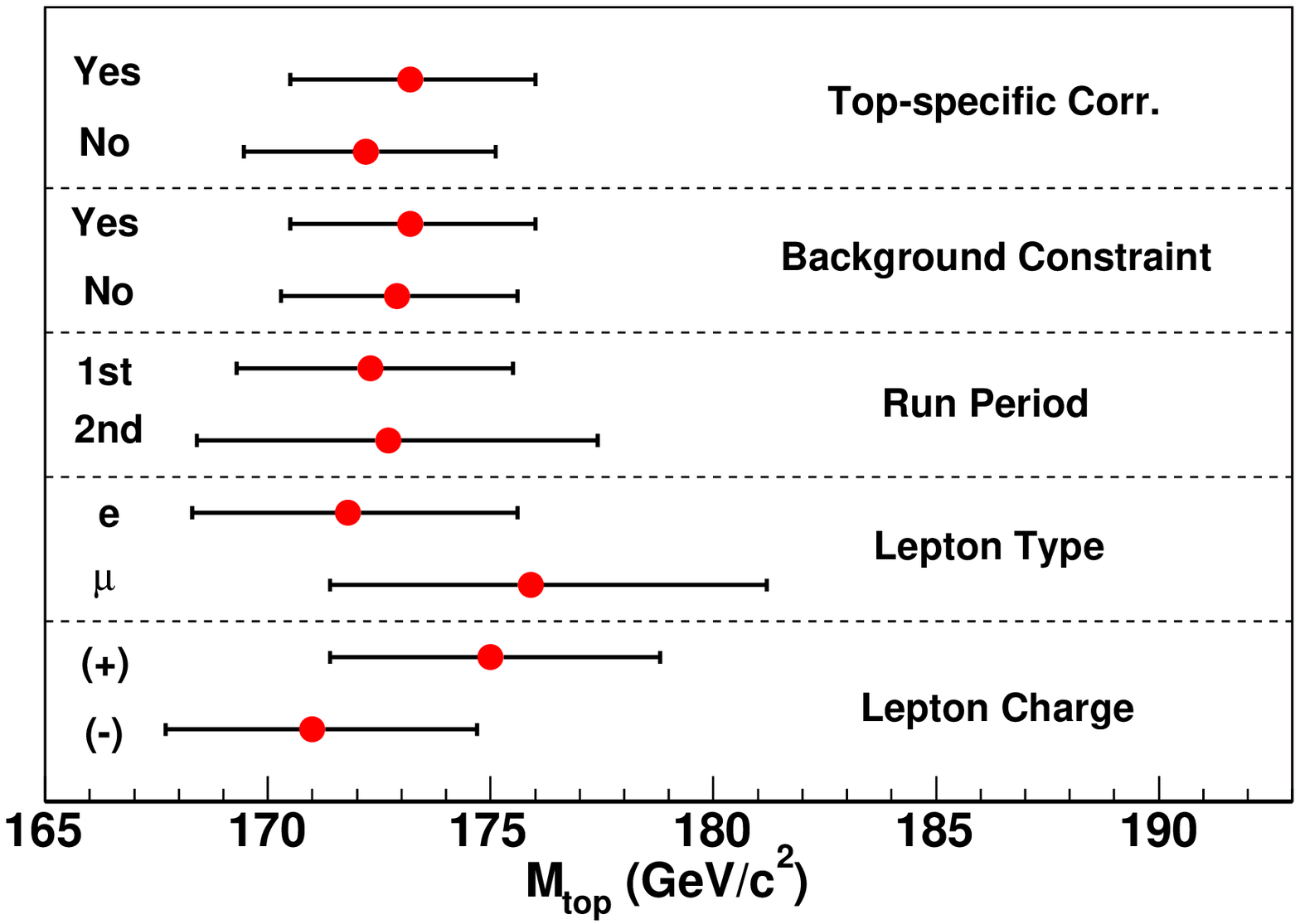}
\caption[Cross-check measurements using different jet corrections,
background constraints, and lepton samples.]
{Top quark mass measurements using the \mtop-only fit are compared for
different assumptions (top-specific corrections (default) vs generic
jet out-of-cone corrections, constrained backgrounds vs unconstrained)
and different ways of subdividing the sample (two different run
periods, electrons vs muons, positive-charge leptons vs
negative-charge leptons).  All results with only statistical errors
are consistent.}
\label{f:xcheck1}
\end{cfigure}

A series of top quark mass measurements using different subsamples of
the data is shown in \fig{f:xcheck2}. The primary effects of the
increasingly tight selection are, first, to increase the sample purity
and thereby decrease sensitivity to modeling of the background
processes; and second, to select events with low extra jet activity,
decreasing sensitivity to modeling of ISR and FSR. The list of samples
used is as follows from top to bottom: all four subsamples (default,
138 events after \chisq cut); \twotag, \onetagt, and \onetagl only (98
events); \twotag and \onetagt only (73 events); \twotag and \onetagt
only with any additional jets required to have $\et<\gev{15}$ (56
events); \twotag and \onetagt only with any additional jets required
to have $\et<\gev{8}$ (38 events).  For the last two cases, top mass
templates are prepared with the additional requirements. Again we find
that all results are consistent, indicating that background kinematics
and extra jet activity are reasonably well modeled.

\begin{cfigure}
\includegraphics{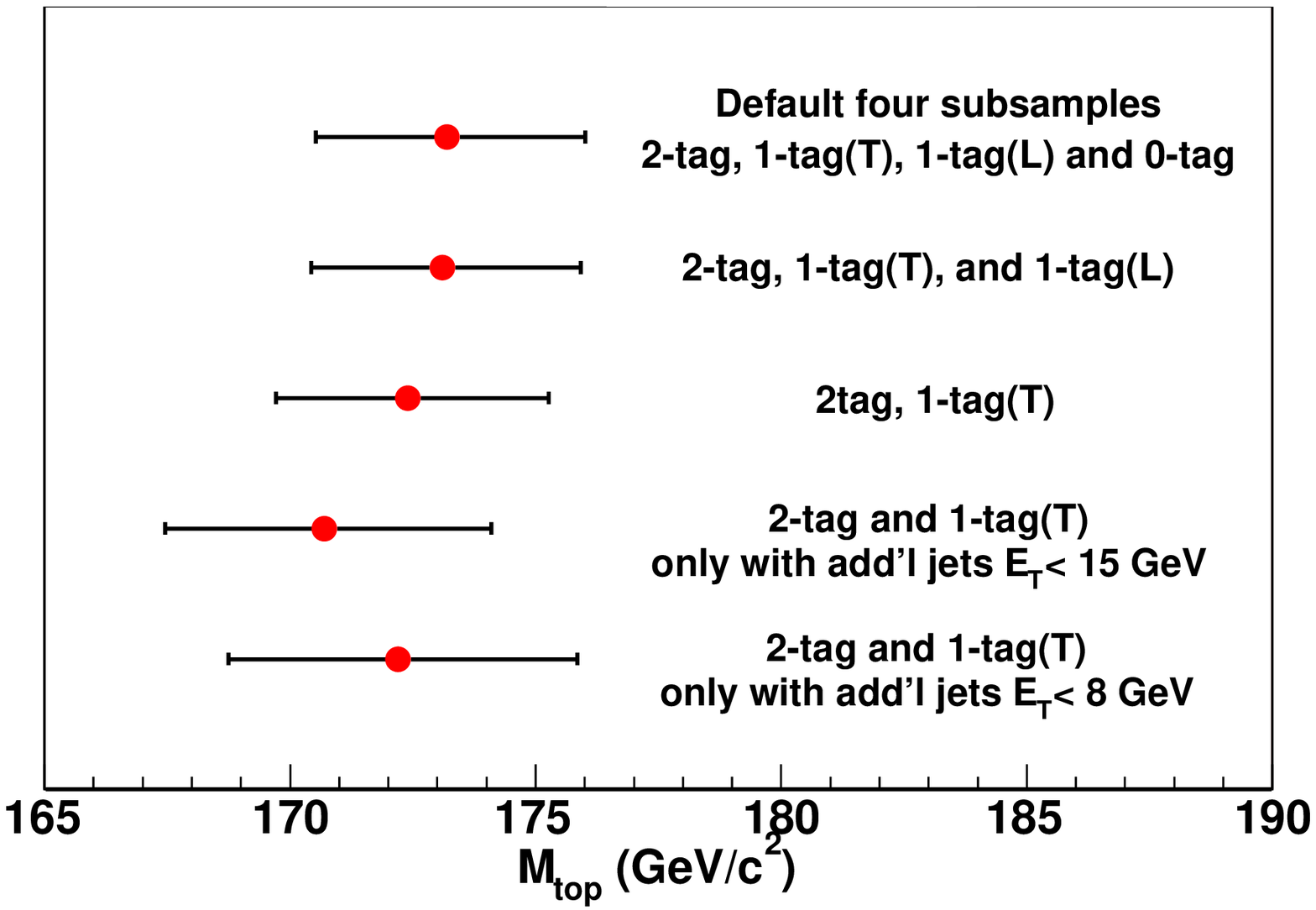}
\caption[Cross-check measurements using different subsamples
and jet requirements.]
{Top quark mass measurements using the \mtop-only fit are compared for
various samples.  From top to bottom: all four subsamples;
\twotag, \onetagt, and \onetagl subsamples;
\twotag and \onetagt subsamples only;
\twotag and \onetagt subsamples only, additional jets $\et<\gev{15}$;
\twotag and \onetagt subsamples only, additional jets $\et<\gev{8}$. 
All results with only statistical errors are found to be consistent.}
\label{f:xcheck2}
\end{cfigure}

\subsection{Kinematic distributions}
\label{ssec:kinematic}

We compare various kinematic distributions for the \ttbar signal
candidate events with the Monte Carlo predictions for combined signal
and backgrounds. Comparisons of kinematic distributions tell us how
well the Monte Carlo models the data, which is very important in this
kinematic analysis.  This information could additionally be used to
test whether the kinematic properties of the top quark we observe are
consistent with standard model predictions.  For these distributions,
we use only \twotag and \onetagt events with $\chisq<9$ (73 events),
in order to increase the signal purity.  All kinematic quantities are
defined using the output of the \chisq fitter, so that both jet-quark
assignments and the \pt of each object are taken at the minimum \chisq
point.

\Fig{f:pt_top} and \fig{f:eta_top} show the \pt and rapidity
distributions of the reconstructed top quarks, respectively. The data
distributions are in agreement with predictions using {\sc herwig}
\ttbar signal events with top quark mass of \gevcc{172.5} and
simulated background events.

\begin{cfigure}
\includegraphics{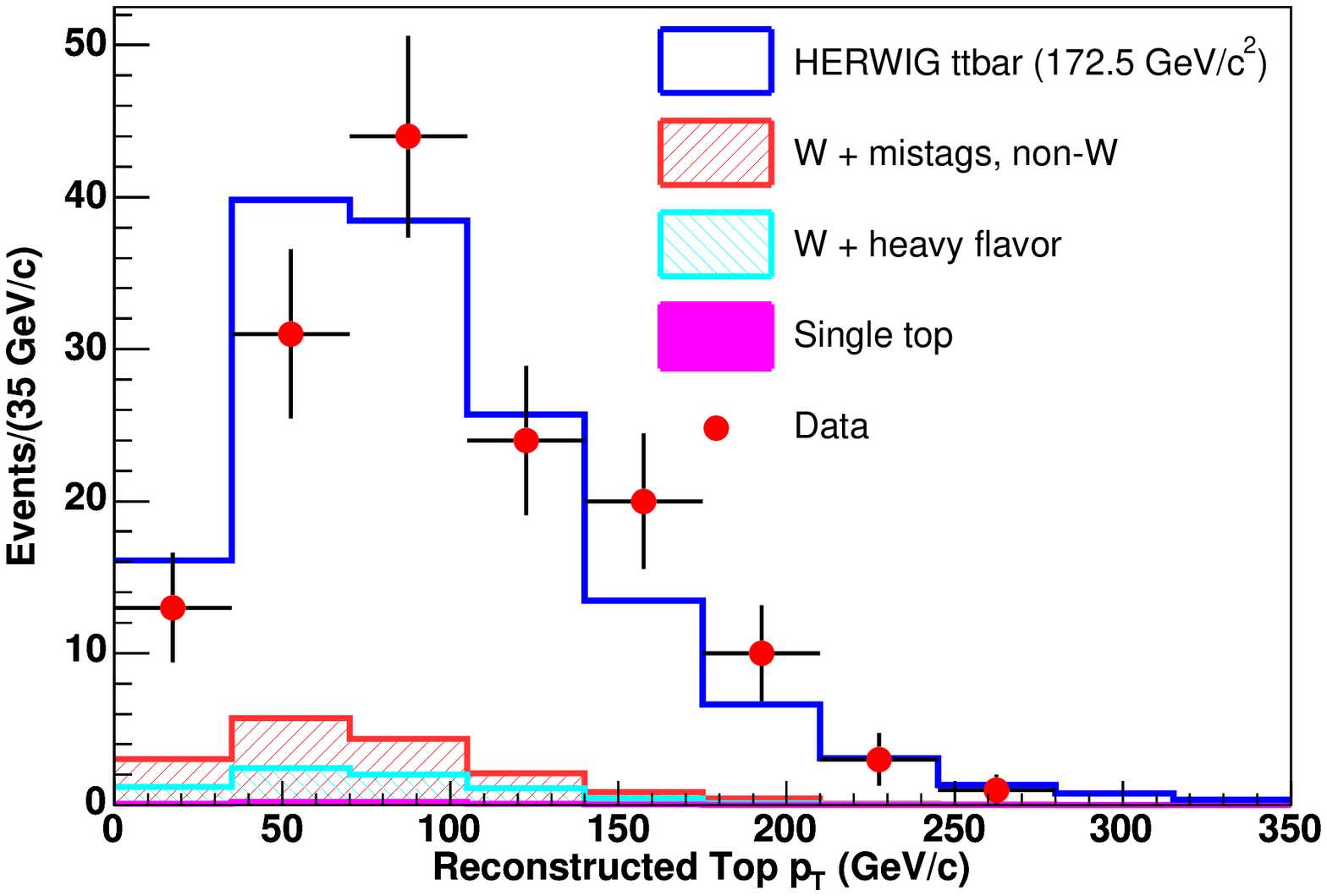}
\caption[Top \pt kinematic distribution.]
{The \pt distribution of the reconstructed top quarks
\kinblah}
\label{f:pt_top}
\end{cfigure}

\begin{cfigure}
\includegraphics{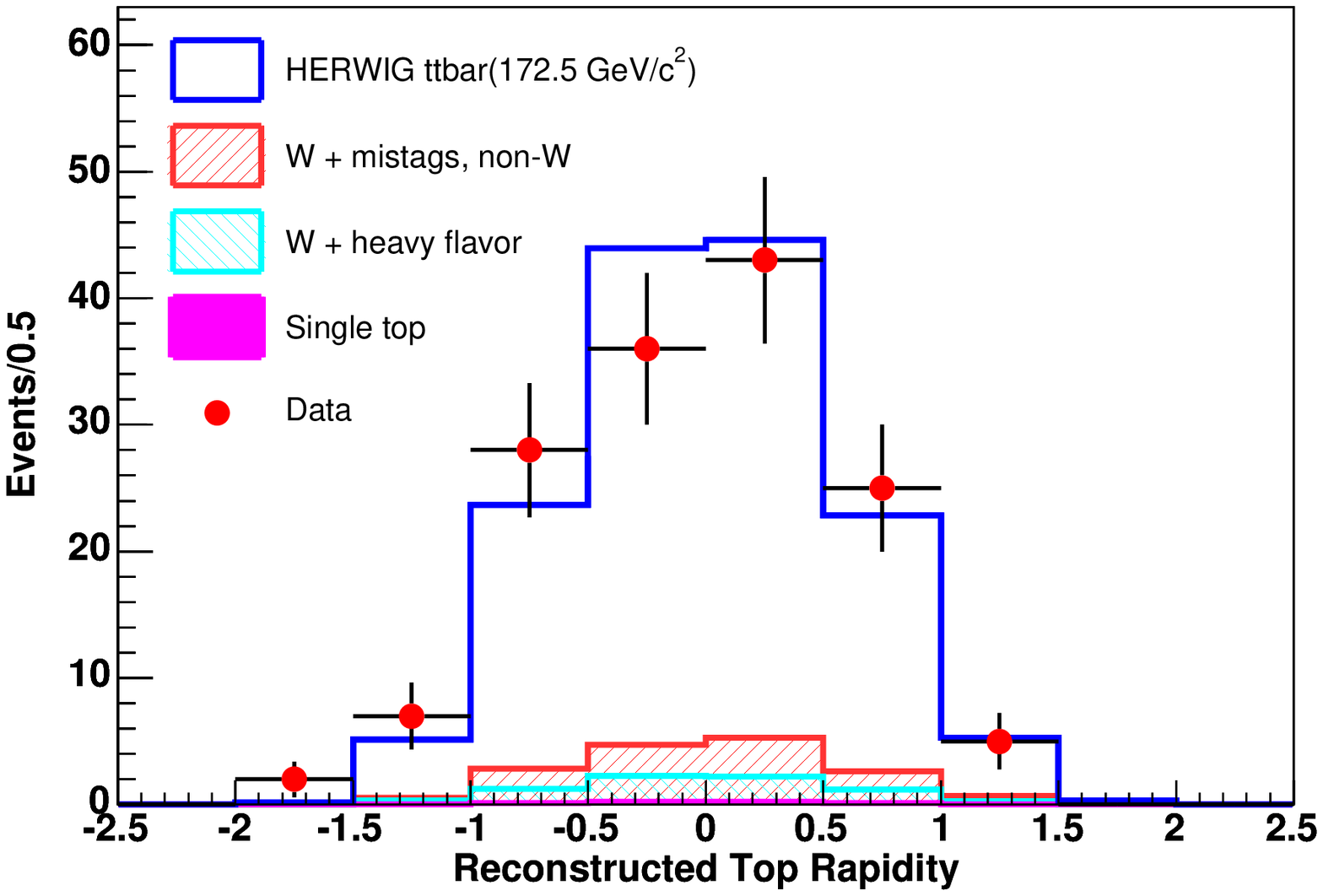}
\caption[Top $\eta$ kinematic distribution.]
{The rapidity distribution of the reconstructed top quarks
\kinblah}
\label{f:eta_top}
\end{cfigure}

We also find good agreement in the \pt distribution of the $b$ jets
from top decays, shown in \fig{f:pt_bjet}. Good modeling of the
$b$-jet spectrum by the Monte Carlo simulation is one of the most
important things for a good determination of the top quark
mass. \Fig{f:pt_w} shows the \pt distribution of the reconstructed $W$
bosons.

\begin{cfigure}
\includegraphics{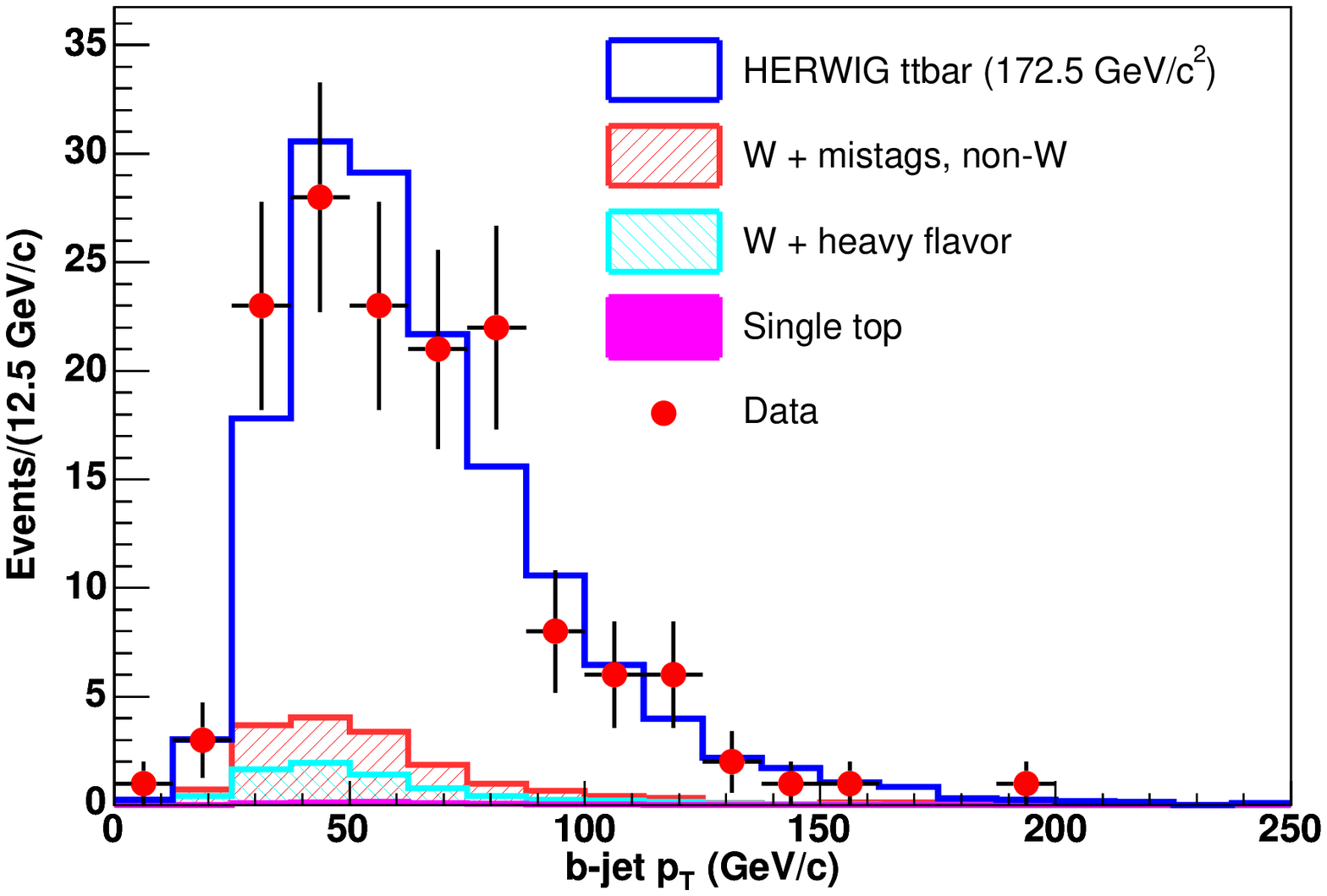}
\caption[Kinematic distribution of $b$-jet \pt.]
{The \pt distribution of the reconstructed $b$ jets
\kinblah}
\label{f:pt_bjet}
\end{cfigure}

\begin{cfigure}
\includegraphics{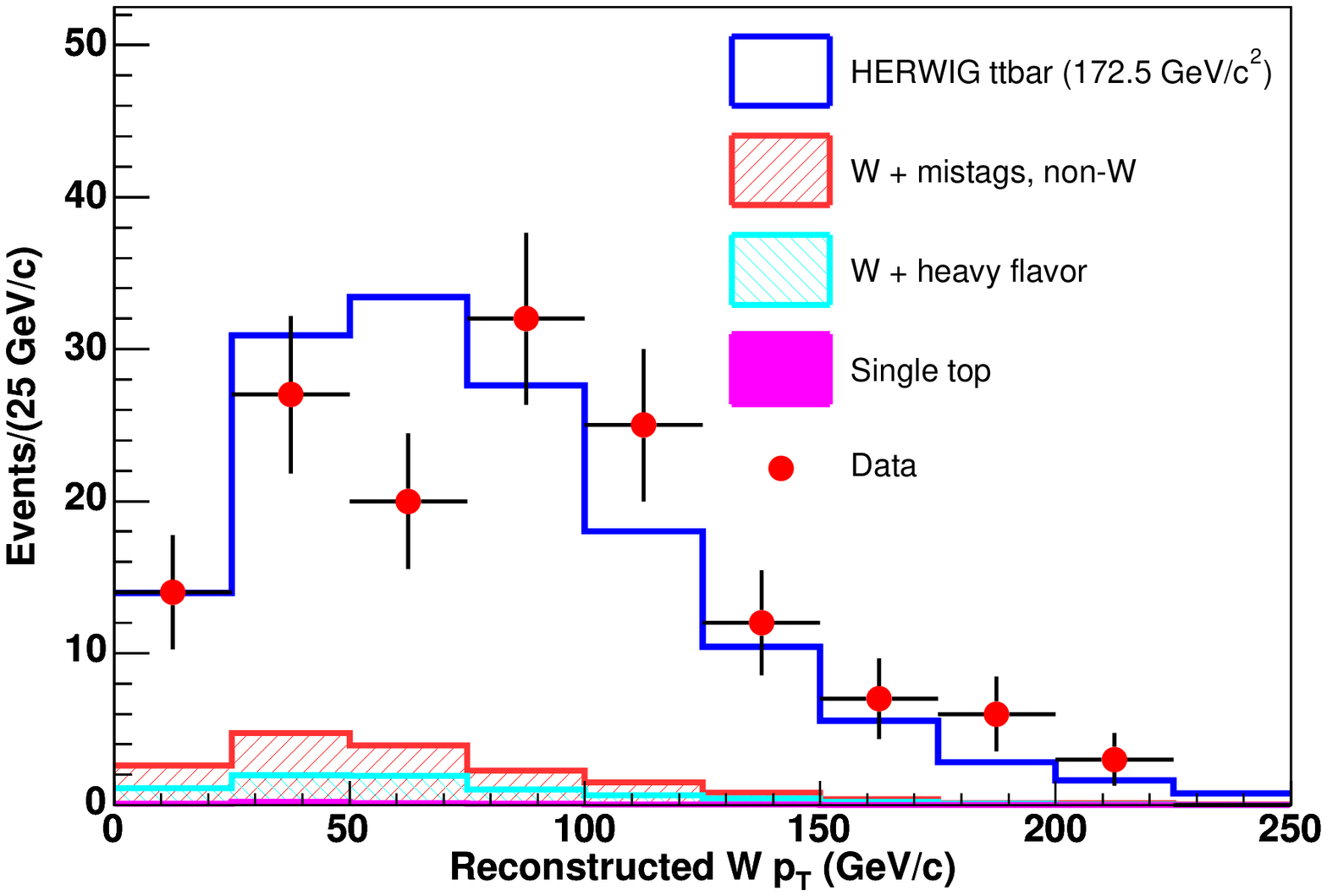}
\caption[Kinematic distribution of $W$ \pt.]
{The \pt distribution of the reconstructed $W$ bosons 
\kinblah}
\label{f:pt_w}
\end{cfigure}

The \pt distribution of the \ttbar system is shown in
\fig{f:pt_ttbar}, which has good agreement between the data and the
prediction from simulated events.  This distribution is sensitive to
the modeling of initial state radiation.  The distribution of the
number of jets from data events is also compared with the prediction
from the Monte Carlo simulation, as shown in \fig{f:njets_ttbar}. To
be counted in this plot, each jet is required to have $\et>\gev{8}$
and $|\eta|<2.0$; note that this distribution is sculpted by the
selection requirements for the \twotag and \onetagt subsamples. The
data and prediction are in good agreement, indicating that the number
of extra jets (from hard initial and final state radiation) is
reasonably well modeled by {\sc herwig}.

\begin{cfigure}
\includegraphics{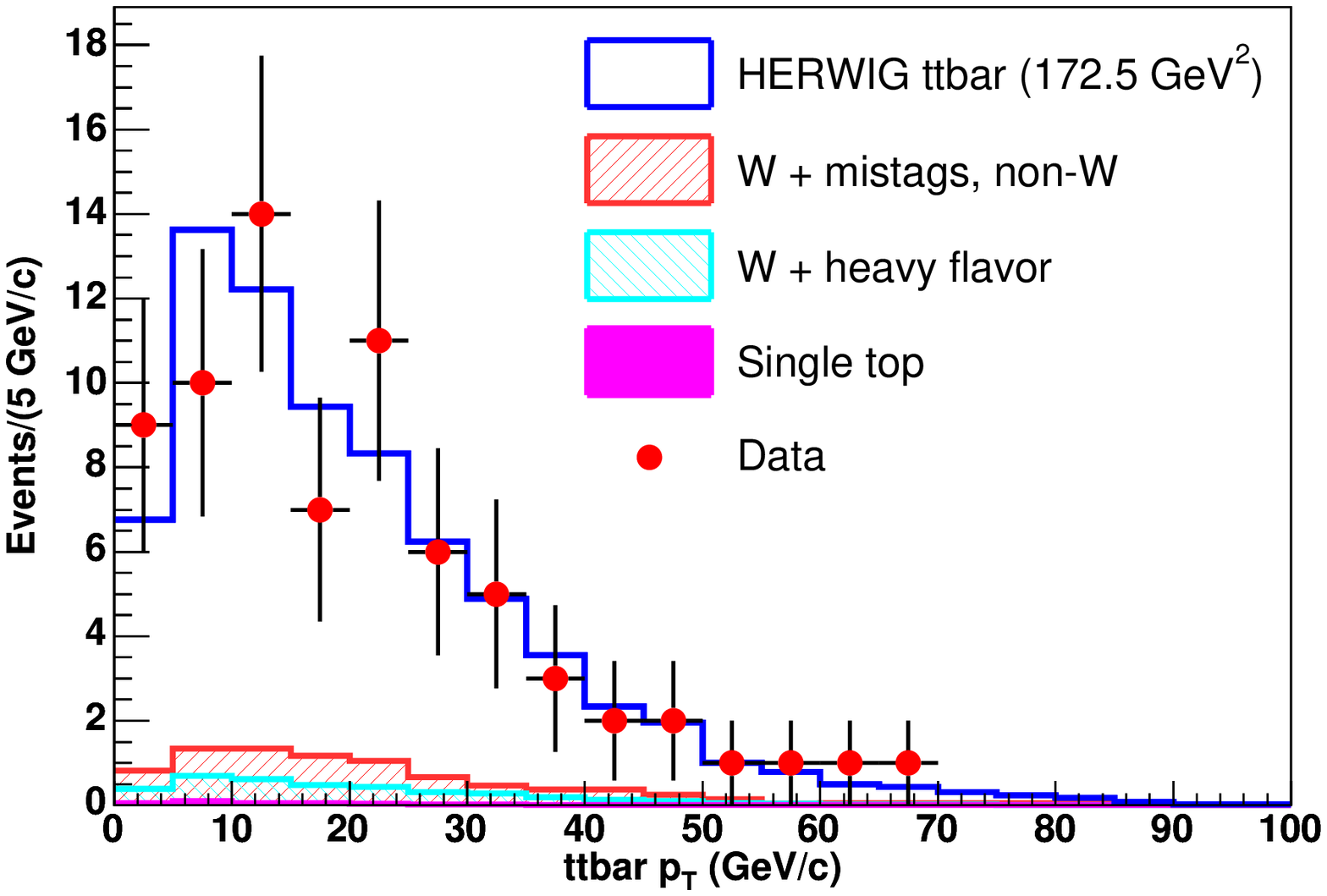}
\caption[Kinematic distribution of \ttbar \pt.]
{The \pt distribution of the reconstructed \ttbar system
\kinblah}
\label{f:pt_ttbar}
\end{cfigure}

\begin{cfigure}
\includegraphics{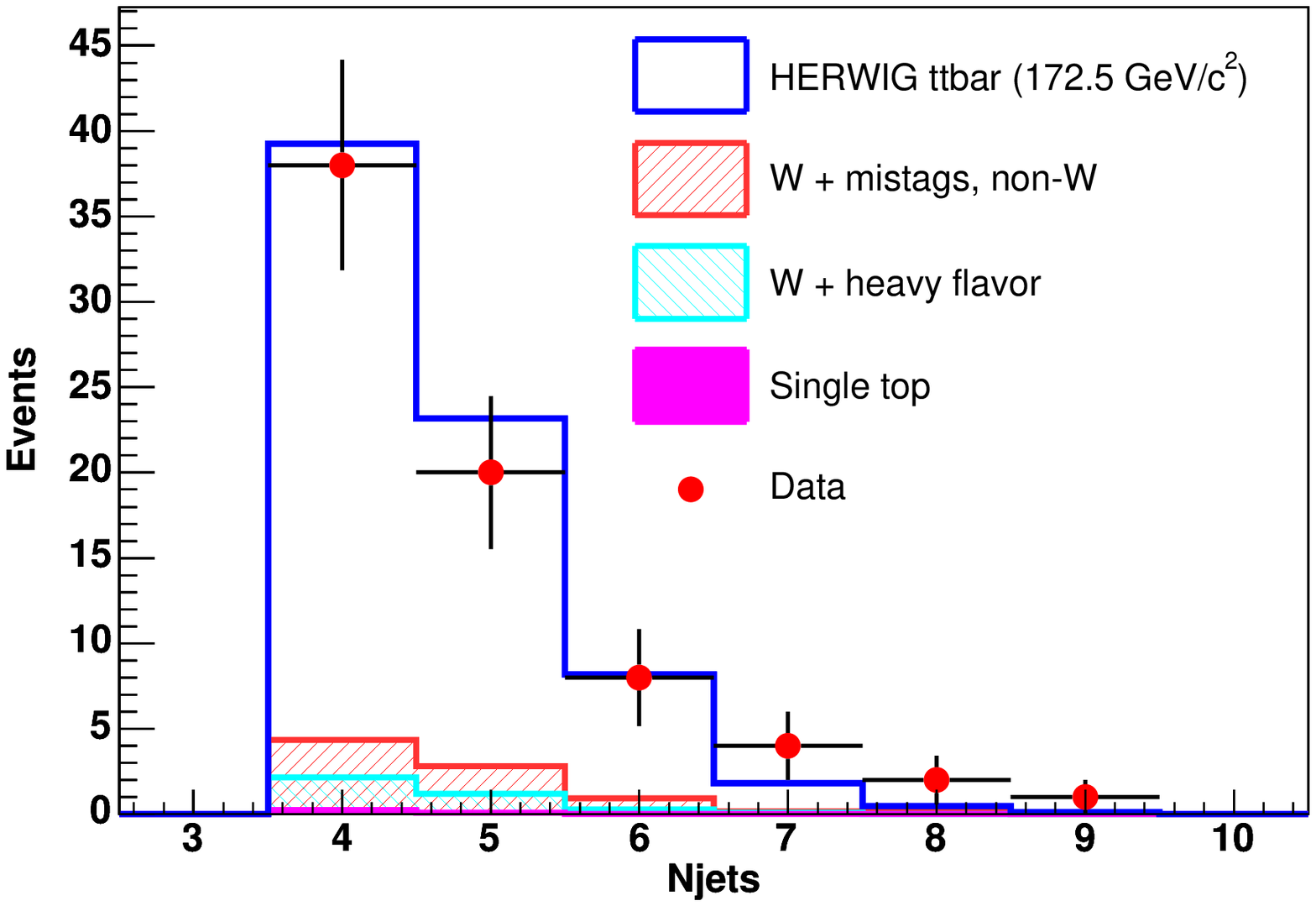}
\caption[Kinematic distribution of the number of jets.]
{The number of jets distribution 
\kinblah \
~~Jets are required to have $\et>\gev{8}$ and $|\eta|<2.0$.}
\label{f:njets_ttbar}
\end{cfigure}

\section{Systematic Uncertainties}
\label{sec:sys}

Systematic uncertainties arise from uncertainties in our understanding
of the detector response, and in the assumptions employed to infer a
top quark mass from the observed data.  The magnitudes of such
uncertainties are estimated using auxiliary data collected for this
purpose, and large samples of Monte Carlo simulated events that allow
us to estimate the sensitivity of the measurements to reasonable
variations in analysis assumptions.

For each source of systematic uncertainty, the relevant quantities or
parameters are varied by $\pm 1 \sigma$, and new \gevcc{178}~\ttbar
signal and background Monte Carlo templates are produced by performing
event selection and mass reconstruction on the modified samples.
Events for pseudo-experiments (see Section~\ref{ssec:methodCheck}) are
taken from these new templates, but the signal and background p.d.f.'s
used in the analysis remain unchanged. The shift in the median fitted
top quark mass for a large ensemble of pseudo-experiments is taken as
the systematic uncertainty associated with a given assumption or
effect.  When the uncertainty on a given systematic shift due to the
statistics of the Monte Carlo sample is larger than the shift itself,
that statistical uncertainty is used for the systematic uncertainty.

\subsection{Systematic Uncertainties Arising from the JES Calibration}
\label{ssec:jesSyst}

The use of the observed $W$ boson mass to constrain the jet energy
scale calibration essentially measures the average energy response of
light quark jets arising from the decay of the colorless $W$ boson.
However, the top quark mass also depends on the energy response to
$b$\ quark jets.  This introduces three possible sources of
uncertainty: i) uncertainties in energy response arising from
uncertainties in the decay properties of bottom quarks, ii)
uncertainties arising from the imperfect knowledge of the
fragmentation properties of bottom quarks, and iii) uncertainties in
energy response arising from the different color flow associated with
bottom quark jets produced in top quark decay.

We varied the B meson semi-leptonic branching ratios by about 10\%
of their values, corresponding to their measurement uncertainties~\cite{r_PDBook},
in our Monte Carlo models to estimate the size of this uncertainty in the
overall energy scale of the bottom quark jet. We found that this
introduced an additional uncertainty in the bottom quark jet energy
scale of 0.4\%, resulting in an uncertainty in the extraction of the
top quark mass of \gevcc{0.4}.
We used the high-statistics measurements of bottom quark fragmentation
observed in $Z\rightarrow\bbbar$ decays at the LEP and SLC colliders
to constrain the fragmentation models in our Monte Carlo calculations.
We found that this variation introduced an additional top quark mass
uncertainty of \gevcc{0.4}.  In order to test the effects of possible
variations in energy response due to different models of ``color
flow'' in the top quark production and decay, we varied the parameters
of the algorithms used to generate this color flow in both {\sc
herwig} and {\sc pythia} and conservatively estimated that this could
result in an uncertainty in the bottom quark jet energy scale of
0.3\%.  This results in an additional uncertainty in the top quark
mass of \gevcc{0.3}.

We add these three contributions in quadrature and include an
additional \gevcc{0.6} systematic uncertainty in the top quark mass
arising from the modeling of the bottom quark jets.

In this analysis, the jet energy scale is assumed to have the same
value for all jets in all events. However, the jet energy systematics
have contributions from many sources, and those component
uncertainties in general have different dependence on, for example,
jet $\pt$ and $\eta$, or on the event environment. To estimate the
uncertainty arising from the assumption of a monolithic jet energy
scale, we produce samples in which the components of the jet energy
systematics are shifted independently and in various combinations. The
typical shift in the top quark mass measurement is \gevcc{0.5}, which
is taken as the largest part of a ``method'' systematic. This
systematic includes the offset of
\sigunit{0.07} observed in the pull distributions of \fig{f:pulls}, which translates to \gevcc{0.3}.

Finally, the effect of the jet energy scale uncertainty on background
events must be treated separately, since it is not included in the
background template parameterization. We find a small uncertainty on
\mtop, \gevcc{0.04}, which we add linearly to the uncertainty due to
the overall jet energy scale since the effects are correlated.

\subsection{ISR/FSR/PDF Systematic Errors}
\label{ssec:isrFsrPdfSyst}

The systematic uncertainties due to initial state radiation, final
state radiation, and parton density functions are summarized in this
section.

Extra jets originating from the incoming partons and outgoing partons
affect the measurement of \mtop when they are misidentified as jets
from the final state partons or change the kinematics of the final
state partons. ISR and FSR are controlled by the same DGLAP evolution
equation that tells us the probability for a parton to
branch~\cite{Dokshitzer:1977sg, Gribov:1972rt, Gribov:1972ri,
Lipatov:1974qm, Altarelli:1977zs}. ISR is studied using Drell-Yan
events in dilepton channels. The advantage of Drell-Yan events is that
there is no FSR, and they are produced by the \qqbar annihilation
process, as are most ($\sim 85\%$) \ttbar pairs.

The level of ISR is measured as a function of the Drell-Yan mass scale
and shows a logarithmic dependence on the Drell-Yan mass squared, as
shown in \fig{f:isrplot}.  By extrapolation, the ISR effect is then
estimated at top pair production energies. Based on this measurement,
two ISR systematic Monte Carlo samples ($+1\sigma_{ISR}$ and
$-1\sigma_{ISR}$) are produced using {\sc pythia}, by varying the
value of $\Lambda_{QCD}$ and scale factor, $K$ to the transverse
momentum scale for ISR showering.  The parameters used are
$\Lambda_{QCD}\text{(5 flavors)}=\mev{292}$, $K=0.5$ for
$+1\sigma_{ISR}$ and $\Lambda_{QCD}\text{(5 flavors)}=\mev{73}$,
$K=2.0$ for $-1\sigma_{ISR}$.  The corresponding curves of Drell-Yan
dilepton $\left<\pt\right>$ vs invariant mass squared are shown in
\fig{f:isrplot}.  Although ISR is also sensitive to the choice of
parton distribution function (PDF), the PDF uncertainty is not
included as a part of the ISR uncertainty.  Because a PDF change
affects not only ISR but also hard scattering kinematics, the PDF
uncertainty is treated separately.  The largest top quark mass shift
between default {\sc pythia} and the two ISR samples, \gevcc{0.4}, is
taken as the ISR uncertainty.

Since ISR and FSR shower algorithms are the same, the same variations
in $\Lambda_{QCD}$ and $K$ are used to generate FSR systematic samples
by varying a set of parameters specific to FSR modeling.
The largest top quark mass shift between default {\sc pythia} 
and the two FSR samples, \gevcc{0.6}, is used as the FSR uncertainty.
We examine the effects of higher order corrections to \ttbar production
using {\sc mc@nlo}~\cite{MCNLO}, a full NLO Monte Carlo.
Based on distributions of the number of jets
and the \ttbar $p_T$, we find that NLO effects are covered
by the ISR/FSR systematics.

\begin{cfigure}
\includegraphics{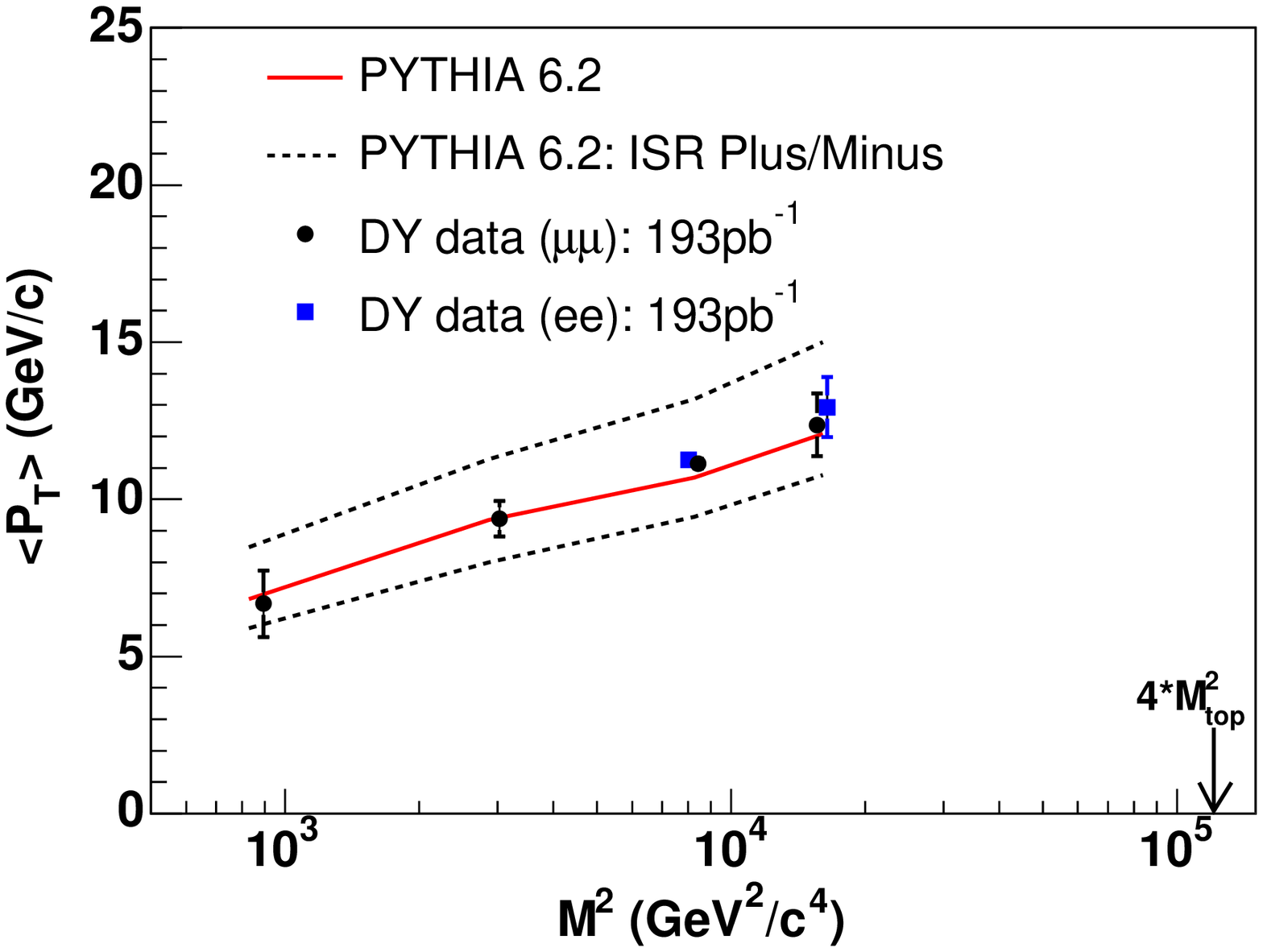}
\caption[Figure supporting ISR uncertainty procedure.]
{The average $\pt$ of the dilepton system, which corresponds to the
level of ISR activity, shows a logarithmic dependence on the dilepton
invariant mass $M_{ll}^2$. The data are compared with the predictions
of {\sc pythia} 6.2 and of the $+1\sigma_{ISR}$ and $-1\sigma_{ISR}$
samples.}
\label{f:isrplot}
\end{cfigure}

The calculation of the top quark invariant mass does not depend
directly on the choice of input PDF. However, changing the PDF changes
the top quark $\eta$ and $\pt$ distributions as well as the size of
ISR.  This results in a change in the jet $\pt$ distributions and in
the probability of selecting the correct jets, both of which affect
the reconstructed top quark mass.

To examine the systematic effect due to PDF uncertainties, 20 pairs of
uncertainty sets based on CTEQ6M are
used~\cite{Pumplin:2002vw,Stump:2003yu}.  These PDFs provide ``$\pm 1
\sigma$'' variations for 20 independent eigenvectors, but do not
include variation in $\Lambda_{QCD}$.  In addition, the MRST
group~\cite{Martin:1999ww} provides PDFs with different assumptions
for the value of $\Lambda_{QCD}$.  The difference between the measured
top quark mass using the MRST72 ($\Lambda_{QCD}=$228 MeV) and MRST75
($\Lambda_{QCD}=$300 MeV) PDFs is taken as an uncertainty, as is the
difference between leading order PDFs CTEQ5L and MRST72.  Instead of
43 different, fully simulated sets of events, a single simulated
sample is used, and mass templates are generated for the different PDF
sets by weighting events according to the probability of observing
their incoming partons using each PDF set. This technique also removes
most of the uncertainty due to limited Monte Carlo statistics.  A
symmetrized uncertainty for each of the 20 pairs of CTEQ6 PDFs
(determined by varying one eigenvector at a time) is added in
quadrature to get one part of the PDF uncerainty, \gevcc{0.20}.  An
additional systematic error of \gevcc{0.22} comes from the variation
of $\Lambda_{QCD}$. This is consistent with the much less precise
estimate using fully simulated samples. Adding a negligible
contribution from the CTEQ--MRST difference, the total PDF uncertainty
comes to \gevcc{0.3}.

In order to check the sensitivity of the top quark mass measurement
to a very different top quark \pt distribution due to a new physics
process, we have used a signal Monte Carlo sample with resonant
\ttbar production, where the resonance occurs at \gevcc{700} and
then top quarks decay according to the standard model. The measured
top quark mass is shifted by only \gevcc{1.5}, demonstrating that this
kinematic top mass fitter is nearly insensitive to the \pt of the top
quark.

\subsection{Other Systematic Errors}
\label{ssec:otherSyst}

The remaining sources of systematic uncertainty are described in this section.

% generator
The difference in the top quark mass between {\sc herwig} and {\sc
pythia} samples is $\gevcc{\measErr{0.2}{0.2}}$. To be conservative,
this difference is taken as another systematic uncertainty, although
the differences in ISR and FSR between the two generators are already
taken into account in the ISR and FSR uncertainties, and fragmentation
effects are accounted for in the jet energy uncertainties.

% background shape
The largest uncertainty in the shape of the reconstructed mass
templates for background events is due to the uncertainty in the $Q^2$
scale that is used for the calculation of the hard scattering and for
the shower evolution. Different background shapes are obtained for
four different $Q^2$ scales ($4M_W^2$, $M_W^2$, $M_W^2/4$, and $M_W^2
+ P_{TW}^2$) using {\sc alpgen} MC samples.  An {\sc alpgen} $W\bbbar$
+ 2 parton Monte Carlo sample is used for the tagged events and a $W$
+ 4 parton sample for the 0-tag and mistagged events.  Half of the
largest difference in top quark mass from pseudo-experiments using
these samples is used as the systematic uncertainty, \gevcc{0.4}.
Smaller contributions to the background shape uncertainty are
estimated by performing sets of pseudo-experiments in which background
events are drawn not from the combined background template but from
templates for one of the individual background processes, or from the
templates derived from QCD-enriched data. Half of the largest
difference observed in these pseudo-experiments is \gevcc{0.3} for the
different background processes, and \gevcc{0.1} for the different
models of the QCD background.  Both of these are taken as additional
systematic uncertainties on the top quark mass due to background shape
modeling.
 
%b tagging
Different $b$-tagging efficiency in data and simulation can introduce
a bias in the top quark mass measurement. The $\etjet$ dependence of
the $b$ tagging in data and simulation agree very well.  But if a
slope on the $\etjet$ dependence (consistent at \sigunit{1} with the
measurement) is introduced in the tagging efficiency, the shift in the
top quark mass is \gevcc{0.1}, which is taken as a systematic
uncertainty.
  
%MC statistics
The analysis can have a systematic bias due to the finite statistics
of Monte Carlo samples that are used to obtain the signal and
background shape parameterizations. For a rough estimate of this
uncertainty, sets of pseudo-experiments are performed with a series of
fluctuated signal and background templates; in each fluctuated
template, each bin is varied randomly according to Poisson statistics.
For each fluctuated template, the median top quark mass measured by
pseudo-experiments is shifted. The typical shift due to these
statistical fluctuations, taken as a systematic uncertainty due to
Monte Carlo statistics, is \gevcc{0.3}.

\subsection{Jet Systematic Errors}
\label{ssec:jetSystOnMtop}

The systematics on jet energy measurements are described in detail in
Section~\ref{ssec:jetSyst}. The primary analysis fits for the jet
energy scale, and the error from the likelihood fit includes a
contribution due to these systematics. For the \mtop-only fits,
however, this systematic uncertainty must be estimated independently.

To determine the systematic uncertainty on the top quark mass
measurement given the various sources of uncertainty on the jet energy
measurements, the mass shifts for $+1\sigma$ and $-1\sigma$
perturbations in the jet energies are extracted, and a symmetric
uncertainty for each source is defined as half the difference between
the two shifts. \Tab{t:jetenergysyst} lists the uncertainties obtained
for the \mtop-only measurement. The total systematic uncertainty in
the top quark mass due to jet energy measurements is \gevcc{3.1} for
the combined measurement. The corresponding systematic uncertainties
for an independent measurement in each subsample are listed for
comparison.

\begin{table}
\caption[Systematic uncertainties due to jet energy systematics.]
{The uncertainties on the \mtop-only top quark mass measurement are
shown for each jet energy systematic error. Estimates are obtained for
the independent subsamples as well as for the combined measurement.}
\label{t:jetenergysyst}
\begin{ruledtabular}
\begin{tabular}{ddddd}
\multicolumn{5}{l}{\textbf{Jet energy systematic}} \\
\multicolumn{5}{c}{$\Delta\mtop$ (\gevccnoarg)} \\
\multicolumn{1}{c}{\twotag} & \multicolumn{1}{c}{\onetagt}
& \multicolumn{1}{c}{\onetagl} & \multicolumn{1}{c}{\zerotag}
& \multicolumn{1}{c}{Combined} \\
\hline
\multicolumn{5}{l}{\textbf{Response relative to central}} \\
0.6 & 0.6 & 0.6 & 1.0 & 0.6 \\
\hline
\multicolumn{5}{l}{\textbf{Modeling hadron jets (absolute scale)}} \\
2.0 & 2.3 & 2.0 & 2.7 & 2.2 \\
\hline
\multicolumn{5}{l}{\textbf{Modeling out-of-cone energy and underlying event}} \\
2.2 & 2.2 & 1.9 & 1.9 & 2.1 \\
\hline\hline
\multicolumn{5}{l}{\textbf{Total systematic due to jet energies}} \\
3.0 & 3.2 & 2.8 & 3.4 & 3.1 \\
\end{tabular}
\end{ruledtabular}
\end{table}

\subsection{Total Systematic Uncertainty}
\label{ssec:totalSyst}

The systematic uncertainties for the combined fit are listed in
\tab{t:uncertainties}.  The total systematic uncertainty is estimated
to be \gevcc{1.3}, exclusive of the uncertainty due to jet energy
scale that is included in the likelihood error.  Also shown in
\tab{t:uncertainties} are the systematic uncertainties on the \jes
measurement (\sigcunit{0.33} total), and the systematic uncertainties
on \mtop for the \mtop-only measurement without (\gevcc{3.3} total)
and with (\gevcc{3.2} total) JPB tags.

\begin{table}
\caption[Systematic uncertainty summary.]
{This table summarizes all systematic uncertainties for the combined
analysis and two alternate fits.}
\label{t:uncertainties}
\begin{ruledtabular}
\begin{tabular}{ldddd}
Method &
\multicolumn{2}{c}{Primary} &
\multicolumn{1}{c}{\mtop-only} &
\multicolumn{1}{c}{\mtop-only} \\
\multicolumn{4}{c}{} &
\multicolumn{1}{c}{+ JPB} \\
&
\multicolumn{1}{c}{$\Delta\mtop$} &
\multicolumn{1}{c}{$\Delta\jes$} &
\multicolumn{1}{c}{$\Delta\mtop$} &
\multicolumn{1}{c}{$\Delta\mtop$} \\
&
\multicolumn{1}{c}{(\gevccnoarg)} &
\multicolumn{1}{c}{($\sigma_c$)} &
\multicolumn{1}{c}{(\gevccnoarg)} &
\multicolumn{1}{c}{(\gevccnoarg)} \\
\hline
Jet Energy 		  & \multicolumn{1}{c}{N/A} & \multicolumn{1}{c}{N/A}  & 3.1 & 3.0 \\
$b$-jet Energy  & 0.6 & 0.25 & 0.6 & 0.6 \\
Method          & 0.5 & 0.02 & \multicolumn{1}{c}{N/A} & \multicolumn{1}{c}{N/A} \\
ISR 			      & 0.4 & 0.08 & 0.4 & 0.3 \\
FSR 			      & 0.6 & 0.06 & 0.4 & 0.6 \\
PDFs 			      & 0.3 & 0.04 & 0.4 & 0.4 \\
Generators 		  & 0.2 & 0.15 & 0.3 & 0.2 \\
Bkgd Shape      & 0.5 & 0.08 & 0.5 & 0.5 \\
$b$ tagging 		& 0.1 & 0.01 & 0.2 & 0.3 \\
MC stats        & 0.3 & 0.05 & 0.4 & 0.4 \\
\hline
Total 			    & 1.3 & 0.33 & 3.3 & 3.2 \\
\end{tabular}
\end{ruledtabular}
\end{table}

\section{Conclusion}
\label{sec:conc}

We have made a new measurement of the top quark mass,
\begin{eqnarray}
&\gevcc{\measAStatJESSyst{173.5}{3.7}{3.6}{1.3}}&\\ \nonumber
&=\gevcc{\measAErr{173.5}{3.9}{3.8}},&
\end{eqnarray} 
using a novel technique that utilizes the jet energy scale information
provided by the hadronically decaying $W$ boson in the top quark
events.  This new top quark mass measurement provides the most precise
single measurement on this important physical parameter.  We have
performed a cross-check of this result using a more traditional fit
that does not use the \emph{in situ} jet energy scale information, and
found excellent agreement in the central value of the top quark mass:
\gevcc{\measAStatSyst{173.2}{2.9}{2.8}{3.3}}.  Finally, by adding an
algorithm to increase the number of tagged $b$ jets, we measure:
\gevcc{\measAStatSyst{173.0}{2.9}{2.8}{3.2}}.

This measurement is part of a rich top physics program at CDF.  As the
luminosity acquired increases from the current \invpb{318} to an
expected 4000--7000 $\invpbnoarg$ for \runii, the statistical
uncertainty on the top quark mass will improve.  Using our technique,
the dominant systematic uncertainty on the measurement, associated with the jet energy scale, will also be reduced with more
data. As we approach total uncertainties of approximately \gevcc{2.0}, the
uncertainties due to initial and final state radiation, as well as the
bottom quark jet energy scale, become comparable to the statistical
uncertainties associated with the top quark mass and jet energy scale
measurement. We expect that these other systematics can
also be improved with more work and more data in the relevant control
samples.  Additional top quark mass results from CDF are expected in
the near future.  We expect that these will continue to provide
important inputs into our understanding of the fundamental fermions
and the nature of the electroweak interaction.

\begin{acknowledgments}
We thank the Fermilab staff and the technical staffs of the
participating institutions for their vital contributions. This work
was supported by the U.S. Department of Energy and National Science
Foundation; the Italian Istituto Nazionale di Fisica Nucleare; the
Ministry of Education, Culture, Sports, Science and Technology of
Japan; the Natural Sciences and Engineering Research Council of
Canada; the National Science Council of the Republic of China; the
Swiss National Science Foundation; the A.P. Sloan Foundation; the
Bundesministerium f\"ur Bildung und Forschung, Germany; the Korean
Science and Engineering Foundation and the Korean Research Foundation;
the Particle Physics and Astronomy Research Council and the Royal
Society, UK; the Russian Foundation for Basic Research; the Comisi\'on
Interministerial de Ciencia y Tecnolog\'{\i}a, Spain; in part by the
European Community's Human Potential Programme under contract
HPRN-CT-2002-00292; and the Academy of Finland.
\end{acknowledgments}

\bibliographystyle{apsrev}
\bibliography{topMassPRD}

\end{document}